\documentclass[twocolumn,floatfix,aps,prd,showpacs]{revtex4}

\usepackage{graphicx}
\usepackage{bm}
\usepackage{epsf}

\newcommand{\beq}{\begin{equation}}
\newcommand{\eeq}{\end{equation}}
\newcommand{\beqn}{\begin{eqnarray}}
\newcommand{\eeqn}{\end{eqnarray}}
\newcommand{\pa}{\partial}

\newcommand{\varep}{\varepsilon}

\begin{document}


\title{Simulating coalescing compact binaries by a new code SACRA}

\author{Tetsuro Yamamoto, Masaru Shibata}
\affiliation{Graduate School of Arts and Sciences, University of
Tokyo, Komaba, Meguro, Tokyo 153-8902, Japan}
\author{Keisuke Taniguchi}
\affiliation{Department of Physics, University of Illinois at
Urbana-Champaign, Illinois 61801, USA}

\begin{abstract}
We report our new code (named SACRA) for numerical relativity
simulations in which an adaptive mesh refinement algorithm is
implemented. In this code, the Einstein equations are solved in the
BSSN formalism with a fourth-order finite differencing, and the
hydrodynamic equations are solved by a third-order high-resolution
central scheme.  The fourth-order Runge-Kutta scheme is adopted for
integration in time. To test the code, simulations for coalescence of
black hole-black hole (BH-BH), neutron star-neutron star (NS-NS), and
black hole-neutron star (BH-NS) binaries are performed, and also,
properties of BHs formed after the merger and gravitational waveforms
are compared among those three cases. For the simulations of BH-BH
binaries, we adopt the same initial conditions as those by Buonanno et
al.  [Phys. Rev.  D {\bf 75}, 124018 (2007)] and compare numerical
results. We find reasonable agreement except for a slight disagreement
possibly associated with the difference in choice of gauge conditions
and numerical schemes. For an NS-NS binary, we performed simulations
employing both SACRA and Shibata's previous code, and find reasonable
agreement between two numerical results for the final outcome and
qualitative property of gravitational waveforms.  We also find that
the convergence is relatively slow for numerical results of NS-NS
binaries, and again realize that longterm numerical simulations with
several resolutions and grid settings are required for validating the
results.  For a BH-NS binary, we compare numerical results with our
previous ones, and find that gravitational waveforms and properties of
the BH formed after the merger agree well with those of our previous
ones, although the disk mass formed after the merger is less than
0.1\% of the total rest mass, which disagrees with the previous
result. We also report numerical results of a longterm simulation
(with $\sim 4$ orbits) for a BH-NS binary for the first time. All
these numerical results show behavior of convergence, and extrapolated
numerical results for time spent in the inspiral phase agree with
post-Newtonian predictions in a reasonable accuracy. These facts
validate the results by SACRA.
\end{abstract}

\maketitle

\section{Introduction} \label{sec:intro}

Coalescence of binary compact objects such as binaries of two neutron
stars (NS-NS), black hole and neutron star (BH-NS), and two black
holes (BH-BH) is the most promising source for kilometer-size
laser-interferometric
gravitational wave detectors such as LIGO, VIRGO,
and LCGT. To detect gravitational waves and to analyze the gravitational
wave signals for extracting physical information of the sources, it is
necessary to prepare theoretical templates of gravitational waves from
the coalescing compact binaries. Motivated by this fact, significant
effort has been paid in the past two decades. For theoretically
computing gravitational waveforms in a relatively early inspiral
phase, post-Newtonian approximations are the robust approach
\cite{Blanchet}. On the other hand, for studying the last inspiral
and merger phases of the coalescing binaries in which general
relativistic effects are significantly strong and any approximation
breaks down, numerical relativity is the unique approach.

In the past decade, in particular in the past three years, a wide
variety of general relativistic simulations have been performed for
the coalescence of NS-NS binaries
\cite{NSNS1,NSNS2,NSNS3,NSNS4,NSNS9,NSNS5,NSNS6,NSNS7,NSNS8} and BH-BH
binaries
\cite{BHBH1,BHBH2,BHBH3,BHBH4,BHBH5,BHBH6,BHBH7,BHBH8,BHBH9,BHBH10,BHBH11,BHBH12,BHBH13,BHBH14,BHBH15,BHBH155,BHBH16}
(see also early-stage results for merger of BH-NS binaries
\cite{BHNS1,BHNS2,BHNS3}).  Since 1999, a variety of simulations have
been performed for the inspiral and merger of NS-NS binaries after the
first success of Shibata and Ury\=u \cite{NSNS1}. Shibata, Ury\=u, and
Taniguchi have then performed simulations focusing mainly on the
merger process and the final fate \cite{NSNS2}. Their simulations were
done for a variety of equations of state (EOSs) as well as for a wide
range of mass of two NSs. They have clarified that the final outcome
of the merger (formation of a BH or a hypermassive neutron star;
hereafter HMNS) depends strongly on the total mass of the system and
on the chosen EOSs. In the latest paper \cite{NSNS3}, they clarified
that with stiff EOSs such as Akmal-Pandharipande-Ravenhall one
\cite{APR}, a BH is not promptly formed even for a system of the total
mass $\sim 2.8M_{\odot}$, but an HMNS is a likely outcome. They also
indicated that the formed HMNSs have an elliptical shape because of
their rapid rotation, and hence, quasiperiodic gravitational waves of
frequency $\sim 3$--4 kHz will be emitted for a long time (for $\sim
100$ cycles) in the absence of dissipative mechanisms except for
gravitational wave emission.  The integrated effective amplitude of
such gravitational waves may be large enough to be detected by
advanced laserinterferometric gravitational wave detectors
\cite{NSNS2,NSNSS}. In the last couple of years, longterm simulations
for the inspiral of NS-NS binaries have been also done. In particular,
in the latest simulations, 3--5 inspiral orbits are stably followed
\cite{NSNS5,NSNS6,NSNS7,NSNS8}, and also, the computations are
continued until the system settles down approximately to a stationary
state even in the case that a BH is the final outcome. Preliminary
simulations for merger of magnetized NSs have been also performed
recently \cite{NSNS6,NSNS9} (although it is not clear whether or not
many of crucial magnetohydrodynamic instabilities are resolved in
these simulations). However, in most of these works, very simple
$\Gamma$-law EOSs is adopted for modeling the NSs, and hence,
realistic simulations with a variety of realistic EOSs have not been
done yet.

The last three years have also witnessed great progress in simulations
of BH-BH binaries, starting with the first stable simulation of
orbiting and merging BHs by Pretorius \cite{BHBH1} and development of
the moving puncture approach \cite{BHBH2,BHBH3} in 2005. Since then, a
large number of simulations have been done on the late inspiral and
merger of BH-BH binaries
\cite{BHBH1,BHBH2,BHBH3,BHBH4,BHBH5,BHBH6,BHBH7,BHBH8,BHBH9,BHBH10,BHBH11,BHBH12,BHBH13,BHBH14,BHBH15,BHBH155,BHBH16}. These
works have clarified that the merger waveforms are universally
characterized by a quasi-normal mode ring-down.  They have also shown
that a large kick velocity is excited at the merger in the cases that
the masses of two BHs are not equal and/or the spin and orbital
angular momentum vectors misalign.  The latest works with a high accuracy
\cite{BHBH12,BHBH13,BHBH14,BHBH15,BHBH155,BHBH16} compare the
numerical gravitational waveforms with post-Newtonian ones and assess
the accuracy of the post-Newtonian waveforms \cite{Blanchet}.  In
particular, the numerical simulation of Ref.~\cite{BHBH16} presents 
highly accurate gravitational waves, which assess the
accuracy of the post-Newtonian gravitational waves with a level much
beyond the previous analysis. They clarify that the so-called Taylor
T4 post-Newtonian gravitational waveforms are very accurate at least
up to the last two orbits before the merger for the equal-mass,
nonspinning BH-BH binaries. This work shows a monumental achievement
of numerical relativity because it demonstrates that numerical
relativity could provide inspiral waveforms for BH-BH binaries more
accurate than the post-Newtonian waveforms.

However, simulations for coalescing compact binaries have been
performed only for a restricted parameter space. Because the ultimate
goal is to prepare a template family which covers gravitational
waveforms for almost all the possible parameters for binary compact
objects, the present status is regarded as a preliminary one from the
view point of gravitational wave astronomy. For example, for BH-BH
binaries, the simulations have been primarily performed for the case
that the spin vector of BHs aligns with the orbital angular momentum
vector and the magnitude of the BH spin is not extremely large. The
simulations for BH-BH binaries of unequal-mass and misaligned spin
have been also performed only for the restricted cases. For NS-NS
binaries, the simulations have been also primarily performed for the
case that masses of two NSs are equal, and the cases of unequal-mass
have been investigated in a small mass range.  Moreover, the
simulations have been performed adapting a few EOSs, mostly a simple
$\Gamma$-law EOS. Because the EOS of NSs is still unknown, it is
necessary to perform simulations choosing a wide variety of EOSs.

To perform a number of simulations for various parameters of compact
objects, an efficient scheme for the numerical simulation is
necessary. For the two-body problem considered here, adaptive
mesh-refinement (AMR) algorithm is well-suited for this purpose
\cite{BO}.  The reason is described as follows: In the two-body
problem, there are three characteristic length scales; the radius of
compact objects, $R$, the orbital separation, $r$, and the gravitational wave
length, $\lambda \approx \pi(r^3/M)^{1/2}$ where $M$ is the total mass
of the system. We have to accurately resolve these three scales. These
scales obey the relation $R < r < \lambda$, and typically, $R \ll
\lambda$. Thus, an issue to be resolved in this problem is to assign
an appropriate resolution for each scale of significantly different
magnitude. To resolve each compact object accurately, the grid spacing
$\Delta x$ in its vicinity has to be much smaller than $R$ ($R/\Delta
x$ should be larger than $\sim 20$). On the other hand, gravitational
waves have to be extracted from the geometric variables in the wave
zone. This implies that the size of the computational region should be
larger than $\lambda$. By simply using a uniform grid, the required
grid number in one direction is $N_g=2 \lambda/\Delta x$ where the
factor 2 comes from the fact that there are plus and minus directions
in each axis. Because of the facts $r \agt 2R$ and $R >M$, the
required value of $N_g$ is larger than several hundreds. To follow the
binary inspiral from $r \sim 5R$, $N_g$ has to be larger than $10^3$.
Even by supercomputers currently available for the general users, it
approximately takes at least a month to perform a simulation of such a
huge grid number. This implies that it is not feasible to perform a
number of simulations for a wide variety of the parameters. 

In the AMR algorithms, one can change the grid spacing and the grid
structure arbitrarily for different scales, preserving the required
grid-resolution for each scale. To accurately resolve each star in a
binary, we need to take $N_g \sim 2 R/\Delta x \sim 100$ to cover the
region in the vicinity of the compact stars. However for other region,
we do not have to take such a small grid spacing. In particular, we
can save the grid number in the distance zone. To follow the
propagation of gravitational waves in the wave zone, the required grid
spacing is $\sim 0.05$--$0.1\lambda$ which is larger than $\Delta x$
by an order of magnitude. Thus, by choosing such a large grid spacing
(and correspondingly, a large time step) in the wave zone, we can
significantly save the grid number for covering the large
computational region as well as computational costs.  Due to this
reason, the AMR algorithms are employed by many numerical relativity
groups now (e.g., \cite{NSNS5,NSNS6,BHBH1,BHBH6}), which have provided
a variety of numerical results recently.

Motivated by the facts mentioned above, we have developed a new code
in which an AMR algorithm is implemented, named SACRA ({\bf S}imul{\bf
A}tor for {\bf C}ompact objects in {\bf R}elativistic {\bf
A}strophysics)
\footnote{SACRA is named after Sakura in Japanese (cherry blossom in
English)}. This code can evolve not only BH-BH binaries but also NS-NS
and BH-NS binaries with a variety of EOSs. In SACRA, the
Einstein equations are solved in a similar AMR technique to that adopted
in Ref.~\cite{BHBH6}. Namely, we adopt a fourth-order finite differencing
scheme for spatial derivatives and a fourth-order Runge-Kutta scheme for
integration forward in time. For the AMR algorithm, six buffer zones
are prepared at the refinement boundaries and for the interpolation at
the refinement boundaries, fifth-order Lagrangian interpolation scheme
in space and second-order Lagrangian interpolation scheme in time are
adopted. For simplicity, the size and the grid spacing of
computational domain for each refinement level are fixed, although the
computational domain can move with the compact objects. We find that
this scheme is so stable that we do not have to introduce the
Kreiss-Oliger-type dissipation which is often necessary in some AMR codes.
For solving the hydrodynamic equations, we adopt a high-resolution
central scheme proposed by Kurganov and Tadmor \cite{KT} with a
third-order interpolation for reconstructing the fluid flux at cell
interfaces.  For implementing the AMR algorithm, six buffer zones are
also prepared as in the gravitational field. Fifth-order and
second-order Lagrangian interpolations are basically adopted in space
and in time, respectively, although a limiter function is applied in
the time interpolation for a region where fluid variables vary
steeply. We also find that with this scheme, a stable longterm 
evolution is feasible for NS-NS and BH-NS binaries.

The paper is organized as follows. In Sec.~\ref{sec:form}, we briefly
describe the basic equations, the gauge conditions, the methods for
extracting gravitational waves, and the quantities used in the
analysis for the numerical results.  We describe an AMR scheme which
we employ in SACRA in Sec.~\ref{sec:AMR}.  In Sec.~\ref{sec:res},
numerical results for the simulation of BH-BH, NS-NS, and BH-NS
binaries are presented separately. The simulations were performed for
a variety of grid resolutions and grid structures.  Convergence of
numerical results shows validity of our code. Section
\ref{sec:summary} is devoted to a summary. Throughout this paper, we
adopt the geometrical units in which $G=c=1$ where $G$ and $c$ are the
gravitational constant and the speed of light.  Latin and Greek
indices denote spatial components ($x, y, z$) and space-time
components ($t, x, y, z$), respectively: $r \equiv
\sqrt{x^2+y^2+z^2}$. $\delta_{ij}(=\delta^{ij})$ denotes the Kronecker
delta.

\section{Formulation} \label{sec:form}

\subsection{Brief Review of Basic Equations}

The fundamental variables for geometry in 3+1 decomposition are
$\alpha$: the lapse function, $\beta^k$: the shift vector, $\gamma_{ij}$:
the metric in a three-dimensional spatial hypersurface, and $K_{ij}$:
the extrinsic curvature. We solve the Einstein evolution equations using a
slightly modified version of the BSSN 
(Baumgarte-Shapiro-Shibata-Nakamura) formalism \cite{BSSN}. In the
original version of the BSSN formalism, one chooses the variables to
be evolved as
\beqn
&& \tilde \gamma_{ij}=e^{-4\phi}\gamma_{ij}, \\
&& \tilde A_{ij}=e^{-4\phi}\Big(K_{ij}-{1\over 3}\gamma_{ij}K\Big), \\
&& \phi={1 \over 12}\ln[{\rm tr}(\gamma_{ij})],\\
&& K = K_{k}^{~k},\\
&& F_i = \delta^{jk}\pa_{j} \tilde \gamma_{ik}~~
{\rm or}~~ \tilde \Gamma^i=-\tilde \gamma^{ij}_{~~,j}.
\eeqn
Note that the condition det$(\tilde \gamma_{ij})=1$ has to be satisfied
(we assume to use Cartesian coordinates). 
In the present approach, we also evolve $\tilde \gamma_{ij}$, $\tilde
A_{ij}$, $K$, and $F_i$ or $\tilde \Gamma^i$, whereas instead of $\phi$, we
evolve $W \equiv e^{-2\phi}$ following Ref.~\cite{BHBH155}.
The primary reason is that we adopt the grid-center-grid in
numerical simulation; when center of a BH is located approximately at a grid
point, $\phi$ becomes too large to compute accurately. With the choice
of $W$, such pathology can be avoided, as first pointed out by
Campanelli et al. \cite{BHBH2}. In this formalism, 
the Ricci tensor with respect to $\gamma_{ij}$ is written as 
\beqn
R_{ij}=\tilde R_{ij} + R_{ij}^W, 
\eeqn
where $\tilde R_{ij}$ is the Ricci tensor with respect to 
$\tilde \gamma_{ij}$ and 
\beqn
&&R_{ij}^W={1 \over W}\tilde D_i \tilde D_j W \nonumber \\
&&~~~~~~+ \tilde \gamma_{ij} \biggl({1\over W} \tilde D_k \tilde D^k W 
-{2 \over W^2} \tilde D_k W \tilde D^k W \biggr),
\eeqn
Here, $\tilde D_i$ is the covariant derivative with respect to 
$\tilde \gamma_{ij}$.
Merits of using $W$ instead of $\chi=e^{-4\phi}$ proposed in
\cite{BHBH2} are that (i) the equation for $R_{ij}^W$ is slightly
simplified and (ii) even for $W \rightarrow 0$,
no singular term appears in the basic equation in which $R_{ij}^W$
always appears in the form of $W^2 R_{ij}^W$. 

In SACRA, we implement both equations for $F_i$ and $\tilde \Gamma^i$.  As we
show in Sec. \ref{sec:res}, numerical results do not depend strongly
on the choice of the variables. 

For the condition of the lapse function $\alpha$ and the shift vector
$\beta^i$, we adopt dynamical gauge conditions. For the case
that we adopt the Shibata-Nakamura-type BSSN formalism (hereafter
$F_i$-BSSN formalism), the gauge equations adopted are \cite{BHNS1}
\beqn
&&(\pa_t -\beta^i \pa_i)\alpha = -2 \alpha K,
\label{lapse} \\
&&\pa_t \beta^i=0.75 \tilde \gamma^{ij} (F_j +\Delta t \pa_t F_j).
\label{shift} 
\eeqn
Here, $\Delta t$ denotes a time step in numerical simulation and the
second term on the right-hand side of Eq. (\ref{shift}) is introduced
for stabilization of numerical computation.
For the Baumgarte-Shapiro-type BSSN formalism (hereafter the
$\tilde \Gamma^i$-BSSN
formalism), we also employ Eq. (\ref{lapse}) for evolution of $\alpha$,
whereas for $\beta^k$, we adopt the so-called $\Gamma$-freezing gauge \cite{AB} 
\beqn
&&(\pa_t -\beta^j\pa_j) \beta^i=0.75B^i, \\
&&(\pa_t -\beta^j\pa_j) B^i =(\pa_t -\beta^j\pa_j) \tilde \Gamma^i
-\eta_s B^i,
\label{shift2} 
\eeqn
where $B^i$ is an auxiliary variable and $\eta_s$ is an arbitrary
constant.  In the present work, we basically choose $\eta_s \approx
1/m$ for BH-BH and BH-NS binaries, and $\approx 3/m$ for NS-NS
binaries. Here, $m$ denotes the irreducible mass for a BH and
mass in the case of isolation for an NS. As shown in
Ref.~\cite{BHBH6}, the coordinate radius of the apparent horizon is
larger for larger value of $\eta_s$.  This implies that the region
near the BH is not well resolved for too small values of $\eta_s$, 
whereas for too large values of $\eta_s$, the BH is not covered only
by the finest level in the AMR algorithm. For $\eta_s=1/m$ and $2/m$,
the coordinate radius of the apparent horizon of a nonspinning BH is
$\sim 0.8m$ and $1.1m$, respectively. 

The adopted spatial gauge condition is different for numerical
simulations with the $F_i$-BSSN and $\tilde \Gamma^i$-BSSN formalisms.
Difference in numerical results computed by both formalisms
results primarily from this difference. 

During evolution, we enforce the following constraints on
$\tilde \gamma_{ij}$ and $\tilde A_{ij}$ at every time step
\beqn
&&{\rm det}(\tilde \gamma_{ij})=1,\\
&&{\rm Tr}(\tilde A_{ij})=0. 
\eeqn
The reason for this is that these constraints are violated slightly
due to numerical error. 
Specifically, we reset, after every time evolution, as
\beqn
&& \tilde \gamma_{ij}
\rightarrow [{\rm det}(\tilde \gamma_{ij})]^{-1/3}
\tilde \gamma_{ij},\\
&& \tilde A_{ij} \rightarrow 
[{\rm det}(\tilde \gamma_{ij})]^{-1/3} \tilde A_{ij}
-{1 \over 3}\tilde \gamma_{ij}{\rm Tr}(\tilde A_{ij}), \\
&& W \rightarrow  [{\rm det}(\tilde \gamma_{ij})]^{-1/6} W,\\
&& K \rightarrow K + {\rm Tr}(\tilde A_{ij}). 
\eeqn
We note that in this adjustment, $\gamma_{ij}$ and $K_{ij}$ are
unchanged. 

We do not add any constraint-violation damping terms in SACRA.
We monitor violation of Hamiltonian and momentum constraints computing 
L2-norm for them, and find that their growth time scales are much longer than
the dynamical time scale even in the absence of the damping terms. 
(Note that an exception is at the formation of BH after the merger
of NS-NS binaries,
at which the degree of constraint violation increases rapidly by an order
of magnitude.) 

The fundamental variables for the hydrodynamics are 
$\rho$: the rest-mass density, 
$\varep$ : the specific internal energy, 
$P$ : the pressure, 
$u^{\mu}$ : the four velocity, and the three velocity defined by 
\beqn
v^i ={dx^i \over dt}={u^i \over u^t}. 
\eeqn
For our numerical implementation of the hydrodynamic equations, we
define a weighted density, a weighted four-velocity, and a specific
energy defined, respectively, by
\beqn
&&\rho_* \equiv \rho \alpha u^t W^{-3}, \\
&&\hat u_i \equiv h u_i, \\
&& \hat e \equiv h\alpha u^t - {P \over \rho \alpha u^t},
\eeqn
where $h=1+\varepsilon+P/\rho$ denotes the specific enthalpy. The general
relativistic hydrodynamic equations are written into a conservative
form for variables $\rho_*$, $\rho_* \hat u_i$, and $\rho_* \hat
e$. Then, we solve these equations using a high-resolution central
scheme \cite{KT,SF}. In our approach, the transport terms such as
$\pa_i (\cdots)$ are computed by the scheme of Kurganov-Tadmor
\cite{KT} with a third-order (piecewise parabolic) spatial interpolation
for reconstructing numerical fluxes.

In the present work, the initial condition for NSs is computed with the
polytropic EOS
\beqn
P=\kappa \rho^{\Gamma},
\eeqn
where $\kappa$ and $\Gamma$ are 
the polytropic constant and the adiabatic index.
Because $\kappa$ is arbitrarily chosen, we set $\kappa=1$ in 
the following. $\Gamma$ is set to be 2 for comparing numerical results
with previous ones \cite{NSNS2,BHNS2}.  
During the numerical simulation, we adopt the $\Gamma$-law EOS 
\beq
P=(\Gamma-1)\rho \varepsilon. 
\eeq
Again, we set $\Gamma=2$. Note that we have already 
implemented a number of EOSs in our code (e.g., \cite{NSNS2,NSNS8}).
In the future, we will perform numerical simulations in such EOSs. 

At each time step, $w=\alpha u^t$ is determined by solving an algebraic
equation derived from the normalization relation $u^{\mu}u_{\mu}=-1$
and EOS. Specifically, the equation is written as
\beqn
w^2=1 + {\gamma^{ij} \hat u_i \hat u_j \over h^2}, \label{eq24}
\eeqn
where in the chosen EOS, $h$ is written as
\beq
h=[\hat e w \Gamma - (\Gamma-1)][w^2 \Gamma - (\Gamma-1)]^{-1}. 
\eeq
After $w$ and $h$ are determined, the
primitive variables such as $\rho$, $\varepsilon$, and $u_i$ are updated
as $\rho=\rho_* W^3/w$, $\varepsilon=(h-1)/\Gamma$, and $u_i=\hat u_i/h$. 

Because any conservation scheme of hydrodynamics is unable to evolve a
vacuum, we have to introduce an artificial atmosphere outside NSs.
Density of the atmosphere should be as small as possible, to
avoid spurious effect due to it. In the present case, we initially
assign a small rest-mass density in vacuum as
\beqn
\rho=
\left\{
\begin{array}{ll}
\rho_{\rm at} & r \leq r_0, \\
\rho_{\rm at}e^{1-r/r_0} & r > r_0, 
\end{array}
\right.
\eeqn
where we choose $\rho_{\rm at}=\rho_{\rm max} \times 10^{-8}$ for
NS-NS binaries and $10^{-9}$ for BH-NS binaries.  Here, $\rho_{\rm
max}$ is the maximum rest-mass density of the NS.  $r_0$ is a
coordinate radius of $\sim 10$--$20M$ where $M$ is the ADM
(Arnowitt-Deser-Misner) mass of the system. With such a choice of
parameters, the total amount of the rest mass of the atmosphere is
about $10^{-5}$ of the rest mass of the NS. Thus, spurious effects due
to the presence of the atmosphere, such as accretion of the atmosphere
onto NS and BH, the resulting dragging effect against orbital motion, 
gravitational effect by the atmosphere, and formation of a
disk around the final outcomes, play a negligible role in the present
context. 

In the presence of a BH, location of apparent horizon is determined by
an apparent horizon finder. In our method, we derive a two-dimensional
elliptic-type equation for the radius of the apparent horizon and
iteratively solve this equation until a sufficient convergence is
achieved. This method is essentially the same as that in
Ref.~\cite{AH}, but in SACRA, we implement a simpler scheme for computing
the source term for the elliptic-type equation. We briefly describe
this method in Appendix \ref{app:ahfinder}. 

\subsection{Formulation for Extracting Gravitational Waves}

Gravitational waves are extracted computing the outgoing component of
the Newman-Penrose quantity (the so-called $\Psi_4$), which is defined by
\beqn
\Psi_4=- {}^{(4)}
R_{\alpha\beta\gamma\delta} 
n^{\alpha} \bar m^{\beta} n^{\gamma} \bar m^{\delta},
\eeqn
where $^{(4)}R_{\alpha\beta\gamma\delta}$ is Riemann tensor
with respect to spacetime metric $g_{\mu\nu}$, and $n^{\alpha}$ and $\bar m^{\beta}$ are
parts of null tetrad ($n^{\alpha},~\ell^{\alpha},~m^{\alpha},~\bar m^{\alpha}$).
Specifically,  $n^{\alpha}$ and $\ell^{\alpha}$ are outgoing and ingoing null
vectors, whereas $m^{\alpha}$ is a complex null vector orthogonal to
$n^{\alpha}$ and $\ell^{\alpha}$. The null tetrad satisfies the conditions
\beqn
-n^{\alpha} \ell_{\alpha}=1=m^{\alpha} \bar m_{\alpha}, 
\eeqn
and $g_{\mu\nu}$ is written as
\beqn
g_{\mu\nu}=-n_{\mu} \ell_{\nu}-n_{\nu} \ell_{\mu}+m_{\mu} \bar m_{\nu}
+m_{\nu} \bar m_{\mu}. 
\eeqn
Denoting $n^{\mu}$ by $n^{\mu}=(N^{\mu}-r^{\mu})/\sqrt{2}$ where
$N^{\mu}$ is unit timelike hypersurface normal
$(\alpha^{-1}, -\beta^i \alpha^{-1})$
and $r^{\mu}$ is a unit radial vector orthogonal to $N^{\mu}$ and $m^{\mu}$,
$\Psi_4$ is rewritten to 
\beqn
\Psi_4&=&-{1 \over 2}\biggl[
{}^{(4)}R_{\alpha\beta\gamma\delta}
N^{\alpha} \bar m^{\beta} N^{\gamma} \bar m^{\delta}
\nonumber \\
&&~~~~~-2 {}^{(4)}R_{\alpha\beta\gamma\delta}
N^{\alpha} \bar m^{\beta} r^{\gamma} \bar m^{\delta}
\nonumber \\
&&~~~~~+ {}^{(4)} R_{\alpha\beta\gamma\delta}
r^{\alpha} \bar m^{\beta} r^{\gamma} \bar m^{\delta}
\biggr]. 
\eeqn
Using the following relations, 
\beqn
&& {}^{(4)}R_{\alpha i \gamma j} N^{\alpha} N^{\gamma}
= R_{ij}-K_{ik}K_{j}^{~k} + K K_{ij} \equiv {\cal E}_{ij},\label{eq25}\\
&& {}^{(4)}R_{\alpha ijk} N^{\alpha} 
= D_j K_{ik}-D_k K_{ij} \equiv {\cal B}_{ijk},\\
&& {}^{(4)}R_{ijkl}
= R_{ijkl}+K_{ik}K_{j}^{~l} - K_{il} K_{jk} \equiv {\cal R}_{ijkl}. 
\eeqn
where $D_i$, $R_{ij}$, and $R_{ijkl}$ are covariant derivative, Ricci tensor,
and Riemann tensor with respect to three-metric $\gamma_{ij}$, 
$\Psi_4$ is written only by geometric variables in 3+1 formalism.
Note that for deriving Eq. (\ref{eq25}), we assume that $\Psi_4$
is extracted in a vacuum region. 
In addition, we have the following identity in three-dimensional space
because of symmetric and antisymmetric relations for ${\cal R}_{ijkl}$: 
\beqn
&&{\cal R}_{ijkl}=\gamma_{ik} {\cal R}_{jl}-\gamma_{il} {\cal R}_{jk}
-\gamma_{jk} {\cal R}_{il}+\gamma_{jl} {\cal R}_{ik} \nonumber \\
&&~~~~~~-{1 \over 2}{\cal R}\Big(\gamma_{ik}\gamma_{jl}
-\gamma_{il}\gamma_{jk}\Big), 
\eeqn
where ${\cal R}_{ik}={\cal R}_{ijk}^{~~~~j}$ and ${\cal R}={\cal R}_k^{~k}$. 
Then, we find
\beqn
{\cal E}_{ij}\bar m^i \bar m^j
={\cal R}_{ijkl} r^i \bar m^j r^k  \bar m^l, 
\eeqn
and obtain a simple formula 
\beqn
\Psi_4=-({\cal E}_{ij} \bar m^i \bar m^j
+{\cal B}_{ijk} \bar m^i r^j \bar m^k). 
\eeqn

For $r \rightarrow \infty$, $\Psi_4$ is written as
\beqn
\Psi_4 =-{1 \over 2} \biggl( \ddot h_+ - i \ddot h_{\times}\biggr),
\eeqn
where $h_+$ and $h_{\times}$ are $+$ and $\times$ modes of
gravitational waves, respectively. Thus, by performing time
integration of $2 \Psi_4$ twice (and by appropriately choosing 
integration constants), one can derive gravitational waveforms.  More
specifically, we decompose $\Psi_4$ into tensor spherical harmonic
modes of $(l,m)$ by surface integral at a sufficiently large radius as
usually done (e.g., see Ref.~\cite{BHBH6} in detail), and pay particular
attention to harmonics of low quantum numbers. In this paper, we
compute the modes with $2 \leq l \leq 4$.

From $\Psi_4$, energy, linear momentum, and angular momentum
dissipation rates by gravitational waves are computed by
\beqn
&&{dE \over dt}= \lim_{r\rightarrow\infty}
\biggl[{r^2 \over 16\pi} \oint_{S} dA \Big| \int \Psi_4 dt \Big|^2 \biggr],\\
&&{dP_i \over dt}= \lim_{r\rightarrow\infty}
\biggl[{r^2 \over 16\pi} \oint_{S} dA {x^i \over r}
\Big| \int \Psi_4 dt \Big|^2 \biggr],\\
&&{dJ_z \over dt}= \lim_{r\rightarrow\infty}
\biggl[{r^2 \over 16\pi} {\rm Re}
\biggl\{\oint_{S} dA \Big( \int \pa_{\varphi} \Psi_4 dt \Big) \nonumber \\
&&~~~~~~~~~~~~~~~~\times
\Big( \int \int \bar \Psi_4 dt dt' \Big)\biggr\}\biggr],
\eeqn
where $\oint dA=\oint d(\cos\theta)d\varphi$ denotes an integral on 
two surface of a constant coordinate radius and $\bar\Psi_4$ is the
complex conjugate of $\Psi_4$. In the actual simulation, gravitational
waves are extracted at finite radii, and then, by an extrapolation,
asymptotic gravitational waveforms should be derived. In such
procedure, we estimate the dissipation rates by exchanging $r$ to a
proper radius approximately defined by $D=r (1+m_0/2r)^2$ where $r$ is
the coordinate radius, $D$ approximately denotes the proper radius,
and $m_0$ is sum of mass of two compact objects (see
Eq. (\ref{eqm0})).

\subsection{Diagnostics} \label{sec:diag}

\subsubsection{Mass, linear momenta, and angular momenta}

We monitor the ADM mass, $M$, the linear momenta, $P_i$, and
the angular momenta, $J_i$, during the evolution.
To do so, we define integrals on two surface of a coordinate radius $r$ 
\beqn
&& M_{\rm ADM}(r)={1 \over 16\pi}
\oint_r \sqrt{\gamma} \gamma^{ij} \gamma^{kl}
(\gamma_{ik,j}-\gamma_{ij,k})dS_l, \\
&& P_{i}(r)={1 \over 8\pi}\oint_r \sqrt{\gamma}
(K_i^{~j}-K \gamma_i^{~j}) dS_j, \\
&& J_{i}(r)={1 \over 8\pi}\epsilon_{ilk}
\oint_r \sqrt{\gamma} x^l (K^{jk}-K\gamma^{jk}) dS_j.  
\eeqn
Then, we extrapolate these quantities for $r \rightarrow \infty$
to obtain the ADM mass $M$, the linear momenta $P_i$,
and the angular momenta $J_i$.
Throughout this paper, the initial values of $M_{\rm ADM}$ and $J_z$ are
denoted by $M_0$ and $J_0$, respectively. 

When simulating a spacetime with NSs, we also monitor the total baryon 
rest mass ($M_*$) 
\beqn
M_*=\int \rho u^t \sqrt{-g} d^3 x.  
\eeqn
In the simulation with a uni-grid domain, it is easy to guarantee that
$M_*$ is conserved by adopting standard schemes of numerical
hydrodynamics (except for a possible slight error associated with an
artificial treatment of atmosphere). In the schemes in which an AMR
algorithm is implemented, it is not straightforward to guarantee that
$M_*$ is conserved when regridding is carried out. In our present
scheme, $M_*$ is not strictly conserved, and it is necessary to
confirm that the violation of the conservation is within an acceptable
level.

\subsubsection{Spin and mass of the formed BH}

For BH-BH and BH-NS binaries, apparent horizons are determined during
the evolution, and thus, we monitor their area. From the area, 
the irreducible mass of each BH is defined by
\beqn
m_i=\sqrt{{A_{\rm AH,i} \over 16\pi}},
\eeqn
where $A_{\rm AH,i}$ is the area of each BH. 
For BH-BH binaries, we define a total mass at $t=0$ as 
\beqn
m_0=m_1 + m_2, \label{eqm0}
\eeqn
and present all the numerical results in units of $m_0$.
(In this paper, $m_0=2m_1$ because we only consider the
equal-mass BH-BH binaries.)

After the merger of compact binary objects, a rotating BH is often formed
in the end.  To determine properties of the formed BH, we analyze
several quantities of the apparent horizon of such BH. Specifically,
we compute the area, $A_{\rm AH}$, polar circumferential length, $C_p$,
and equatorial circumferential length, $C_e$, of the apparent
horizon. If the formed BH is a Kerr BH and the system relaxes to a
stationary state, the area obeys the relation of
\beqn
A_{\rm AH}=8\pi M_{\rm BHf}^2 (1+ \sqrt{1-a^2}), \label{AAH}
\eeqn
where $M_{\rm BHf}$ and $a$ are mass and spin parameter of the
Kerr BH, respectively. Also, $C_e$ should be $4\pi M_{\rm BHf}$ and
$C_p/C_e$ is a known function composed only of $a$ as
\beqn
{C_p \over C_e}={\sqrt{2  \hat r_+} \over \pi}
E(a^2/2 \hat r_+), \label{eq:cpce}
\eeqn
where $\hat r_+=1+\sqrt{1-a^2}$ and $E(z)$ is an elliptic
integral defined by
\beq
E(z)=\int^{\pi/2}_0 \sqrt{1-z\sin^2\theta} d\theta.
\eeq
In the analysis of numerical results, we determine the spin parameter,
$a$, from Eq. (\ref{eq:cpce}) providing $C_p/C_e$ by direct measurement
from numerical results.  Then, $M_{\rm BHf}$ can be determined either
from Eq. (\ref{AAH}) or from $C_e/4\pi$. We calculate the BH mass
using both methods and check that two results agree well. In addition,
we can infer the final ADM mass of the system from the initial value
of the ADM mass and the total radiated energy by gravitational waves,
and the final angular momentum of the system from the initial angular
momentum and the total radiated one.  These values have to also agree
with the mass and angular momentum of a system finally formed,
due to the presence of conservation laws.

\subsubsection{Gravitational waves}

We compare inspiral orbital trajectories with the results by 
the so-called Taylor T4 post-Newtonian formula for two point masses
in quasi-circular orbits (see, e.g., Refs.~\cite{BHBH12,BHBH16} for a 
detailed description of various post-Newtonian formulas). Recent
high-accuracy simulations for equal-mass (nonspinning or corotating)
BH-BH binaries have proven that the Taylor T4 formula provides their
orbital evolution and gravitational waveforms with a high accuracy
at least up to about one orbit before the merger. In this formula,
the angular velocity, $\Omega$, is determined by solving \cite{BHBH12}
\beqn
{dX \over dt}&=&{64\eta X^5 \over 5M_0} \biggl[
1-{743+924 \eta \over 336} X
+ \Bigl(4\pi -{47 \over 3} \chi \Bigr) X^{3/2} \nonumber \\
&& +\biggl({34103 \over 18144}+{13661\eta \over 2016}
+{59\eta^2 \over 18}\biggr)X^2 \nonumber \\
&& -\biggl\{ {4159+15876 \eta \over 672}\pi \nonumber \\
&&~~~~
+\Bigl({31811 \over 1008}-{5039 \over 84}\eta\Bigr)\chi \biggr\} X^{5/2} \nonumber \\
&& +\biggl\{ {16447322263 \over 139708800}-{1712\gamma_E \over 105}
     +{16 \pi^2 \over 3} \nonumber \\
&&~~~~ +\biggl({-56198689 \over 217728}+{451 \over 48}\pi^2
     \biggr)\eta+{541 \over 896}\eta^2 \nonumber \\
&&~~~~ -{5605 \over 2592} \eta^3
     -{856 \over 105}\log(16 X)\biggr\}X^3 \nonumber \\
&& +\biggl({-4415 \over 4032}+{358675 \over 6048}\eta
       +{91495 \over 1512} \eta^2\biggr)\pi X^{7/2}\biggr], \label{T4}
\eeqn
where $X=[m_0\Omega(t)]^{2/3}$ is a function of time,  
$\eta$ is a ratio of the reduced mass to the total mass $m_0$,
$\gamma_E=0.577 \cdots$ is the Euler constant, and
$\chi \equiv S/4m_0^2$ which is defined from the sum of spin angular momentum
of BHs, $S$. The spin is present for each BH of BH-BH binaries
considered in this paper. In Eq. (\ref{T4}), we omit to write
terms associated with the difference in spins, because we only consider
the case that the spins of two BHs are equal. 

From $X(t)$, gravitational waveforms are determined from
\beqn
&&h_+(t)={4\eta m_0 X \over D} A(X) \cos [\Phi(t) + \delta],\\
&&h_{\times}(t)={4\eta m_0 X \over D} A(X) \sin [\Phi(t) + \delta],
\eeqn
where $A(X)$ is a nondimensional function of $X$ for which
$A(X)\rightarrow 1$ for $X \rightarrow 0$, $\delta$ is an
arbitrary phase, and 
\beqn
\Phi(t)=2 \int \Omega(t) dt. 
\eeqn
For $A(X)$, we adopt the 2.5 post-Newtonian formula (e.g., Ref.~\cite{BHBH16}).

\section{Adaptive mesh refinement} \label{sec:AMR}

\subsection{Adaptive Mesh Refinement for the Einstein Equations}
\label{sec:AMR-Einstein}

Our AMR algorithm for solving the Einstein evolution equations are very
similar to that described in Ref.~\cite{BHBH6}: We employ the
Berger-Oliger-type AMR algorithm \cite{BO} with the centered
fourth-order finite-differencing in space for evaluating spatial
derivatives and with the lop-sided fourth-order finite-differencing
for advection terms like $\beta^i\pa_i W$.  For integration forward in
time, the fourth-order Runge-Kutta scheme is adopted. There are also
slight differences between our scheme and the scheme of
Ref.~\cite{BHBH6}. Main difference comes from the choice of grid
structure; we adopt the grid-center-grid whereas the code of
Ref.~\cite{BHBH6} adopts the cell-center-scheme. The reason of our
choice is simply that we felt that with the grid-center-grid, it is
easier to implement the interpolation and extrapolation required to be
carried out at refinement boundaries in any AMR algorithm. Because of
the difference in the grid structure, our interpolation and
extrapolation schemes around the refinement boundaries are different
from those of Ref.~\cite{BHBH6}. In order to clarify the difference,
we describe our method in detail in the following. 

As in the code of Ref.~\cite{BHBH6}, the whole numerical domain is
composed of a hierarchy of nested Cartesian grids.  The hierarchy
consists of $L$ levels of refinement domains of indices $l=0, 1,
\cdots, L-1$.  Here, $l=0$ is the coarsest level whereas $l=L-1$ is
the finest one. Each refinement level consists of one or two 
domains. For coarser levels of $l \leq L_1$ where $L_1 (<L-1)$ is a
constant, the number of the refinement domain is one, and their grid 
locations are fixed throughout numerical simulation. We call this type
of domain the coarser domain in the following. On the other hand, for
finer levels with $l > L_1$, the number of the refinement domain is two,
each of which covers a region near the center of two compact
objects. We call this type of domain the finer domain. For the levels
composed of only one domain, we initially choose the grid for which
the center agrees (approximately) with mass center of the system. For 
the levels composed of two domains, the grid center is chosen to agree 
approximately with the center of the compact objects at $t=0$. 

Each domain is in general composed of $(2N+1)\times (2N+1) \times
(2N+1)$ grid points for the $x$-$y$-$z$ axis directions, where $N$ is
an even integer and it is the same value for all the domains.  Note that in
counting the grid number, the number of buffer zone (see below) is not
included.  When symmetries are imposed, the grid number is
appropriately saved.  For example, when the equatorial-plane symmetry
is imposed, the grid number is $(2N+1)\times (2N+1) \times (N+1)$ for
the $x$-$y$-$z$ axis directions.  The grid spacing in each level is
fixed to be uniform and denoted by $h_l$ for the $l$-th level. For
simplicity, a refinement of factor 2 is adopted, i.e., $h_l=h_0/2^{l}$
where $h_0$ is the largest grid spacing. Thus, the length of a side of
each cube is $2 N h_l$ for the $l$-th level.

Specifically, the center of any finer domain is arranged to agree
approximately with mass center of a compact object. To guarantee this
arrangement during time evolution, regridding is necessary as
the compact objects move.  Following Ref.~\cite{BHBH6}, we use the
shift to track the position of BH centers by integrating
\beqn
\pa_t x^i_{\rm BH}=-\beta^i(x^j_{\rm BH}), \label{bpunct}
\eeqn
where $x^i_{\rm BH}$ denotes the center of a BH. The shift vector
at $x^i_{\rm BH}$ is determined by the linear interpolation of
$\beta^i$ in the finest refinement levels. The time integration of Eq. 
(\ref{bpunct}) is performed with the fourth-order Runge-Kutta
scheme. For the case of NSs, coordinate position 
of the center, $x^i_{\rm NS}$, is determined by searching for the 
local maximum density at every time step. Here, the maximum density
implies the maximum of $\rho_*$ (not $\rho$) in SACRA.

During time evolution, the finer domains are moved for the $i$-th axis
direction, whenever the following condition is satisfied:
\beqn
&& |x^i_{\rm BH}-x^i_{l0}| \geq 2h_l~~~{\rm for~~BHs}, \\
&& |x^i_{\rm NS}-x^i_{l0}| \geq 2h_l~~~{\rm for~~NSs}, 
\eeqn
where $x^i_{l0}$ denotes the center of a finer domain in the $l$-th level.
Then, we translate the finer domain by $2h_l$ toward the $x^i$ axis 
direction. Here, the factor 2 comes from the requirement that 
a refinement boundary surface of a domain of level $l$ (hereafter,
referred to as ``child domain'') always overlaps with a surface
of a domain of level $l-1$ (hereafter ``parent domain''), which 
is defined by $x^i=$const in the parent domain. 

We arrange that each child domain of level $l~(\geq 1)$ is guaranteed
to be completely covered by its parent domain of level $l-1$.  Here,
we determine that each child has only one parent. If there are two
domains in the same level, say ($l-1$)-th level, we refer to one of
two as the parent and to the other as the uncle. For more specific
description, let us denote the location of grid points for the child
and the parent, respectively, by $(i_c, j_c, k_c)$ and $(i_p, j_p,
k_p)$ for $(x,y,z)$, where $i_c$, $j_c$, $k_c$, $i_p$, $j_p$, and
$k_p$ are all in the range between $-N$ and $N$.  We arrange that
refinement boundary surfaces of the child domain, i.e., $i_c=-N$,
$i_c=N$, $j_c=-N$, $j_c=N$, $k_c=-N$, and $k_c=N$, always overlap with
surfaces of $x^i=$const in the parent domain (here, $x^i$ denotes $x$
or $y$ or $z$). Namely, the surfaces of $i_c=-N$ and $i_c=N$ overlap
with the parent's surfaces of $i=i_{p1}=$ const and $i=i_{p2}=$ const,
respectively. (This is also the case for $j_c=\pm N$ and $k_c=\pm N$.) 
Typically, the following conditions are satisfied: $i_{p1} \approx -
N/2$ and $i_{p2} \approx N/2$. By this arrangement, our refinement
procedure becomes very simple: Assigning finer quantities of the child
level to its one-coarser level is straightforward because the grid
points of the parent domain for $i_{p1} \leq i \leq i_{p2}$ overlap
with those of the child domain.

A parent domain overlap not only with its child domain (level $l$)
but also may overlap with other domain of level $l$ (we call this
nephew).  We have to copy values of the nephew to the parent 
in the same procedure as described above (from the viewpoint of the
child, values of the child are copied to its uncle).  To carry out this
procedure, we have to check status of overlapping for all the 
levels composed of two domains at each time step. We note that copying
the finer quantities to the coarser ones is carried out at each time
step that the quantities of the coarser levels are defined.
(Note that the time step of the child domain is always half of
that of the parent domain; cf. Eq. (\ref{eqdt}).)

To evolve quantities near the refinement boundaries of a child domain,
we have to prepare buffer zones and to assign an approximate value on
them. Following Ref.~\cite{BHBH6}, we prepare six buffer zone points
along each axis (e.g., for the $x$ axis direction, extra regions of
$-N-6 \leq i \leq -N-1$ and $N+1 \leq i \leq N+6$ are prepared as the
buffer zones).  The quantities at the buffer zones are provided from
the corresponding parent domain by the following procedure: (i) If a
buffer-zone's grid point of the child domain overlaps with its
parent's grid point, we simply copy the value, and (ii) if a
buffer-zone's grid point of the child domain is located between its
parent's grid points, the fifth-order centered Lagrangian 
interpolation is applied using nearby six parent's grid points.
Actual three-dimensional procedure is carried out by successive 
one-dimensional procedures. 

The interpolation procedure from the parent to the child for the
child's buffer zone, which is described above, is carried out in the
straightforward manner whenever the time-step level coincides between
two levels. However, it does not, in general, coincide because the
time step of the parent level $\Delta t_{l-1}$ is twice larger than
that of the child level $\Delta t_l$ in typical AMR algorithms
\cite{BO}.  Specifically, the time-step level does not agree (i) at a
child's time step of odd number, and (ii) at each Runge-Kutta sub time
step. For the interpolation at such time step, we employ the following
method: (I) For the inner three buffer-zone's points (e.g., $-N-3 \leq
i \leq - N-1$ and $N+1 \leq i \leq N+3$ for the $x$ axis direction),
we evolve all the quantities using the fourth-order
finite-differencing scheme. Because there are sufficient number of
buffer-zone's points to solve the evolution equations in the inner
three buffer-zone's points, no interpolation is necessary; (II) For
the fourth buffer-zone's point (e.g., $i=\pm (N+4)$ for the $x$ axis
direction), all the quantities are evolved using the second-order
finite-differencing scheme with no interpolation: (III) For the outer
two buffer-zone's points (e.g., $-N-6 \leq i \leq -N-5$ and $N+5 \leq
i \leq N+6$ for the $x$ axis direction), the second-order Lagrangian
interpolation of the parent's quantities in time is carried out to
determine the values of the parent level at the corresponding child's
time-step level as a first step, and then, the fifth-order Lagrangian
interpolation in space is carried out.

Because there are two domains in the finer levels, they often overlap
with each other. In such case, the values of all the quantities should
agree with each other. However, the evolution equations for those two
domains are solved independently, and consequently, the values do not
always agree. To guarantee that they agree, we simply take average 
of two values as $Q_1 \rightarrow (Q_1 + Q_2)/2$ and
$Q_2 \rightarrow (Q_1+Q_2)/2$ where $Q_1$ and $Q_2$ denote the values
of two domains of the same refinement level.  When a buffer-zone's point
of one of the two domains overlaps with a point in the main region of
the other domain, the values at the point of the main region are
copied to those at the buffer-zone's point. When two buffer zones 
overlap at some points, the simple averaging, described above, is
again used.

At the outer boundaries of the coarsest refinement level, an outgoing
boundary condition is imposed for all the geometric variables.  The
outgoing boundary condition is the same as that suggested by Shibata
and Nakamura \cite{BSSN}.

It is possible to add artificial dissipation terms. Following 
Ref.~\cite{BHBH6}, we tried to add the sixth-order Kreiss-Oliger-type 
dissipation term as
\beqn
Q_l \rightarrow Q_l - \sigma h_l^6 Q^{(6)}_l 
\eeqn
where $Q_l$ is a quantity in the $l$-th level, $Q^{(6)}_l$ is the sum
of the sixth derivative along the $x$, $y$, and $z$ axis directions, and
$\sigma$ is a constant of order 0.1.  We performed simulations for BH-BH
binaries with $\sigma=0$ and 0.1, and found that the simulations
proceed with no instability even for $\sigma=0$ and that the numerical
results depended very weakly on the dissipation.  The numerical
dissipation, however, spuriously accelerated the merger process for
the nonzero value of $\sigma$, and as a result, the merger time was
shortened slightly. Thus, in this paper, we do not add any dissipation
term in the simulation. Even in hydrodynamic simulations, we do not
have to add it.

Numerical simulations for BH-BH binaries reported in this paper are
performed with nine or ten refinement levels, which include five or
six coarser levels composed of one domain and four finer levels
composed of two domains.  Simulations for NS-NS binaries are performed
with seven or eight refinement levels, i.e., three or four coarser
levels composed of one domain and four or five finer levels composed
of two domains. For BH-NS binaries, simulations are done with eight
refinement levels, i.e., four coarser levels composed of one domain
and four finer levels composed of two domains.

Time step for each refinement level, $dt_l$, is determined by the
following rule:
\beqn
dt_l=\left\{
\begin{array}{ll}
h_{l_c} /2 & {\rm for}~0 \leq l \leq l_c \\ 
h_l /2 & {\rm for}~l_c < l \leq L-1,  \\ 
\end{array}
\right. \label{eqdt}
\eeqn
where $l_c=4$ for simulations of BH-BH binary and $l_c=2$ for NS-NS
and BH-NS binaries. Namely, the Courant number is 1/2 for the finer
refinement levels with $l \geq l_c$, whereas for the coarser levels,
it is smaller than 1/2. The reason why the small Courant number is
chosen for the small values of $l$ is that with a high Courant number
such as 1/2, the numerical instability occurs near the outer
boundaries for the coarsest refinement level.

\subsection{Adaptive Mesh Refinement for the Hydrodynamic Equations}

When a hydrodynamic simulation is performed employing an AMR
algorithm, first of all, we have to determine for which variables the
interpolation from coarser to finer levels and the copy from finer to
coarser levels are carried out. In the present work, we choose
$\rho_*$, $\hat u_i$, and $h$ for the interpolation and copying
procedures.  The copying procedure is totally the same as that for
geometric variables (see Sec.~\ref{sec:AMR-Einstein}).  The
interpolation procedure is basically the same as that for geometric
variables; if grid points of the child and parent domains overlap, we
simply copy the values of the parent to the child, whereas if they do
not overlap, we adopt the fifth-order Lagrangian interpolation. However,
for the fluid variables such as $\rho_*$ and $h$, this interpolation
scheme could fail in particular in the vicinity of surface of NSs for
which $\rho_*$ is small and steeply varies. The reason for this
possible failure is that the interpolation may give a negative value
of $\rho_*$ (and also $h-1$) which is unphysical.  Thus, in case that
$\rho_* < \rho_{\rm min}$ or $h <1$ are results of the fifth-order
interpolation, we adopt the first-order scheme for the interpolation
(i.e., linear interpolation). Here, $\rho_{\rm min}$ is chosen to be
$\rho_{\rm max}/10^8$ for NS-NS binaries and $\rho_{\rm max}/10^9$ for
BH-NS binaries in the present case where $\rho_{\rm max}$ is the
initial maximum value of $\rho_*$. We have found that the linear
interpolation is too dissipative to adopt for the whole
interpolation. Therefore, this is used only in case.

We also modify the scheme of interpolation in time which is necessary
for the interpolation procedure in the buffer zone (see
Sec.~\ref{sec:AMR-Einstein}).  For geometric variables, we always use
the second-order interpolation scheme as described in
Sec.~\ref{sec:AMR-Einstein}.  Specifically, we determine an
interpolated value at a child's time step from values at three time
levels of its parent, say, $n-1$, $n$, and $n+1$. Here, the
interpolation is necessary for determining the values at a time $t$
that satisfies $t^n < t < t^{n+1}$. For the fluid variables, we
basically adopt the same interpolation scheme as that for the
geometrical variables. However, for maintaining numerical stability,
we modify it when the following relation holds: \beq
(Q^{n+1}-Q^n)(Q^n-Q^{n-1}) < 0.  \eeq Here, $Q$ is $\rho_*$ or $\hat
u_i$ or $h$, and $Q^n$ denotes $Q$ at $t^n$. In this case, we adopt the
first-order interpolation scheme, only using $Q^{n+1}$ and
$Q^n$. Namely, a limiter procedure is introduced. We have found that
this prescription is robust for stabilizing numerical computation.

After the interpolation or the copy is carried out, we have to determine
values of primitive variables such as $\rho$, $u^t$, and $\varep$. In
the present choice of the variables to be interpolated or copied
($\rho_*$, $\hat u_i$, and $h$), this procedure is quite simple. From
$h$ and $\hat u_i$, $w$ is determined from Eq. (\ref{eq24}). Then,
$\rho$ is computed by $\rho_* W^3/w$. Because the relation,
$h=1+\Gamma \varep$, holds, $\varep$ is also immediately obtained.
Even if we adopt more complicated EOSs, $\rho$ and $w$ are immediately
calculated. In general EOSs, $h$ is a complicated function of $\rho$
and $\varep$. Thus, the procedure for getting $\varep$ may be
much more complicated.  However, $\rho$ is obtained very easily, and hence,
$\varep$ should be obtained by simply solving a one-dimensional
equation for $h=h(\varep)$.

\subsection{Extracting Gravitational Waves in AMR} \label{sec3c}

During inspiraling and merging of binary compact objects,
gravitational wavelength gradually decreases (the frequency
increases).  Propagation of gravitational waves is accurately
computed only in the case that the grid spacing is at least by one
order of magnitude smaller than the wavelength.  Thus, the
required grid resolution changes during the evolution. In the late
inspiral phase in which $m_0\Omega=0.03$--0.1, the wavelength is 
\beqn
\lambda = {\pi \over \Omega}
\approx 105 \biggl({m_0\Omega \over 0.03}\biggr)^{-1} m_0. 
\eeqn
This implies that the grid spacing should be smaller than $\sim 10 m_0$
for $m_0\Omega=0.03$ and $\sim 3m_0$ for $m_0\Omega=0.1$. 
By contrast, in the merger phase in which $m_0\Omega$ can be as large as
$\sim 0.3$, the grid spacing has to be smaller than $\approx m_0$.

Another requirement for accurate computation of gravitational waves is
that wave extraction has to be done in a wave zone. Thus, inspiral 
gravitational waveforms should be extracted in a region far from the
source. On the other hand, merger waveforms may be extracted at a
distance of $\sim 20m_0$ because the gravitational wavelength at
the merger phase is 10--$15 m_0$ (see Sec.~\ref{sec:res}).

Taking into account these requirements, we extract gravitational
waves in the following manner. For the inspiral gravitational
waveforms, the radius of the extraction is chosen to be 50--$70 m_0$
in the present paper. The grid spacing at such radius is
$\sim 2$--$3m_0$ in the present grid setting. For the merger gravitational
waveforms, the radius of the extraction is $\sim 20$--$30 m_0$. More
specifically, the inspiral waveforms are extracted
for $t_{\rm ret} \leq t_{\rm sep}$, whereas the merger ones are done 
for $t_{\rm ret} \geq t_{\rm sep}$. Here, $t_{\rm ret}$ denotes
retarded time defined by
\beqn
t_{\rm ret} \equiv t - r - 2m_0 \log(r/m_0),
\eeqn
where $r$ is the coordinate radius of the extraction and
we assume $r \gg m_0$ for defining this retarded time. 
$t_{\rm sep}$ denotes a retarded time at which the orbital
angular velocity of the binary motion becomes $m_0\Omega \sim 0.1$.
In Sec.~\ref{sec:res}, we will show that this strategy is acceptable. 

\section{Numerical results}\label{sec:res}

\begin{figure*}[t]
\begin{center}
\epsfxsize=3.2in
\leavevmode
(a)\epsffile{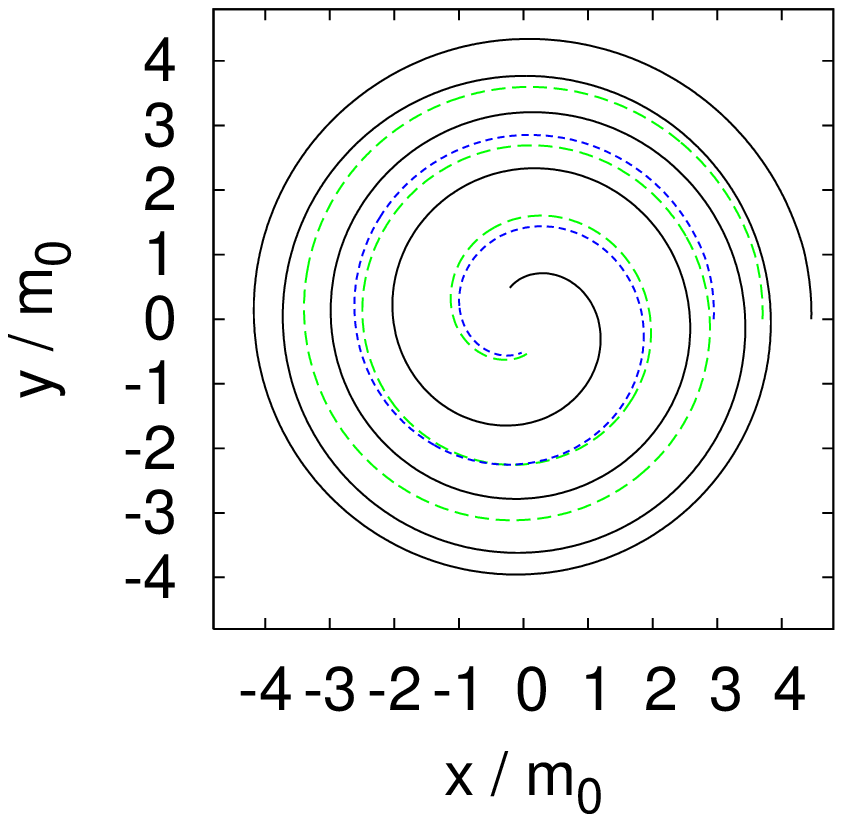}
\epsfxsize=3.2in
\leavevmode
\hspace{-10mm}(b)\epsffile{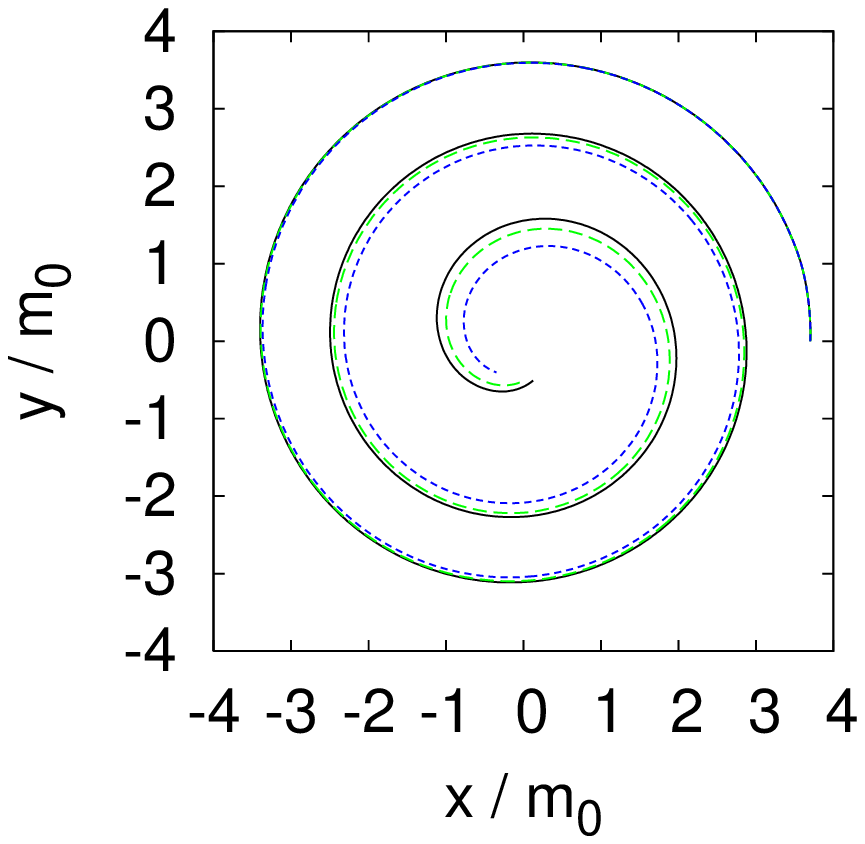} \\
\epsfxsize=3.2in
\leavevmode
(c)\epsffile{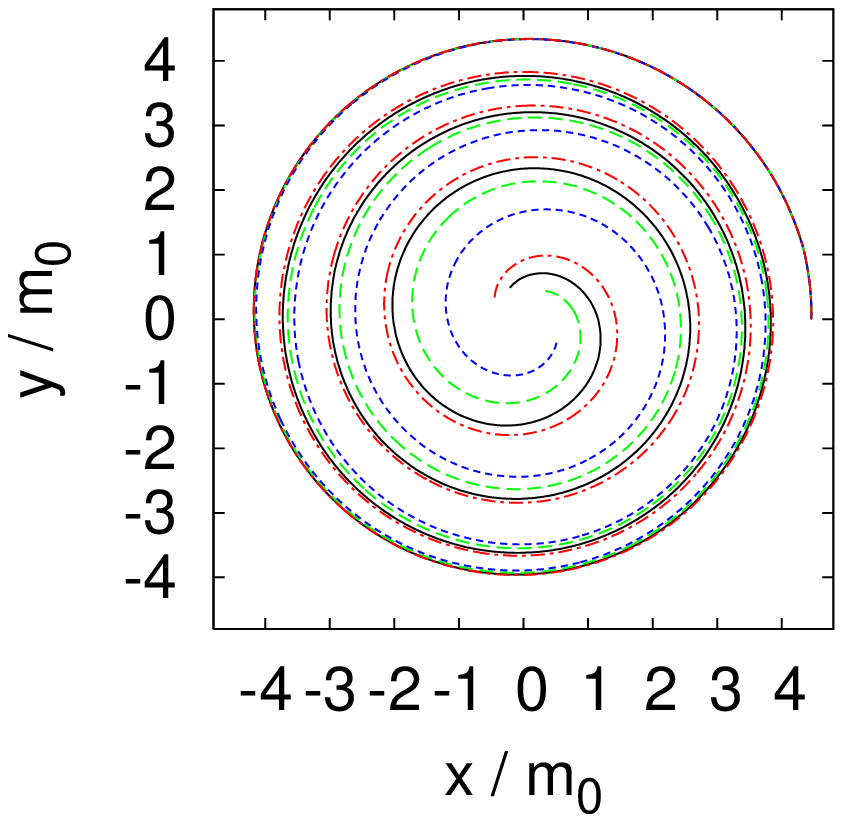}
\epsfxsize=3.2in
\leavevmode
\hspace{-10mm}(d)\epsffile{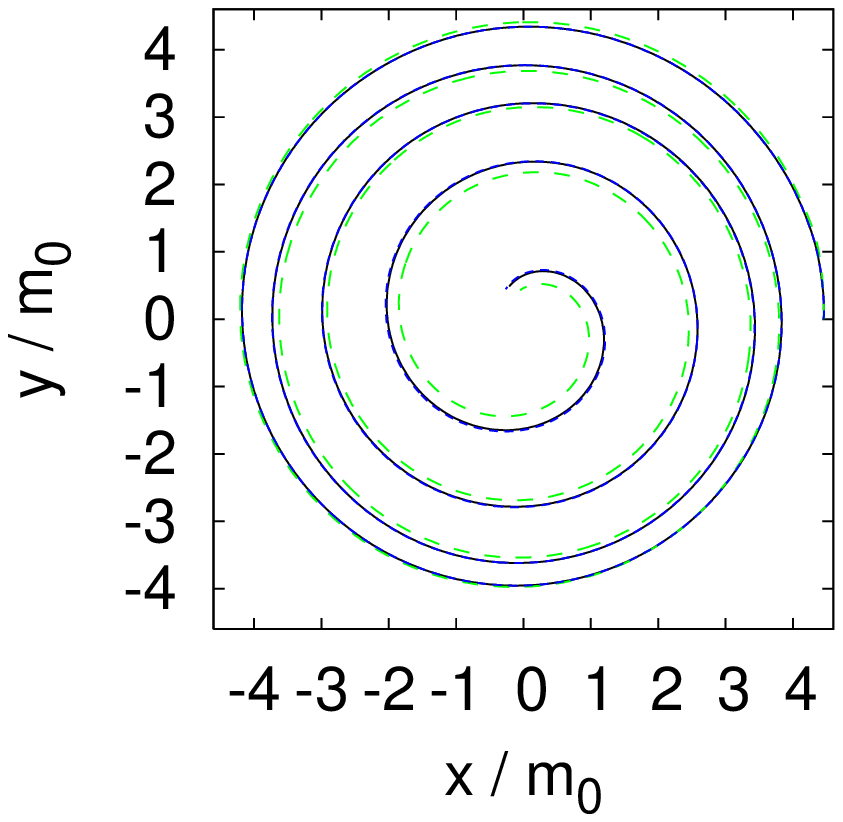}
\end{center}
\vspace{-4mm}
\caption{(a) Coordinate trajectories of a BH from $t=0$ to the
time at which common apparent horizon is first formed for runs
19a (solid curve), 16a (long-dashed curve), and 13a (dashed curve).
(b) The same as (a) but for runs 
16a (solid curve), 16c (long-dashed curve), and 16d (dashed curve).
(c) The same as (a) but for runs 
19a (solid curve), 19b (dotted-dashed curve),  
19c (long-dashed curve), and 19d (dashed curve).
(d) The same as (a) but for runs 19a (solid curve), 19aF (long-dashed curve),
and 19f (dashed curve). The orbits for runs 19a and 19f overlap
approximately. 
\label{FIG1}}
\end{figure*}

In the following three subsections
\ref{sec:res-bhbh}--\ref{sec:res-bhns}, we separately report numerical
results for BH-BH, NS-NS, and BH-NS binaries, respectively.  All the
numerical results obtained by SACRA were performed on
personal computers which possesses 2.4 or 2.6 or 3.0 GHz Opteron processor
and 4 or 8 GBytes memory. Many of numerical simulations were performed
both in the $F_i$-BSSN and $\tilde \Gamma^i$-BSSN formalisms. Although
both formalisms give similar results, slight quantitative difference
is also found.  (The difference results primarily from the difference
in the gauge conditions adopted in both formalisms.) In each following
section, we basically present the results in the $\tilde
\Gamma^i$-BSSN formalism. In the presence of remarkable quantitative
difference between two results, we will notice the difference.

\subsection{BH-BH Binaries} \label{sec:res-bhbh}

\begin{table}[t]
\caption{Parameters for BH-BH binaries in quasicircular states.  We
list the ADM mass ($M_0$), angular velocity ($\Omega_0$), angular
momentum ($J_0$), and a spin parameter of binary ($\chi$). All these
quantities are scaled with respect to $m_0$ which is sum of
irreducible mass of two BHs at $t=0$.}
\label{bhbh_ini}
\begin{tabular}{ccccc} \hline
~~~~~$d$~~~~~ & $M_0/m_0$ & $m_0\Omega_0$ & $J_0/m_0^2$ & $\chi$ \\ \hline
13 & ~0.9858~ & ~0.05617~ & ~0.875~ &~0.054\\ \hline
16 & ~0.9875~ & ~0.04164~ & ~0.911~ &~0.040\\ \hline
19 & ~0.9890~ & ~0.03245~ & ~0.951~ &~0.032\\ \hline
\end{tabular}
\end{table}

\begin{table*}[th]
\caption{Parameters of grid structure for simulations of BH-BH
binaries.  In the column named ``Levels'', the number of total refinement
levels is written.  (In the bracket, the numbers of coarser and finer
levels are written.)  $\Delta x$ is minimum grid spacing, $\Delta
x_{\rm QNM}$ the grid spacing at which quasi-normal mode of
gravitational waves are extracted, $\Delta x_{\rm ins}$ the grid
spacing at which inspiral gravitational waves are extracted, $m_1$ the
irreducible mass of each BH, $L$ the location of outer 
boundaries along each axis, and $\lambda_0$ the gravitational
wavelength at $t=0$.  Run names with ``F'' denote that the simulations were
performed with the $F_i$-BSSN formalism, and otherwise, the simulations
were performed with the $\tilde \Gamma^i$-BSSN formalism.
\label{BHBHGRID}}
\begin{tabular}{cccccccc} \hline
Run & ~~~''$d$''~~~ & Levels & ~~$N$~~ & ~$\Delta x/m_1$~ & ~$L/m_0~(L/\lambda_0)$~
& ~$\Delta x_{\rm QNM}/m_0$~ & ~$\Delta x_{\rm ins} /m_0$~ \\ \hline
~13a, aF~&13&9 (5+4)& 24 & 0.0566 &  174 ~(3.1) & 0.46, 0.91 & 1.82 \\ \hline
13b    & 13 & 9 (5+4) & 20 & 0.0680 &  174 ~(3.1) & 0.55, 1.09 & 2.18 \\ \hline
13c    & 13 & 9 (5+4) & 16 & 0.0850 &  174 ~(3.1) & 0.68, 1.36 & 2.72 \\ \hline
13d    & 13 & 9 (5+4) & 24 & 0.0708 &  217 ~(3.9) & 0.57, 1.13 & 2.24 \\ \hline
16a, aF& 16 & 9 (5+4) & 30 & 0.0578 &  222 ~(2.9) & 0.46, 0.92 & 1.84 \\ \hline
16b, bF& 16 & 9 (5+4) & 24 & 0.0578 &  178 ~(2.3) & 0.46, 0.92 & 1.84 \\ \hline
16c, cF& 16 & 9 (5+4) & 24 & 0.0723 &  222 ~(2.9) & 0.58, 1.16 & 2.32 \\ \hline
16d, dF& 16 & 9 (5+4) & 20 & 0.0868 &  222 ~(2.9) & 0.70, 1.39 & 2.78 \\ \hline
19a, aF& 19 & 9 (5+4) & 30 & 0.0587 &  225 ~(2.4) & 0.47, 0.94 & 1.88 \\ \hline
19b, bF& 19 & 9 (5+4) & 24 & 0.0587 &  180 ~(1.9) & 0.47, 0.94 & 1.88 \\ \hline
19c, cF& 19 & 9 (5+4) & 24 & 0.0733 &  225 ~(2.4) & 0.59, 1.18 & 2.35 \\ \hline
19d, dF& 19 & 9 (5+4) & 20 & 0.0880 &  225 ~(2.4) & 0.71, 1.41 & 2.82 \\ \hline
19e, eF& 19 & 10(6+4) & 24 & 0.0587 &  360 ~(3.8) & 0.47, 0.94 & 1.88 \\ \hline
19f    & 19 & 9 (5+4) & 36 & 0.0587 &  270 ~(2.9) & 0.47, 0.94 & 1.88 \\ \hline
19g    & 19 & 9 (5+4) & 24 & 0.0880 &  270 ~(2.9) & 0.71, 1.41 & 2.82 \\ \hline
\end{tabular}
\end{table*}

The first step is to validate the Einstein equations solver of
SACRA. For this purpose, we performed simulations of BH-BH binaries of
equal mass. Because many simulations have been already performed for
the equal-mass binary in the past three years (see
Sec.~\ref{sec:intro} for review), it is possible to compare our
numerical results with the previous ones and to check the validity of
our code.

\subsubsection{Initial condition}

Following Ref.~\cite{BHBH12}, as initial conditions, we adopt 
quasiequilibrium states of BH-BH binaries in corotating circular
orbits, which are computed by Cook and Pfeiffer \cite{CP} (see also
\cite{PKST,CCGP}) in the conformal-thin sandwich framework. The data can be
obtained from http://twww.black-holes.org/researcher3.html/.  Cook, 
Pfeiffer, and their collaborators have computed a wide variety of
quasiequilibrium states by a spectral method with a high
accuracy. Among many quasiequilibria they computed, we pick up the
corotating models with labels $d=13$, 16, and 19 (see Table
\ref{bhbh_ini} for key quantities of these initial conditions)
following a previous work \cite{BHBH12}.  These initial conditions are
computed in an excision method \cite{CP}, and hence, no data is
present inside apparent horizons. We simply adopt a third-order Lagrangian
interpolation to provide a spurious data inside the apparent horizons. 
As shown in Refs.~\cite{BSSTDHP,EFLSB}, this quite simple method is
acceptable because the spurious information inside the apparent
horizons does not propagate outward. Indeed, no trouble was found also
in our simulations.  As shown in Ref.~\cite{BHBH12}, BH-BH binaries
orbit for about 1.5, 2.5, and 4.5 times before formation of common
apparent horizon for $d=13$, 16, and 19, respectively. 


\subsubsection{Setting}

The simulations were performed changing the grid resolution and grid
structure for a wide range, to examine convergence of the numerical
results as well as to check dependence of the results on locations of
outer and refinement boundaries (see Table \ref{BHBHGRID} for the key
parameters of the grid structure). The numerical experiments were 
extensively performed, in particular, for $d=19$.  For all the cases,
the grid spacing in the vicinity of BHs is between $\approx m_1/12$
and $\approx m_1/18$ ($m_1$ is the irreducible mass of each BH), and
the outer boundaries along each axis are located at 2--4 times of
gravitational wavelength at $t=0$ (which is denoted by $\lambda_0$).

Instead of employing the solution of quasiequilibrium states
for $\alpha$ and $\beta^k$ at $t=0$, we initially give 
\beqn
\alpha=W~{\rm and}~\beta^k=0. \label{gauge0}
\eeqn
We also performed simulations with the quasiequilibrium gauge as
initial condition for a few models (see Appendix \ref{app:res}).
Switching the initial condition for $\alpha$ from Eq. (\ref{gauge0})
to the quasiequilibrium one does not change the numerical results
significantly.  By contrast, using the quasiequilibrium solution 
for $\beta^k$, the orbital trajectory of BHs
(in coordinate description) are significantly modified for the case
that the $\tilde \Gamma^i$-BSSN formalism is employed (see also
\cite{EFLSB}).  Specifically, the orbit becomes elliptical in the
coordinate description.  By contrast, for the case that the $F_i$-BSSN
formalism is employed, numerical results depend only weakly on the
initial condition. In both cases, physical results (e.g.,
gravitational waveforms and state of the BH finally formed) depend
very weakly on the initial condition. The results are briefly
presented in Appendix \ref{app:res}.

The elliptical orbit in the $\Gamma$-freezing gauge is likely to result 
simply from a gauge effect. However, the gauge could affect the
physical results (see discussion below), and hence, it is better to
fix the condition for studying convergence of the numerical results
for different grid resolutions. In the present paper, we employ the
gauge condition of Eq. (\ref{gauge0}) at $t=0$ following \cite{EFLSB},
and discuss the convergence and dependence of numerical results on the
grid structure fixing the initial condition for $\alpha$ and $\beta^i$. 

Most of the simulations were performed with $N=30$ or 24. Required
memories for runs with $N=30$ and 24 are at most about 2.8 and 1.6
GBytes, respectively, when the $\tilde \Gamma^i$-BSSN formalism is
employed.  When the $F_i$-BSSN formalism is employed, we do not have
to introduce the auxiliary variable $B^i$, and hence, the memory is
slightly saved. In both cases, the simulations are feasible on
inexpensive personal computers of 4 GBytes memory. A few simulations
were performed for $N=36$, but it is still feasible by personal
computers of 8 GBytes memory. The computation time required for run
19a, for which binary orbits for about 4.5 times before the onset of
merger, is about two weeks on 2.4 GHz Opteron machine, even with no
parallelization.  For $d \leq 16$, the required computational time is
at most 10 days even for $N=30$.

\begin{table*}[th]
\caption{Numerical results for simulations of BH-BH binaries.  We list
the time at which common apparent horizon is formed ($T_{\rm
AH}$), final value of the irreducible mass for the common apparent
horizon ($M_{\rm irr}$), final value of the ratio of the polar
circumferential length to the equatorial one ($C_p/C_e$), final BH
mass estimated from the equatorial circumferential length
($C_e/4\pi$), BH mass estimated from $M_{\rm irr}$ and $C_p/C_e$
($M_{\rm BHf}$), final spin parameter of the BH estimated from
$C_p/C_e$, and energy and angular momentum carried away by
gravitational waves ($\Delta E$ and $\Delta J$). ``---'' denotes that
the values of the area and $C_p/C_e$ do not relax to constants because
of the poor grid resolution.  The last column denotes the refinement
level in which the common apparent horizon is determined. ``1'' and
``2'' are the finest and second-finest levels, respectively. For runs
16dF, 19cF, 19d, and 19dF, the area and the circumferential length of the
apparent horizon vary with time and the values are not determined with a
good accuracy. 
\label{BHBHRES}}
\begin{tabular}{cccccccccc} \hline
~Run~ & ~$T_{\rm AH}/m_0$~ & ~$M_{\rm irr}/m_0$~ & ~$C_p/C_e$~
& ~$C_e/(4\pi m_0)$~& ~$M_{\rm BHf}/m_0$~ & ~$a$~ & ~$\Delta E/m_0$~
& ~$\Delta J/J_0$~ & ~Level~ \\ \hline
13a & 125.3 & 0.873       & 0.887 & 0.946 & 0.946 & 0.712 &0.034&0.24& $L$-2 \\ \hline
13aF& 125.7 & 0.873       & 0.884 & 0.948 & 0.948 & 0.720 &0.034&0.24& $L$-1 \\ \hline
13b & 123.8 & 0.872       & 0.888 & 0.946 & 0.945 & 0.710 &0.034&0.24& $L$-2 \\ \hline
13c & 122.3 & $\sim 0.873$&  ---  & 0.944 &  ---  &  ---  &0.033&0.23& $L$-2 \\ \hline
13d & 123.8 & 0.871       & 0.884 & 0.947 & 0.946 & 0.720 &0.033&0.23& $L$-1 \\ \hline \hline
16a & 256.8 & 0.876       & 0.889 & 0.948 & 0.948 & 0.707 &0.035&0.27& $L$-1 \\ \hline
16aF& 253.6 & 0.876       & 0.888 & 0.948 & 0.949 & 0.709 &0.035&0.26& $L$-1 \\ \hline
16b & 257.1 & 0.876       & 0.889 & 0.947 & 0.948 & 0.707 &0.035&0.27& $L$-2 \\ \hline
16b'& 255.4 & 0.876       & 0.889 & 0.948 & 0.949 & 0.707 &0.035&0.27& $L$-1 \\ \hline
16bF& 268.6 & 0.876       & 0.888 & 0.948 & 0.948 & 0.709 &0.035&0.26& $L$-1 \\ \hline
16c & 250.2 & 0.876       & 0.888 & 0.948 & 0.949 & 0.710 &0.034&0.26& $L$-1 \\ \hline
16cF& 245.3 & 0.877       & 0.889 & 0.949 & 0.949 & 0.707 &0.033&0.25& $L$-1 \\ \hline
16d & 237.0 & 0.876       & 0.889 & 0.948 & 0.948 & 0.707 &0.035&0.27& $L$-1 \\ \hline
16dF& 220.7 & $\approx 0.881$ & $\approx 0.895$ & $\approx 0.949$ & $\approx 0.949$ & $\approx 0.69$ & 0.029 & 0.23& $L$-1 \\ \hline \hline
19a & 516.7 & 0.879       & 0.891 & 0.949 & 0.950 & 0.702 &0.036 &0.29& $L$-1 \\ \hline
19aF& 499.3 & 0.878       & 0.891 & 0.949 & 0.950 & 0.703 &0.035& 0.28& $L$-1 \\ \hline
19b & 535.9 & 0.878       & 0.892 & 0.947 & 0.948 & 0.699 &0.036 &0.29& $L$-2 \\ \hline
19bF& 582.8 & 0.877       & 0.890 & 0.949 & 0.949 & 0.706 &0.035 &0.29& $L$-1 \\ \hline
19c & 491.6 & 0.878       & 0.890 & 0.949 & 0.949 & 0.705 &0.036 &0.28& $L$-1 \\ \hline
19cF& 488.9 &$\approx 0.882$&0.893 &$\approx 0.950$&0.951&0.696&0.034 &0.27& $L$-1 \\ \hline
19d & 456.8 &$\approx 0.878$&0.891 &$\approx 0.948$&0.949&0.701&0.034 &0.27& $L$-1 \\ \hline
19dF& 449.1 & $\approx 0.884$ & $\approx 0.898$ & $\approx 0.950$
& $\approx 0.950$ & $\approx 0.68$ &0.030 &0.24& $L$-1 \\ \hline 
19e & 535.9 & 0.878       & 0.892 & 0.947 & 0.948 & 0.699 &0.036 &0.29& $L$-2 \\ \hline
19eF& 582.8 & 0.877       & 0.890 & 0.949 & 0.949 & 0.705 &0.035 &0.29& $L$-1 \\ \hline
19f & 517.8 & 0.879       & 0.891 & 0.949 & 0.950 & 0.703 &0.036 &0.29 & $L$-1 \\ \hline
19g & 452.3 & 0.878       & 0.890 & 0.949 & 0.950 & 0.704 &0.034 &0.27 & $L$-1 \\ \hline
\end{tabular}
\end{table*}

\subsubsection{Evolution of BHs and final outcome}

Figure \ref{FIG1}(a) plots orbital trajectories of one of two BHs for
runs 19a, 16a, and 13a. The trajectories from $t=0$ to the time at
which common apparent horizon is first formed are drawn. This shows
that for $d=19$, 16, and 13, the BH-BH binaries orbit approximately
for 4.3, 2.75, and 1.75 times, respectively.  The result for $d=19$ 
approximately agrees with that of Ref.~\cite{BHBH12}, whereas for $d=16$
and 13, our results are by about a quarter orbit longer (see
discussion below). As pointed out in Ref.~\cite{BHBH12}, the trajectory 
for $d=19$ looks slightly eccentric, whereas for $d=16$ and 13, the
eccentricity is not very outstanding. 

Figure \ref{FIG1}(b) and (c) are the same as Fig. \ref{FIG1}(a) but
for runs 16a, 16c, 16d and for runs 19a, 19b, 19c, and 19d,
respectively. These two figures compare the trajectories in different
grid resolutions but with the same arrangement for locations of
refinement boundaries. They show that for the finer grid resolutions,
the number of orbits increases, i.e., the time at which common
apparent horizon is first formed (hereafter referred to as the merger
time, $T_{\rm AH}$) is longer. The reason for this feature is that
numerical dissipation is larger for the simulations with poorer grid
resolutions, and as a result, the decrease rate of orbital separation
is spuriously enhanced. However, Fig. \ref{FIG1}(b) indicates that the
difference in the merger time is not very large for $d=16$, and
suggests that the numerical results are close to convergence. For
$d=13$ and 16, we infer that in the best-resolved runs, the merger
time is determined within an error of $\sim 2m_0$ and $10m_0$,
respectively. By contrast, for $d=19$, the merger time may be
underestimated by $\sim 50m_0$ even for run 19a. This point will be
revisited in Sec.~\ref{sec5a3}.

The trajectory of BHs for run 16b is very similar to that for 16a (we
do not plot it because it approximately agrees with that for run 16a). 
By contrast, the trajectory for run 19b does not agree well with that
for run 19a (see also Table \ref{BHBHRES} for the merger time which
shows that the difference in the merger time is $\sim 20m_0$). This
indicates that for the simulations started from small initial orbital
separations ($d\leq 16$), our choice for the location of outer and
refinement boundaries and for the grid structure is appropriate. On
the other hand, for a simulation started from a large initial
separation as $d=19$, a careful choice of the grid structure is
necessary. In addition, the trajectory and merger time depend on the
gauge condition; see comparison between the results with $F_i$-BSSN
and $\tilde \Gamma^i$-BSSN formalisms, for which the chosen spatial
gauges are different (see Sec.~\ref{sec5a3}).

\begin{figure*}[t]
\epsfxsize=3.in \leavevmode (a)\epsffile{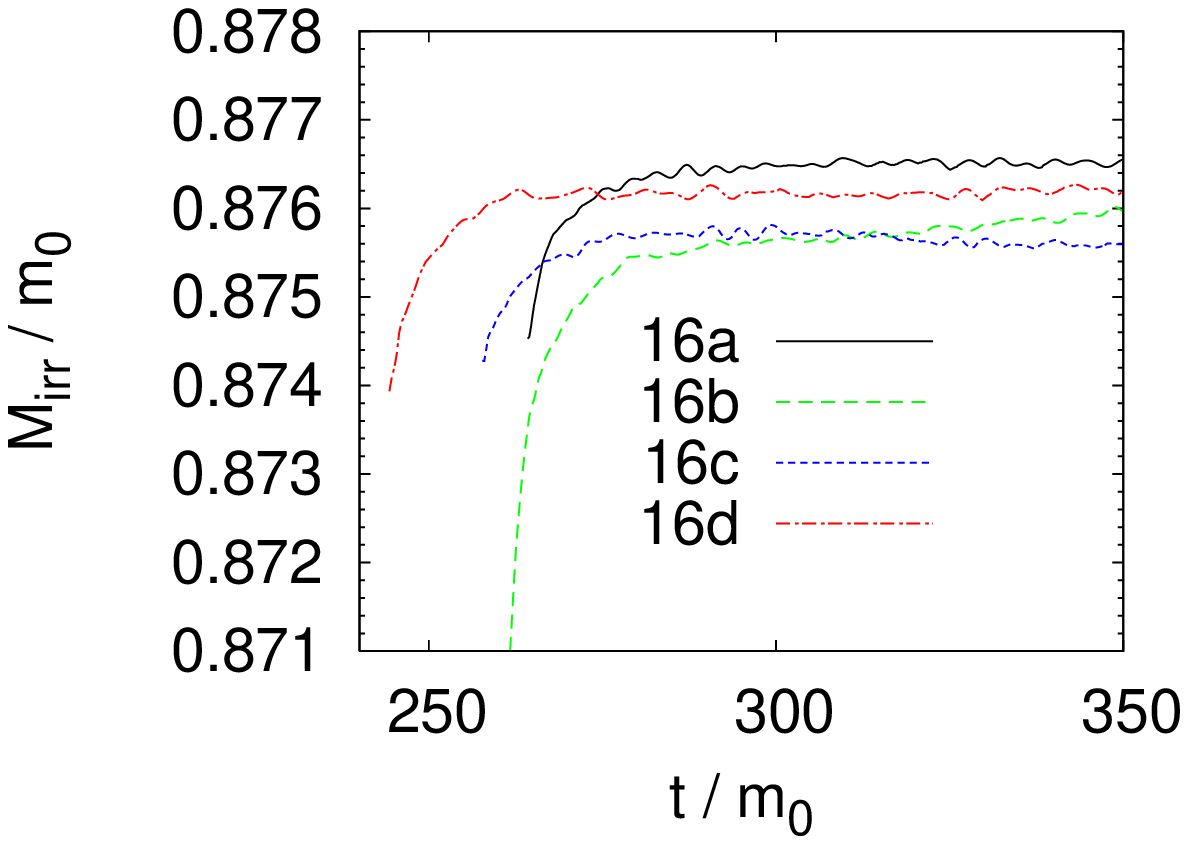} \epsfxsize=3.in
\leavevmode ~~~~~(d)\epsffile{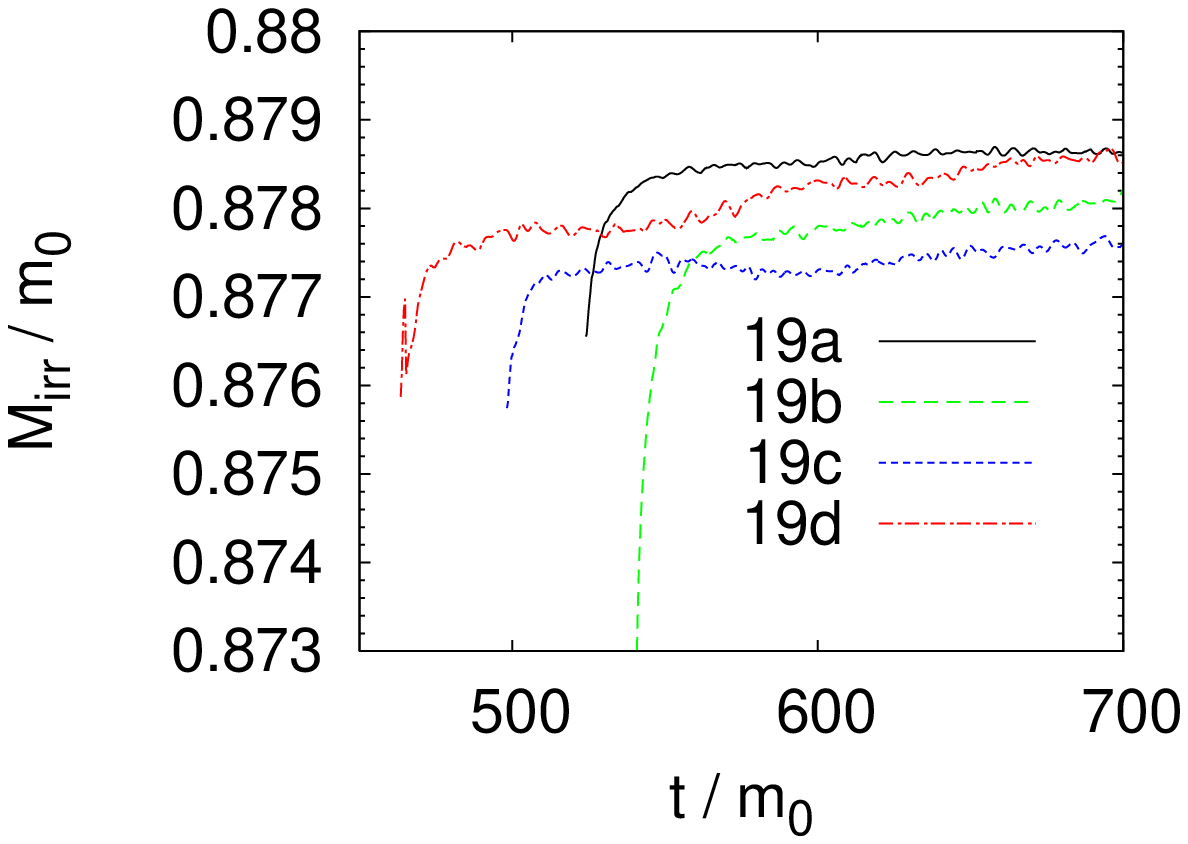} \\ \epsfxsize=3.in \leavevmode
(b)\epsffile{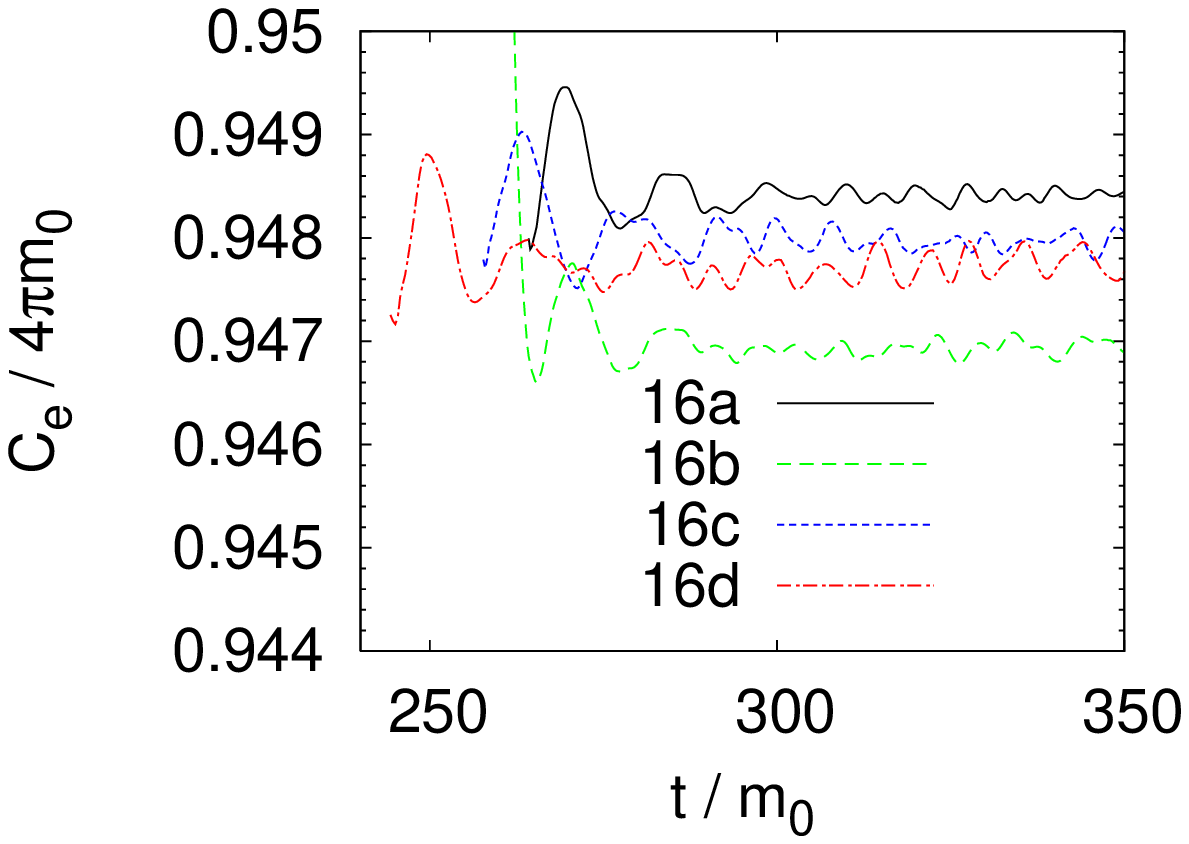} \epsfxsize=3.in \leavevmode ~~~~~
(e)\epsffile{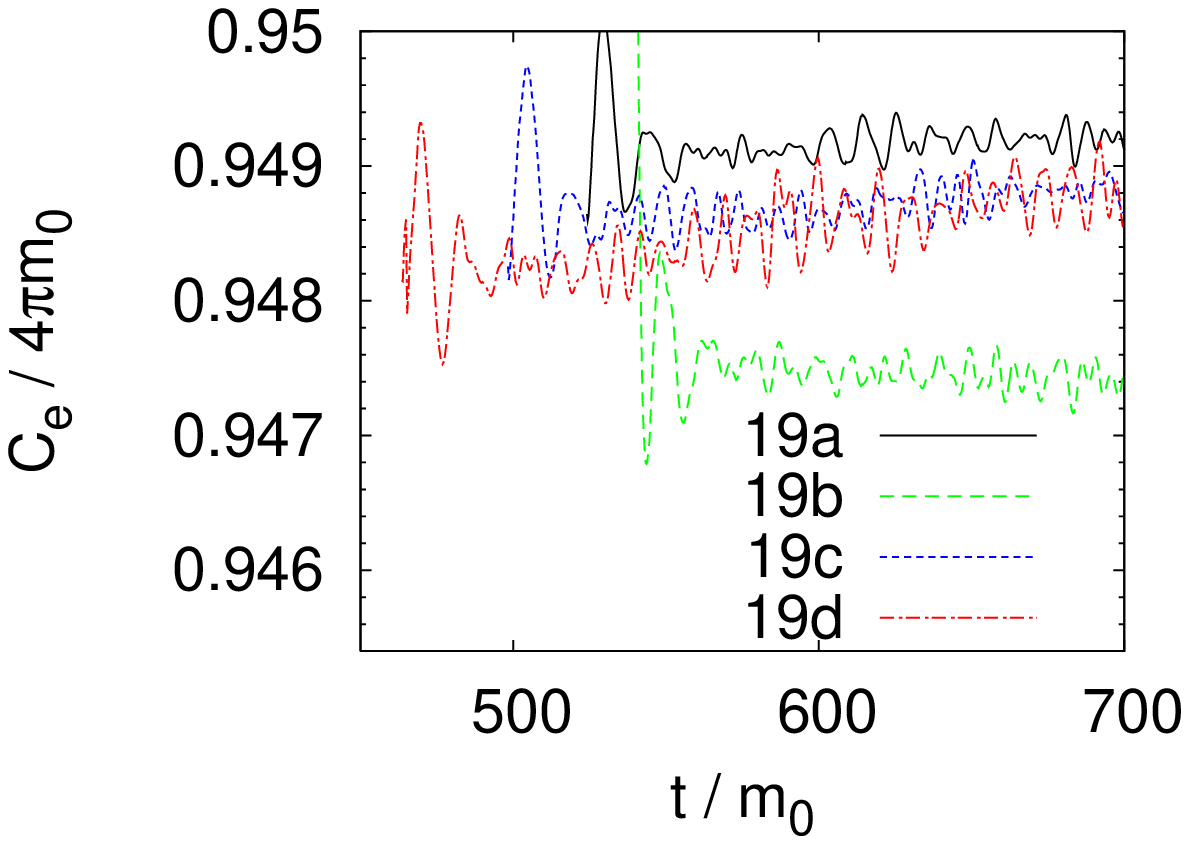} \\ \epsfxsize=3.in \leavevmode
(c)\epsffile{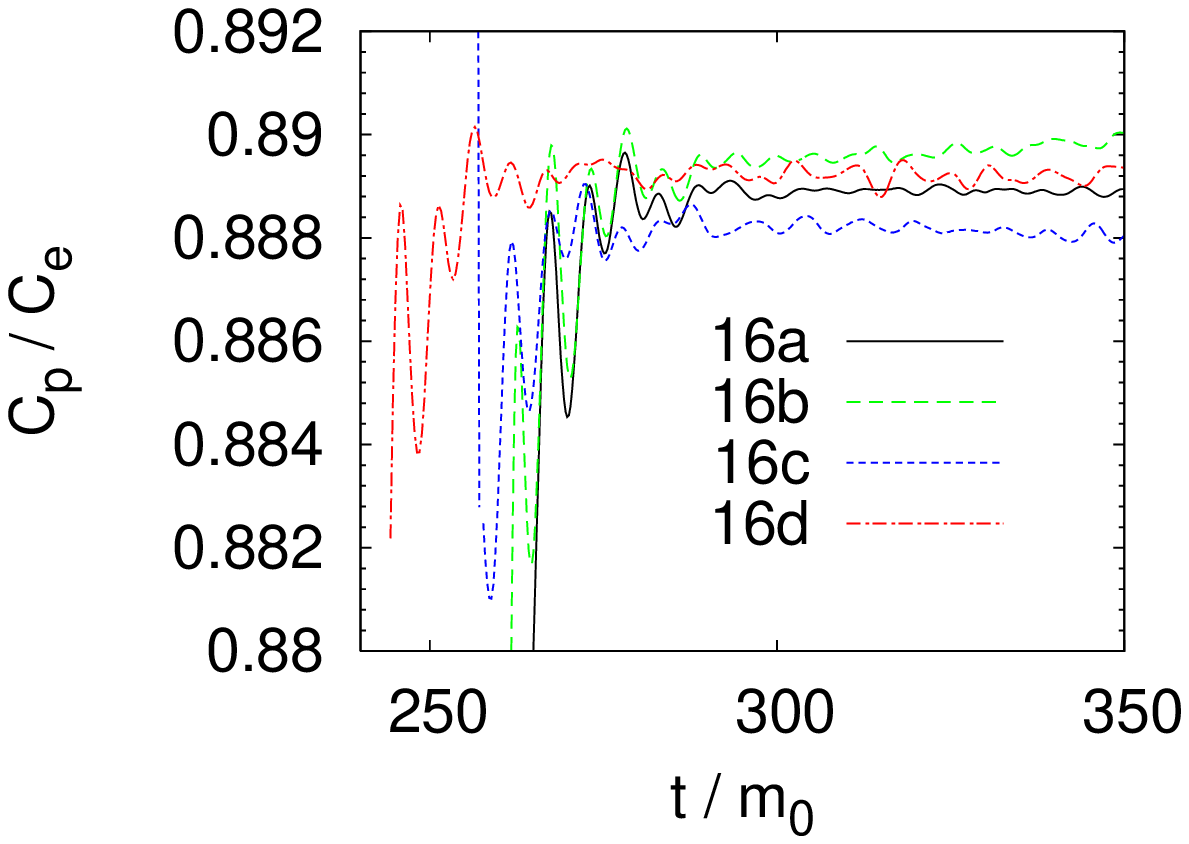} \epsfxsize=3.in \leavevmode ~~~~~
(f)\epsffile{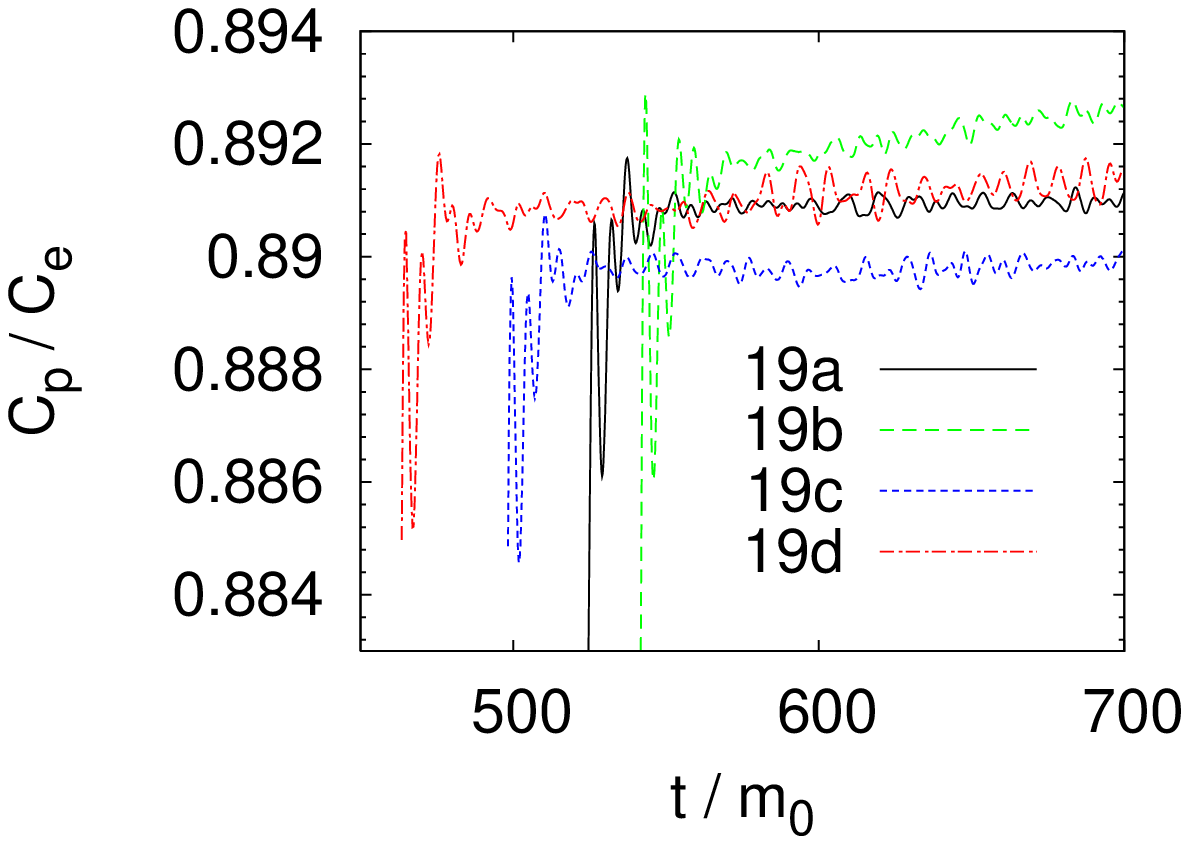}
\vspace{-2mm}
\caption{(a) $M_{\rm irr}/m_0$ as a function of time for a BH formed
after merger for run 16a--16d.  (b) The same as (a) but for
$C_e/4\pi m_0$. (c) The same as (a) but for $C_p/C_e$.  (d) The same as
(a) but for runs 19a--19d. (e) The same as (b) but for runs
19a--19d. (f) The same as (c) but for runs 19a--19d. We note that 
numerical error of the results for runs 16b and 19b is larger than
those for runs 16a and 19a because the apparent horizon for these runs
is determined from the data of the second-finest AMR level. 
\label{FIG2}}
\end{figure*}

Figure \ref{FIG2} plots $M_{\rm irr}/m_0$, $C_e/4\pi m_0$, and
$C_p/C_e$ as functions of time for common apparent horizon for $d=16$
and 19. The asymptotic values of these quantities characterize
properties of the final state of the formed BHs, as described in
Sec.~\ref{sec:diag}. Figure \ref{FIG2} shows that all the quantities
approach approximately to constants and the formed BHs relax to a
stationary state irrespective of initial orbital separation. An
oscillation associated with numerical error is seen, but the amplitude
of such oscillation is within $\sim 0.1\%$. Thus, the final stationary
state of the BHs is determined with a small error of $\alt 0.1\%$
(except for runs performed with a poor grid resolution such as runs
16dF, 19d, and 19dF, for which values for these quantities do not
approach to constants).

In Table \ref{BHBHRES}, we summarize key numerical results about the
formed BHs.  We note that the last column of Table \ref{BHBHRES}
denotes the refinement level for which the properties of the common
apparent horizon are determined; ``$L$-1'' and ``$L$-2 denote the
finest and second-finest levels, respectively. For the case that
volume of the finest refinement domain is so small that the radius of
the common apparent horizon is larger than the domain size, we have to
determine it in the second-finest one for analyzing the properties of the
BH. Because its resolution is poorer than that of the finest one, we
have to keep in mind that systematic error for the results marked
with ``$L$-2'' is larger than that with ''$L$-1''. In particular, a
substantial error appears to be always present for the estimated mass;
by comparing the results determined in the finest and second-finest
levels, we find that the mass is underestimated by $\sim 0.2\%$ when
the results in the second-finest level is used.

Although such systematic error is present, Table \ref{BHBHRES} shows
that the results for the properties of the BH finally formed depend 
weakly on the grid resolution, grid structure, chosen formalism, and
gauge condition: The final mass determined both from $M_{\rm irr}$
and $C_e$ is $(0.948 \pm 0.001) m_0$ for $d=13$ and 16, and $(0.949
\pm 0.001) m_0$ for $d=19$. The final spin determined from $C_p/C_e$ 
is $0.71 \pm 0.01$ for $d=13$ and 16 and $0.70\pm 0.01$ for $d=19$. 
These results agree with those of \cite{BHBH12} within estimated
numerical error.

In our results, the final masses of the BHs computed both from $C_e$
and Eq. (\ref{AAH}) agree within $\approx 0.1\%$ error. Because two
values are determined by two independent methods, this agreement also
indicates that the BH mass is determined within $\sim 0.1\%$ error.

Another point worth noting is that the final mass and spin depend very
weakly on the initial orbital separation. This is natural because the
merger should start at an approximately unique point in the vicinity
of an innermost stable orbit at which the energy and angular momentum
of the binary system is approximately identical independent of the
initial orbital separation. Note that slight difference in spin of
individual BHs could cause a slight difference of the location of the
innermost stable orbit. However, the magnitude of the spin is small
and the effect is minor. Hence, after the merger sets in, the
evolution path toward the final state and the final outcome should 
depend only weakly on the initial separation.

\begin{figure*}[t]
\epsfxsize=3.2in
\leavevmode
(a)\epsffile{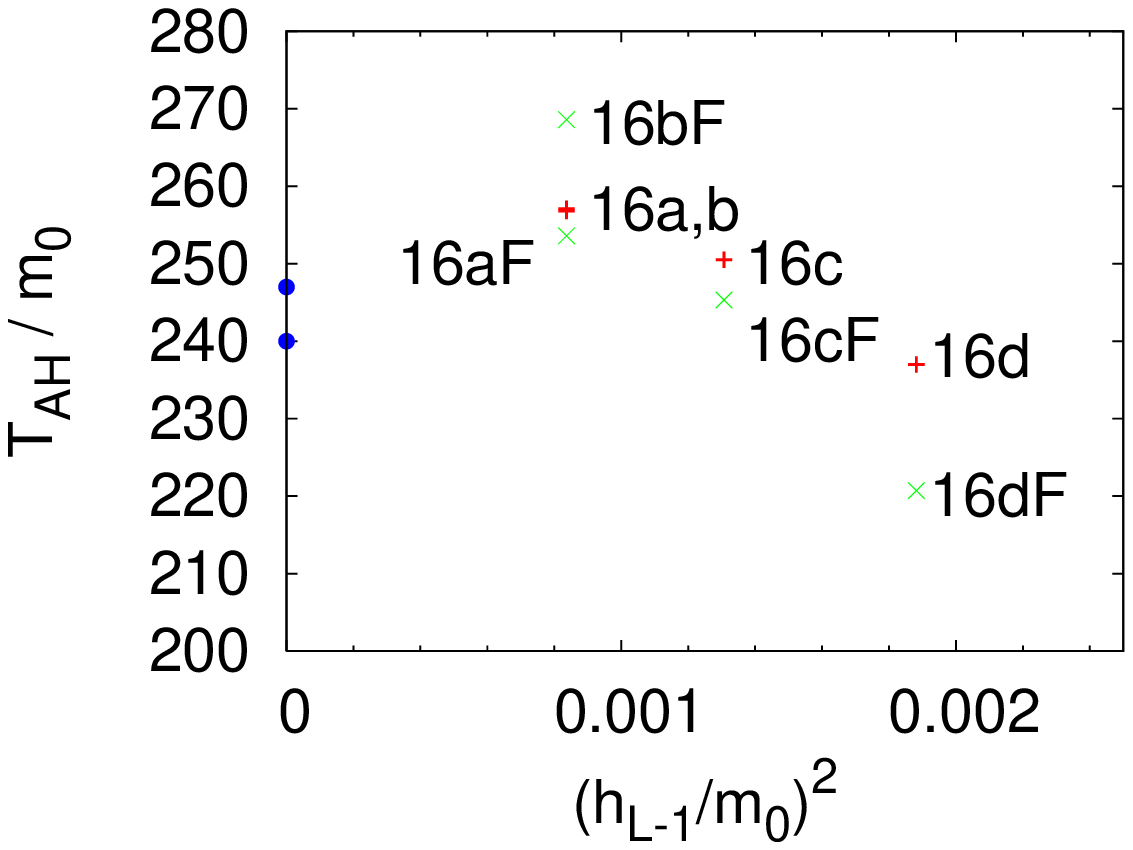}
\epsfxsize=3.2in
\leavevmode
~~~(b)\epsffile{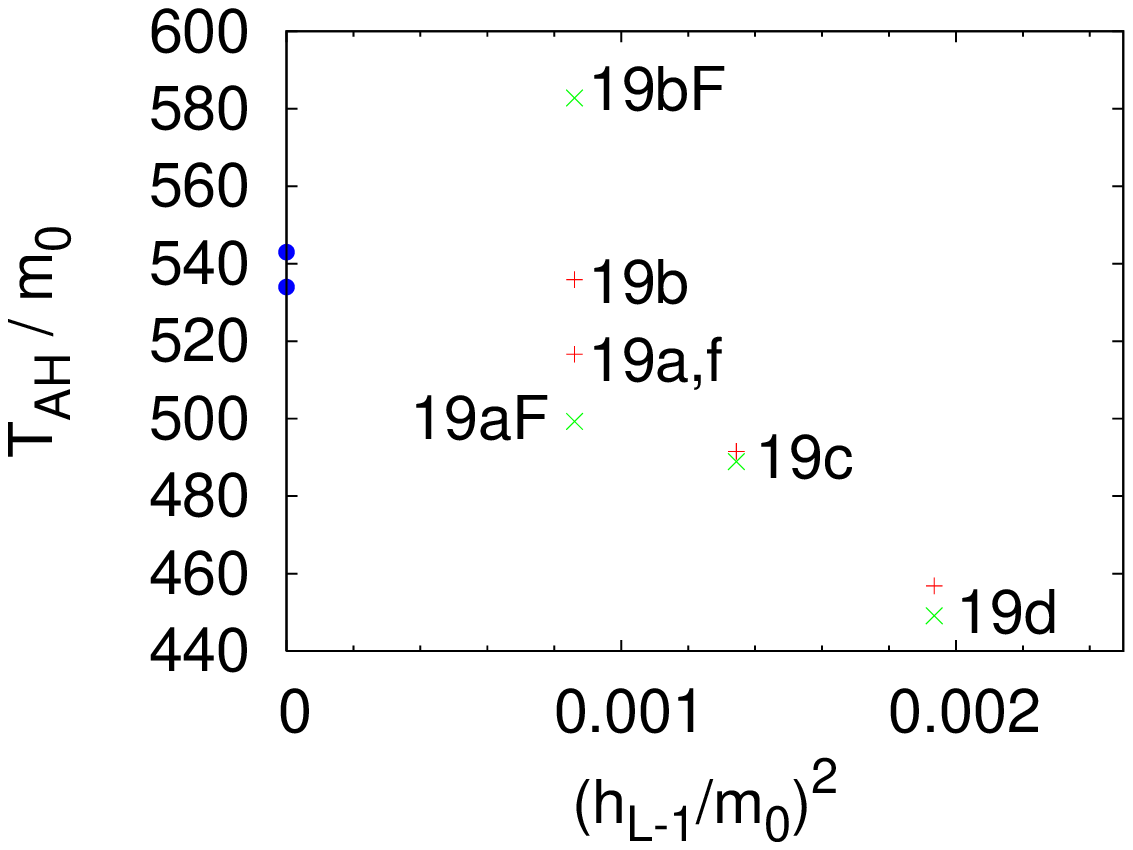}
\vspace{-2mm}
\caption{Merger time $T_{\rm AH}/m_0$ as a function of the square of
the finest grid resolution $h_{L-1}^2$ (a) for $d=16$ and (b) for
$d=19$. The plus and cross denote the results in the $\tilde
\Gamma^i$- and $F_i$-BSSN formalisms, respectively.  Note that the
plots for runs 16a and 16b and for runs 19a and 19f approximately
overlap.  The filled circles on the vertical axis denote a merger time
predicted by the Taylor T4 formalism: The larger value denotes the
value derived including the spin effect of BHs, whereas the smaller
one is the value derived in the assumption of zero spin.
\label{FIG3}}
\end{figure*}

\begin{figure*}[thb]
\epsfxsize=3.3in
\leavevmode
(a)\epsffile{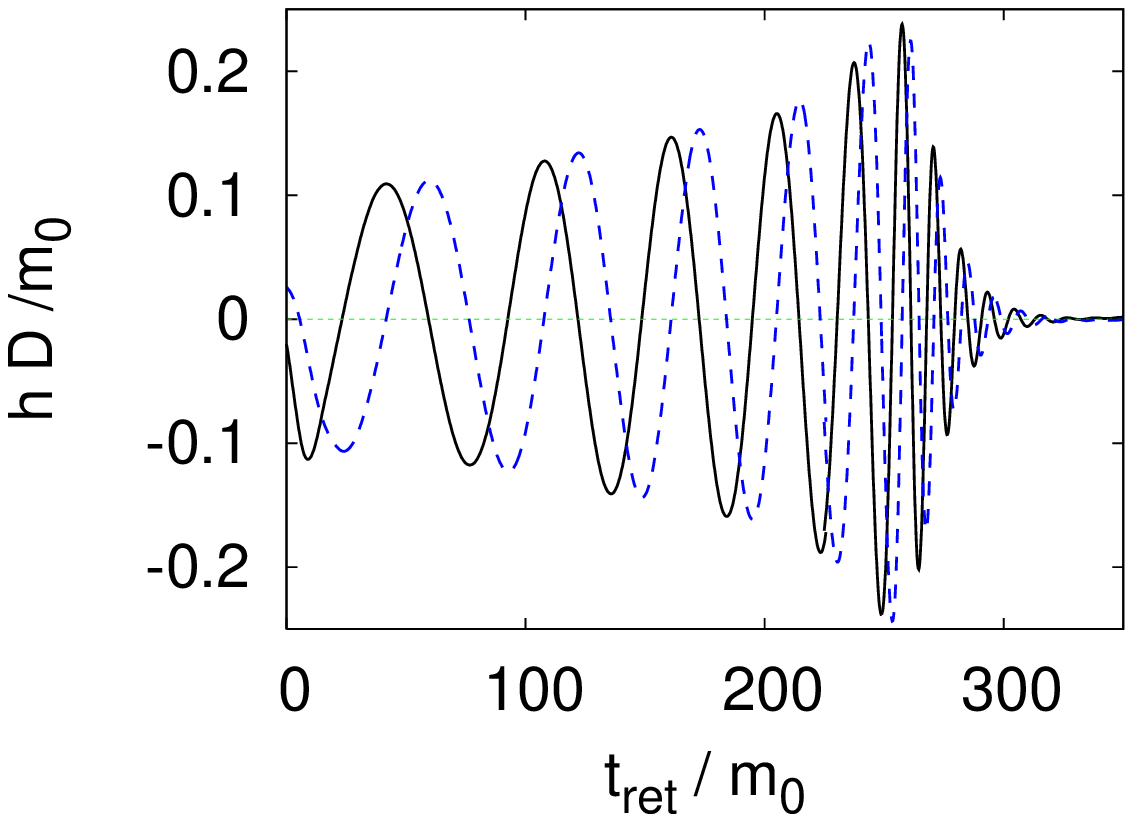}
\epsfxsize=3.3in
\leavevmode
(b)\epsffile{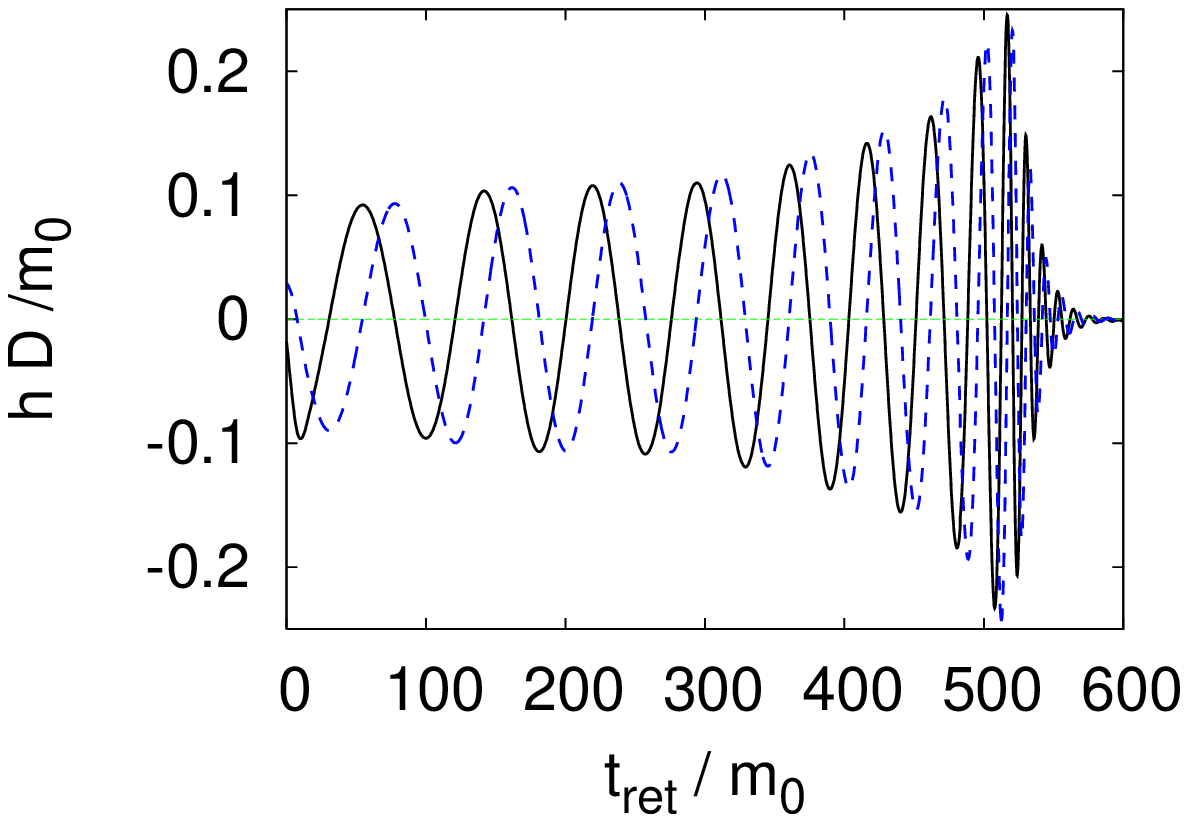}\\
\epsfxsize=3.3in
\leavevmode
(c)\epsffile{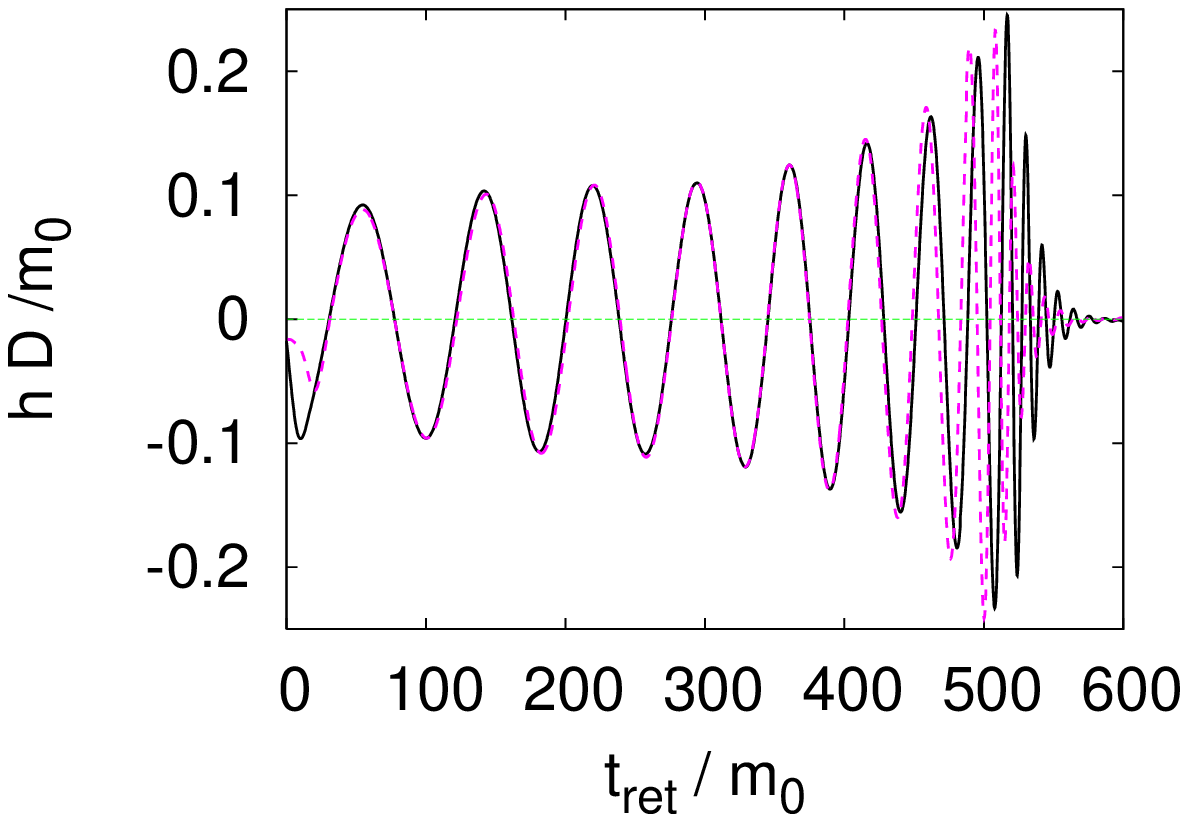}
\epsfxsize=3.3in
\leavevmode
(d)\epsffile{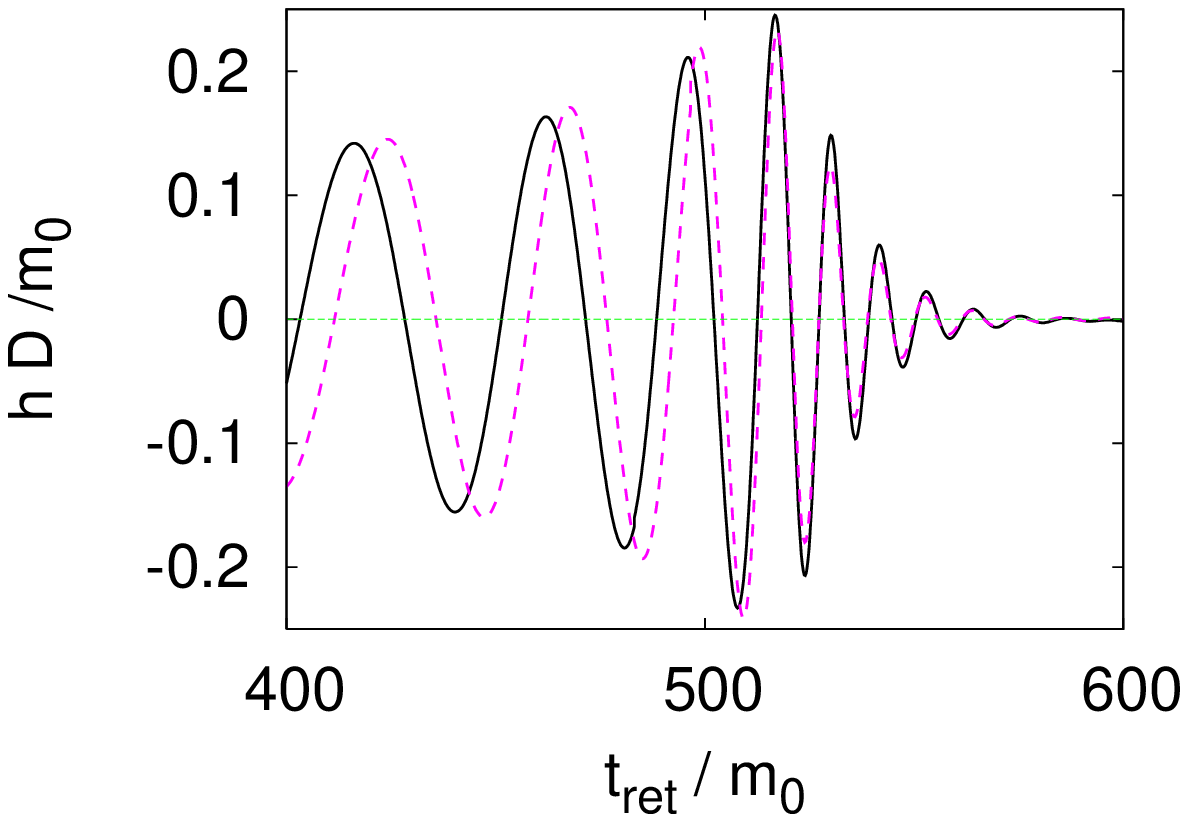}
\vspace{-2mm}
\caption{Gravitational waveforms (a) for run 16a and (b) for run
19a. The solid and dashed curves are plus and cross modes,
respectively. $D$ denotes distance from the source to an observer.
(c) Plus mode of gravitational waves for runs 19a (solid curve)
and 19c (dashed curve). For run 19c, the results are appropriately
shifted to match the inspiral waveform (see text). 
(d) The same as (c) but here we compare ringdown waveforms. 
\label{FIG4}}
\end{figure*}

\subsubsection{Merger time}\label{sec5a3}

In contrast to the results for the mass and spin of the BHs finally
formed, the merger time depends on the grid resolution for $d=16$ and
in particular for $d=19$. Because it increases systematically with
improving the grid resolution, the smaller merger time is a result due
to the fact that numerical dissipation is larger for the poorer grid
resolutions. To see the dependence of the merger time on the grid
resolution, we plot $T_{\rm AH}$ as a function of $h_{L-1}^2$ in
Fig. \ref{FIG3}. It is found that for a given location of outer and
refinement boundaries (compare the results for 16a, 16c, 16d and for
19a, 19c, 19d, respectively), $T_{\rm AH}$ systematically increases in
a manner better than second-order convergence.

The merger time for runs 16a and 16b (and also for runs 13a and 13b),
for which the finest grid resolution is the same whereas the locations
of outer and refinement boundaries at each level are different,
agrees approximately with each other. This implies that the grid
structures for these runs are well-suited for an accurate simulation;
the outer boundaries are located far enough to exclude spurious
effects associated with the finite size of computational domain, and
also, the refinement boundaries and the domain size of each level are
appropriately chosen. We can conclude that the results depend primarily
on the finest grid resolution as long as $N \geq 24$ and $L \agt
2\lambda_0$ for $d=13$ and 16.

In contrast to the results for $d=13$ and 16, the merger time for runs
19a and 19b does not agree well with each other.  This implies that
the orbital evolution of BHs depends either on the location of outer
boundaries or on the location of refinement boundaries. For $d=19$,
the BHs orbit for $\sim 4.5$ times. For such a long run, a small error
is likely to be accumulated, leading to a nonnegligible error. This
disagreement gives us a caution that careful choice of the grid
structure is necessary for the longterm evolution.

To clarify sources of the error in the merger time, we performed
additional simulations for $d=19$; runs 19e, 19f, and 19g (cf. Table
\ref{BHBHGRID}).  For run 19e, the location of refinement boundaries
is the same as that for run 19b, although outer boundaries are located
twice far away from the center. We found that numerical results for run 19e
agree very well with those for run 19b. This implies that the
numerical results do not depend on the location of outer boundaries
but on the location of refinement boundaries.

Additional runs 19f and 19g were performed to clarify the dependence 
of numerical results on the location of refinement boundaries, i.e.,
on the domain size of each refinement level. For these runs,
the domain size of each refinement level is 1.2 times as large as that 
for runs 19a, 19c, and 19d, whereas the grid resolution for runs 19a and 19f
and runs 19d and 19g are identical, respectively. We find that the
results for runs 19f and 19g agree well with those for runs 19a and
19d, respectively (see, e.g. Fig. \ref{FIG1}(d) for the trajectories
of runs 19a and 19f). By these results, we confirm that the location
of refinement boundaries (and the size of domain) for run 19a is
appropriately chosen: The error in the merger time comes primarily from
the grid resolution. In any case, the present numerical results show
that for simulations with a large initial orbital separation, a large
domain size of the refinement levels is required. 


The merger time for runs 19b and 19bF and for runs 16b and 16bF does
not agree, although that for 19a and 19aF (see, e.g.,
Fig. \ref{FIG1}(d)) and for 16a and 16aF, respectively, agrees in a
much better manner. Note that the domain size of each refinement level
for runs 16b and 16bF (19b and 19bF) is smaller than that for run 16a
and 16aF (19a and 19aF); see Table \ref{BHBHGRID}. This indicates that
if the outer boundaries are too close or the domain size of each
refinement level is too small, numerical results depend on the spatial
gauge condition and/or the formulation. To check the dependence on the
spatial gauge, we also performed a simulation in the $\tilde
\Gamma^i$-BSSN formalism with $\eta_s \approx 0.5/m_1$ for $d=16$ (run
16b').  With this change, the merger time changes by $\sim 2m_0$ (see
Table \ref{BHBHRES}), which is a fairly large difference. This
indicates that the difference in the spatial gauge seems to be the 
primary reason for discrepancy in the merger time. 

Because the spatial gauge condition does not affect the slicing, one
may think that the merger time should not depend on it.  However, this
is not correct in numerical computation because the spatial gauge
condition determines physical grid spacing between two grid points
even if the coordinate separation is the same. Namely, it affects the grid
resolution physically, and hence, determines magnitude of numerical
dissipation. Therefore, the merger time should depend on the chosen spatial
gauge condition in general. 

Due to the same reason, the physical location (not coordinate
location) of outer and refinement boundaries depends on the spatial gauge
condition. In particular, the physical size of the finest refinement 
level is likely to be sensitive to it. Thus, the magnitude of
numerical error (and resulting merger time), in particular around BHs
where the curvature is large, depends on the spatial gauge condition. 

Another characteristic feature in the simulation with the $F_i$-BSSN
formalism is that the merger time depends more strongly on the grid
resolution than the simulation in the $\tilde \Gamma^i$-BSSN
formalism.  For the poor-resolution simulations such as runs 16dF and
19dF, the merger time is much shorter than that for the corresponding
finer-resolution simulations, and the quantities for the formed BHs
after the merger are not determined accurately. Probably, this is also
due to the fact that in the chosen spatial gauge condition, the BHs
are not resolved well.

All these results suggest that with the $F_i$-BSSN formalism,
systematic errors associated with a finite location of outer
boundaries and/or finite grid resolutions are larger. However, in the
case that the appropriate location of the outer boundaries and
the appropriate
grid resolution are chosen, both the $F_i$-BSSN and $\tilde
\Gamma^i$-BSSN formalisms provide approximately the same result. 

Extrapolating the value of $T_{\rm AH}$ to the limit $h_{L-1}
\rightarrow 0$ for runs 16a and 16c and for runs 16aF and 16cF
assuming that the error in $T_{\rm AH}$ is proportional to
$h_{L-1}^2$, the true value of $T_{\rm AH}/m_0$ is estimated to be
$\approx 260$. Thus, for the best-resolved runs 16a and 16aF, the
merger time is computed with $\approx 2\%$ error.  Extrapolating to
$h_{L-1} \rightarrow 0$ for runs 19a and 19c, the true value of
$T_{\rm AH}/m_0$ is estimated to be $\approx 560$. Thus, even for the
best-resolved runs 19a and 19f, the merger time is underestimated by
$\approx 40m_0$. For such a longterm simulation, a better-resolution
is obviously required.

\begin{figure*}[thb]
\epsfxsize=3.3in
\leavevmode
(a)\epsffile{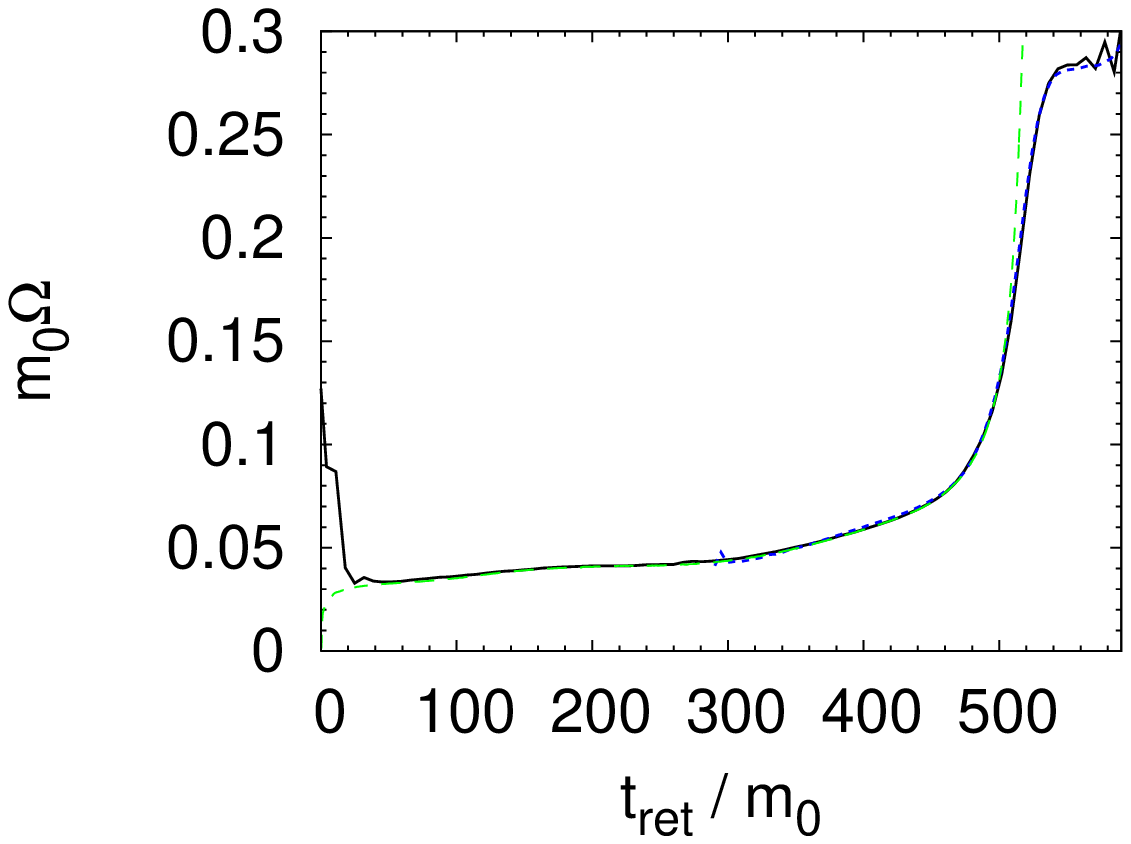}
\epsfxsize=3.3in
\leavevmode
(b)\epsffile{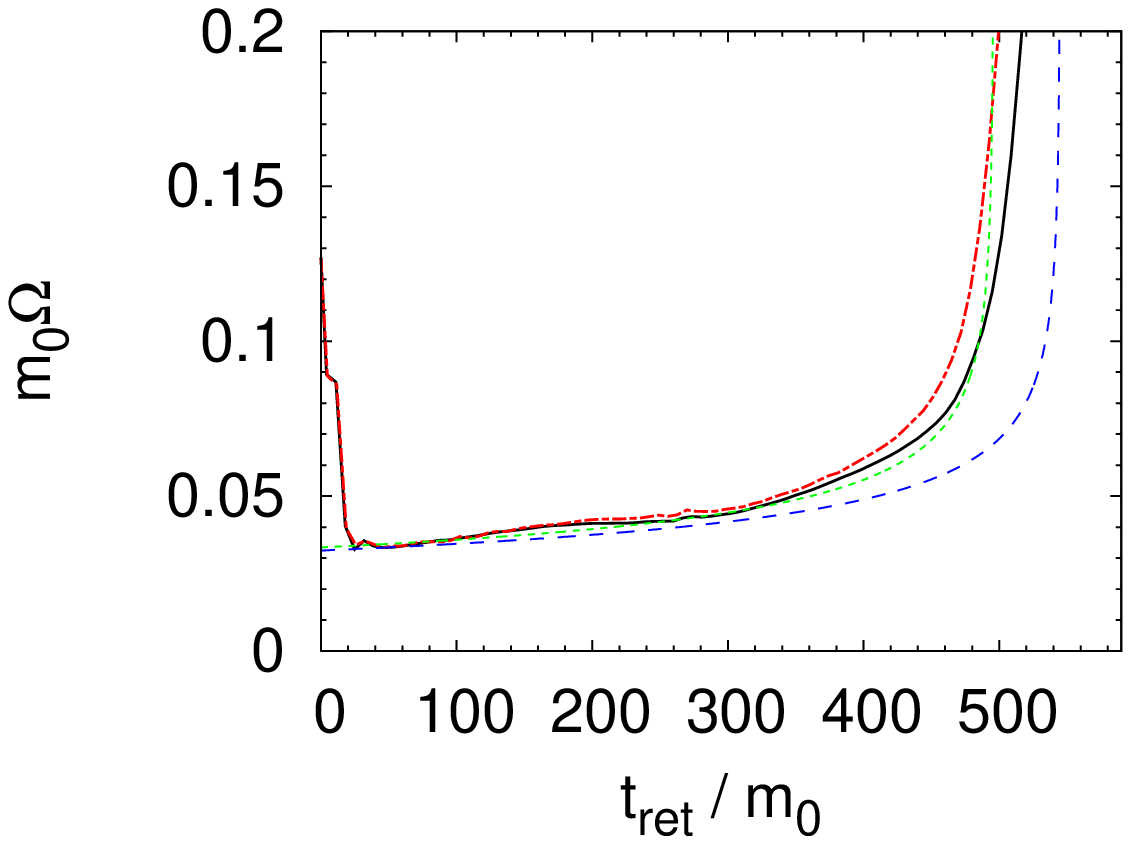}
\vspace{-2mm}
\caption{(a) $m_0\Omega$ computed from gravitational waveforms for runs 
19a (solid curve) and 16a (dashed curve).  For run 16a, we plot
$m_0\Omega$ as a function of $t_{\rm ret} + 258m_0$.  The long dashed
curve denotes the results derived from the orbital motion of one of
BHs. (b) Comparison of $m_0\Omega$ computed from gravitational
waveforms with those derived from the Taylor T4 formalism for runs 19a
(solid curve) and 19aF (dotted-dashed curve). The long and short 
dashed curves are results derived by the Taylor T4 formalism, drawn 
for the nonspinning equal-mass binary with $m_0\Omega(t=0) \approx
0.03245$ and 0.03345 at $t_{\rm ret}=0$, respectively.
\label{FIG5}}
\end{figure*}

On the vertical axis of Fig. \ref{FIG3}, we plot time at the onset of
merger that is predicted by the Taylor T4 formalism. Here, we assume
that the merger sets in when $m_0 \Omega$ reaches 0.2. (The initial
condition is chosen to be the orbit with $\Omega=\Omega_0$ for each
model.) Thus, this value may be slightly smaller than $T_{\rm AH}$
because it takes time from the onset of merger to formation of common
apparent horizon. We also note that the Taylor T4 formalism is not a
good approximation for the orbital evolution near the innermost stable
circular orbit \cite{BHBH16}.

For $d=16$ and $d=19$, the predicted merger time by the Taylor T4
formalism is $247m_0$ and $544m_0$, respectively.  Therefore, the
merger time determined by the extrapolation of the numerical results
for $h_{L-1} \rightarrow 0$ agrees with the predicted value within
error of $20m_0$. The predicted merger time is smaller than the
numerical results. This seems to be reasonable because the definition
of the merger time for the numerical results and for the Taylor T4
formalism is different, as mentioned above. Nevertheless, the error is
not so large that we conclude that the Taylor T4 formalism provides a
good approximate value for the merger time which can be a guideline
for analyzing the numerical results.

The derived merger time ($\sim 125m_0$ for $d=13$, $\sim 260m_0$ for
$d=16$, and $\sim 560m_0$ for $d=19$) is slightly longer than the
results reported in Ref.~\cite{BHBH12} in which $T_{\rm AH}/m_0=109 \pm 4$
for $d=13$, $228 \pm 16$ for $d=16$, and $529 \pm 22$ for $d=19$. A
part of the reason is that the slicing is different between two
groups. The possible other reasons may be that (i) our code is fully
fourth-order-accurate whereas the code of Ref.~\cite{BHBH12} is not, and
(ii) our code does not include Kreiss-Oliger-type dissipation term whereas
in the simulation of Ref.~\cite{BHBH12}, it is included and the 
dissipative effect may spuriously enhance the decrease rate of
the orbital separation.

\begin{figure}[t]
\epsfxsize=3.3in
\leavevmode
\epsffile{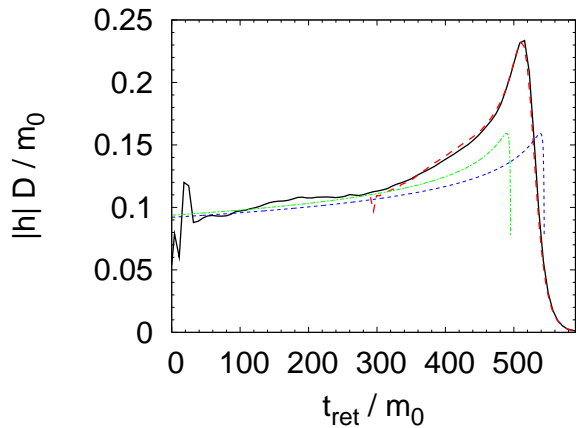}
\vspace{-2mm}
\caption{Evolution of gravitational wave amplitude defined by
$(h_+^2+h_{\times}^2)^{1/2}$ as a function of retarded time for run 19a 
(solid curve). For comparison, amplitude for run 16a (dashed curve) is
plotted as a function of $t_{\rm ret} + 258m_0$.  The long dashed and
dotted-dashed curves denote wave amplitudes derived from the Taylor T4
formalism. These are drawn for the nonspinning equal-mass binary with
$m_0\Omega(t=0) \approx 0.03245$ and 0.03345 at $t_{\rm ret}=0$,
respectively.
\label{FIG6}}
\end{figure}

\subsubsection{Gravitational waves}

Figure \ref{FIG4}(a) and (b) plot plus and cross modes of
gravitational waves for runs 16a and 19a. As described in
Sec.~\ref{sec3c}, the waveforms in the early inspiral phase are
extracted at large radii $\approx 70 m_0$, and those in the late
inspiral and merger phases are at small radii ($\approx 30 m_0$) of a
small grid spacing.  Then, we match two waveforms at a retarded time
$t_{\rm ret}=t_{\rm sep}$.  Specifically, we match the waveforms at
$t_{\rm sep} \approx 225 m_0$ for run 19a and $\approx 103 m_0$ for
run 16a. The phase of gravitational waves depends slightly on the
extracted radii, and a small phase difference between two waveforms
extracted at different radii is present for both runs; for runs 19a
and 16a, the phase difference is $\approx 2.5 m_0$ and $2.9 m_0$,
respectively. We correct these phase differences to constitute smooth
waveforms shown in Fig.~\ref{FIG4}.  This figure shows that our
strategy can produce waveforms of a good quality.

As we noted in Sec.~\ref{sec5a3}, the merger time for runs 16a and 19a
would be shorter than the true values, determined by extrapolation, by
$\sim 5m_0$ and $\sim 40m_0$, respectively.  Thus, in the waveforms
shown in Fig.~\ref{FIG4}, such phase error is included. As mentioned
above, to derive a waveform with sufficiently small phase error for
run 19a, a simulation with a finer resolution is necessary.

To clarify the properties of the error associated with finite grid
resolution, we generate Fig.~\ref{FIG4}(c) and (d). 
In these figures, we compare plus mode of gravitational waves 
for runs 19a and 19c. To match the wave phases of two numerical 
results, we plot
\beq
h_+\cos(0.3\pi)+h_{\times}\sin(0.3\pi) \label{eqhhh}
\eeq
as a function of $t_{\rm ret}+14m_0$ for the result of run 19c in Fig. 
\ref{FIG4}(c). It is found that the waveforms in the inspiral orbit
for two runs agree well except for those in the last inspiral orbit.
This indicates that for accurately computing gravitational waveforms
in the early inspiral phase (in this case, from about 0.5th orbit to
about 3rd orbit), the present choice of the grid resolution is acceptable.

Figure \ref{FIG4}(d) compares the waveforms in the final inspiral and
merger phases.  In this figure, the waveform defined by
Eq. (\ref{eqhhh}) as a function of $t_{\rm ret}+23m_0$ is plotted for
run 19c to match the ringdown waveforms. The figure shows that the
phase error is rapidly accumulated near the last inspiral phase.
Also, we can see that the amplitude of the ringdown phase is
underestimated for run 19c (by contrast, Fig. \ref{FIG4}(c) shows that
the amplitude in the inspiral phase depends weakly on the grid
resolution). Thus, we conclude that in a run with a poor grid
resolution, (i) the time duration for the inspiral phase near the last
inspiral orbit is underestimated and (ii) the amplitude of the
ringdown waveform is underestimated.

Figure \ref{FIG5}(a) plots angular velocity computed from
gravitational waveforms for runs 16a and 19a. Here, 
the angular velocity is derived from $\Psi_4$ by 
\beqn
\Omega(t)
={1 \over 2} {|\Psi_4(l=m=2)| \over \displaystyle \Big|\int dt \Psi_4(l=m=2)\Big|},
\label{gwangv}
\eeqn
where $\Psi_4(l=m=2)$ is the $l=m=2$ mode of $\Psi_4$. We also derive
it from the orbital motion of BHs for run 19a (the short dashed curve of
Fig. \ref{FIG5}), and this result agrees well with that derived from
Eq. (\ref{gwangv}). Thus, in this case, the coordinate trajectory 
approximately represents the physical trajectory (but this is not 
always the case; see Appendix \ref{app:res}). 

The curves for runs 16a and 19a agree approximate with each other,
indicating that gravitational waveforms in the late inspiral phase
depend very weakly on the initial condition as far as the initial
value of $m_0\Omega_0 \alt 0.041$. Figure \ref{FIG5}(a) also shows
that the angular velocity does not increase monotonically in the early
stage for run 19a. This implies that an eccentricity is present in the
early stage. This is also pointed out in Ref. \cite{BHBH12} in which the
estimated eccentricity is $\sim 0.02$. The curve of Fig. \ref{FIG5} is
similar to that reported in Ref. \cite{BHBH12}: Initially, $m_0\Omega
\approx 0.033$, and then, it reaches a local maximum of $m_0\Omega
\approx 0.040$. These results reconfirm that the eccentricity of the
initial condition would be $\sim 0.02$. 


We compare the numerical results for $m_0\Omega(t)$ for runs 19a and
19aF with those derived from the Taylor T4 formalism in
Fig. \ref{FIG5}(b).  Because we adopt corotating binary BHs as the
initial conditions, the spin of each BH is not zero \cite{CCGP} and,
thus, we take into account the spin effects in this analysis. In
Fig. \ref{FIG5}(b), the results by the Taylor T4 formalism are plotted
for the case $m_0 \Omega(t=0) =0.03245(=m_0\Omega_0)$ and 0.03345.
The figure shows that the numerical results agree approximately with
the Taylor T4 curve of $m_0 \Omega(t=0)=0.03345$ besides a modulation
associated with an elliptical orbital motion, but not very well with
the curve of $m_0 \Omega(t=0)=0.03245$, which is approximately equal
to the initial angular velocity for $d=19$. There are at least two
reasons for this discrepancy. The primary reason is that numerical
dissipation associated with finite-differencing spuriously enhances
the decrease rate of the orbital separation. Indeed, the merger time
derived from the Taylor T4 formalism with $m_0 \Omega(t=0)=0.03245$ is
by $\approx 50 m_0$ longer than that with $m_0
\Omega(t=0)=0.03345$. The error of $50m_0$ agrees approximately with
the possible error size for run 19a (cf. Sec.~\ref{sec5a3}). The other
is that the initial condition is not exactly in a circular orbit but
in an elliptical orbit for which the initial averaged angular velocity
is not equal to $m_0 \Omega_0 \approx 0.03245$ but slightly larger
than it.

In the final phase of merger, ringdown gravitational waves associated 
with quasi-normal modes are emitted. Perturbation studies predict 
their angular velocity and damping time scale for the
nonaxisymmetric fundamental mode with $l=m=2$ as \cite{leaver}
\beqn
&& M_{\rm BHf} \Omega_{\rm QNM} \approx 1.0
[ 1-0.63(1-a)^{0.3}], \label{eqbh} \\
&& t_d \approx {4 (1-a)^{-0.45} \over \Omega_{\rm QNM}}. \label{QNMf}
\eeqn
For $a=0.70$, $M_{\rm BHf}\Omega_{\rm QNM} \approx 0.56$. 
Because $M_{\rm BHf} \approx 0.95m_0$, the predicted value is
$m_0\Omega_{\rm QNM} \approx 0.59$. Figure \ref{FIG5}
shows that the numerical result of this value is $\approx 0.57$
(note that the angular velocity of gravitational waves is $2\Omega$).
Thus, the frequency of the quasi-normal mode is computed
with $\sim 3\%$ error. 

Figure \ref{FIG6} plots time evolution of gravitational wave amplitude
defined by $(h_+^2+h_{\times}^2)^{1/2}$ as a function of the retarded time
for run 19a. For comparison, the amplitude for run 16a and those derived
by the Taylor T4 formalism are shown together (see the figure caption
for details). This figure shows that the amplitude for run 19a agrees
with that in the Taylor T4 formalism with $\approx 10$--20\% error for
$t_{\rm ret} \alt 300 m_0$. The amplitude modulates with time in the
early phase because of the presence of the orbital eccentricity. In
the late phase, the amplitude of the numerical results is much larger
than that in the Taylor T4 formalism. This is also seen in other
numerical results (e.g., \cite{BHBH12,BHBH16}), and our results are
consistent with the previous results. The possible reason for this
large amplitude is that the tidal deformation of BHs, which is not taken
into account the Taylor T4 formalism, increases the attraction force
between two BHs. This leads to the acceleration of inward motion and
consequently to the speed-up of the orbital motion, resulting in the
amplification of the gravitational wave amplitude.

\begin{table*}[t]
\caption{List of several quantities for irrotational binary NSs in
quasiequilibrium circular orbits. We show the compactness of each NS
in isolation ($M_{\rm NS}/R_{\rm NS}$), gravitational mass of each NS
in isolation ($M_{\rm NS}$), maximum density for each star ($\rho_{\rm
max}$), total baryon rest mass ($M_*$), ADM mass at $t=0$ ($M_{0}$),
nondimensional angular momentum parameter ($J_0/M_{0}^2$), and angular
velocity in units of $M_0^{-1}$ ($M_0\Omega_0$). All these quantities
are shown in units of $c=G=\kappa=1$; in other words, they 
are normalized by $\kappa$ appropriately to be dimensionless. We note
that the mass, the radius, and the density can be rescaled to
desirable values by appropriately choosing $\kappa$.
\label{TNSNS}}
\begin{tabular}{cccccccc} \hline
~~~Model~~~ &  ~~~~$M_{\rm NS}/R_{\rm NS}$~~~~ & ~~~~~~$M_{\rm NS}$~~~~~~
& ~~~~~~~$\rho_{\rm max}$~~~~~~~ & ~~~~$M_*$~~~~  & 
~~~~$M_0$~~~~ & ~~$J_0/M_0^2$~~ &  ~~$M_0\Omega_0$~~  \\ \hline
NS1616 &0.160, 0.160 & 0.1478, 0.1478 & 0.152, 0.152
&0.3200 & 0.2924 & 0.9584& 0.0305 \\ \hline 
NS1416 &0.140, 0.160 & 0.1363, 0.1478 & 0.118  0.152
&0.3061 & 0.2810 & 0.9685& 0.0289 \\ \hline
\end{tabular}
\end{table*}

\begin{table*}[t]
\caption{The same as Table \ref{BHBHGRID} but for simulations of
models NS1616 and NS1416. $\Delta x$ is the minimum grid spacing,
$R_{\rm diam}$ the coordinate length of the semi-major diameter of
NSs, $L$ the location of outer boundaries along each axis, $\lambda_0$
the gravitational wavelength at $t=0$, and $\Delta x_{\rm gw}$ the
grid spacing at which gravitational waves are extracted. $M_{\rm NS}$
is the ADM mass of larger NS in isolation which is 0.1478 in the 
present units.  For models NS1616a--NS1616c, simulations are performed
both in the $F_i$-BSSN and $\tilde \Gamma^i$-BSSN formalisms.
\label{NSNSGRID}}
\begin{tabular}{ccccccc} \hline
Run & ~~Levels~~ & ~~$N$~~ & $\Delta x/M_{\rm NS}$ & ~$R_{\rm diam}/\Delta x$~
& ~$L/M_0~(L/\lambda_0)$~ & ~$\Delta x_{\rm gw}/M_0$~  \\ \hline
NS1616s     & 8~(3+5) & 36 & 0.068 & 130 & 158~(1.53)  & 1.09  \\ \hline
NS1616a, aF & 8~(3+5) & 30 & 0.081 & 108 & 158~(1.53)  & 1.31  \\ \hline
NS1616b, bF & 8~(3+5) & 24 & 0.101 & 87 & 158~(1.53)  & 1.64  \\ \hline
NS1616c, cF & 8~(3+5) & 20 & 0.122 & 67 & 158~(1.53)  & 1.97 \\ \hline
NS1616d & 7~(3+4) & 30 & 0.135 & 65 & 131~(1.27)  & 1.09   \\ \hline
NS1616e & 7~(3+4) & 24 & 0.169 & 52 & 131~(1.27)  & 1.37   \\ \hline
Shibata & --- & --- & 0.147 & 60 & 106~(1.03)  & 1.09  \\ \hline \hline
NS1416a & 8~(3+5) & 30 & 0.081 & 100 & 164~(1.51)  & 1.37  \\ \hline
NS1416b & 8~(3+5) & 24 & 0.101 & 87  & 164~(1.51)  & 1.71  \\ \hline
NS1416c & 8~(3+5) & 20 & 0.122 & 67  & 164~(1.51)  & 2.05  \\ \hline 
\end{tabular}
\end{table*}

\subsubsection{Radiated energy and angular momentum, and their conservation}

Total energy and angular momentum radiated by gravitational waves are
listed in Table \ref{BHBHRES} (see $\Delta E$ and $\Delta J$). It is
found that the radiated energy depends very weakly on the initial
condition.  This implies that most of the energy is radiated in the
final merger phase; during inspiral from $m_0\Omega \approx 0.032$
(initial condition for model d19) to $\approx 0.056$ (that for model
d13), the energy is radiated only by $\sim 0.002$--$0.003 m_0$. This
fact is easily inferred from small difference in the ADM mass among
three models of $d=13$, 16, and 19 (see Table I).  Indeed, Table I
shows that the difference in the ADM mass is $\sim 0.003 m_0$ between
the results for $d=13$ and $d=19$, which agrees approximately with the
estimated radiated energy during the inspiral phase.  The angular
momentum is also radiated most efficiently in the final merger
phase. However, it is also radiated by several percents in the late
inspiral orbits in contrast to the energy. This is simply because the
angular momentum of the binary system depends on the orbital
separation more strongly than the energy. 

The total radiated energy derived here is significantly different from
the results of Ref.~\cite{BHBH12} in particular for $d=16$ and 19.  In
their results, it depends strongly on the initial condition. However,
we believe that our results are more reliable because of the following
reasons: (i) As mentioned above, the total radiated energy should not
depend strongly on the initial condition.  Our results are consistent
with this expectation; (ii) The sum of the BH mass finally formed and the
total radiated energy should be equal to the initial ADM mass.
Namely, the following relation should hold;
\beqn
M_{\rm BHf} + \Delta E = M_0. \label{cons}
\eeqn
In our results, the left-hand side is $\approx 0.983 m_0$ for $d=16$
and $\approx 0.985 m_0$ for $d=19$, whereas the right-hand side is
0.9875 and 0.9890, respectively. Thus, magnitude of the error is
$\approx 0.5\%$. The left-hand side of Eq. (\ref{cons}) is
systematically smaller than $M_0$, and hence, $\Delta E$ is likely to
be underestimated due to numerical dissipation by $\sim 0.004m_0$.  On
the other hand, in the results of Ref.~\cite{BHBH12}, the left-hand
side of Eq. (\ref{cons}) is $(0.997 \pm 0.009)m_0$ for $d=16$ and
$(1.004 \pm 0.009)m_0$ for $d=19$. Thus, these are larger than the
left-hand side ($M_0=0.9875$ and 0.9890 for $d=16$ and 19) by
1--1.5\%, and magnitude of the error increases with the increase of
$d$. The reason seems to be that the total radiated energy is
systematically overestimated for larger values of $d$.  (As pointed
out above, $\Delta E$ should not depend strongly on the initial
condition, but in their results, $\Delta E$ steeply increases with
increasing the value of $d$.)

The conservation relation for angular momentum is written by 
\beqn
M_{\rm BHf}^2 a + \Delta J = J_0. \label{consJ}
\eeqn
For $d=13$, 16, and 19, the error in the conservation relation defined
by $1-(M_{\rm BHf}^2 a + \Delta J)/J_0$ is $\sim$ 2\%, 3\%, and 4\%,
respectively, for the best-resolved runs. Thus, magnitude of the error
is larger than that for the energy conservation.  The left-hand side
of Eq. (\ref{consJ}) is always smaller than the initial value,
$J_0$. This implies that either $a$ or $\Delta J$ is
underestimated. As mentioned above, $\Delta E$ is underestimated.
Thus, error in $\Delta J$ is likely to be the primary source of the
underestimation.

\subsection{NS-NS Binaries} \label{sec:res-nsns}

\begin{figure}[thb]
\epsfxsize=3.05in
\leavevmode
(a)~~~~~~~~~\epsffile{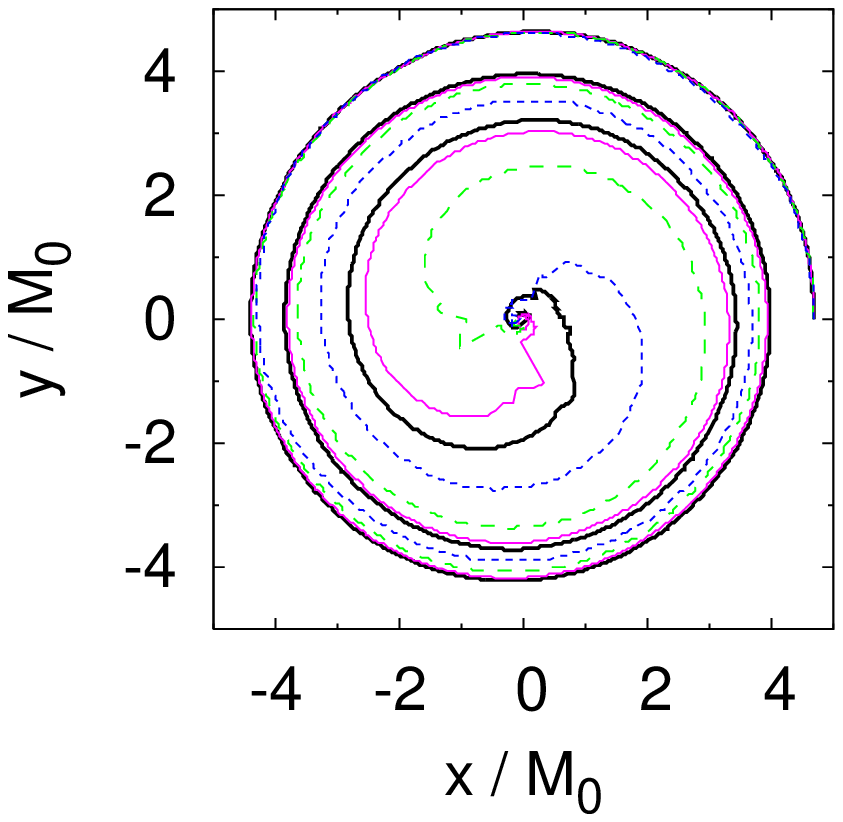} \\
\epsfxsize=3.05in
\leavevmode
(b)~~~~~~~~~\epsffile{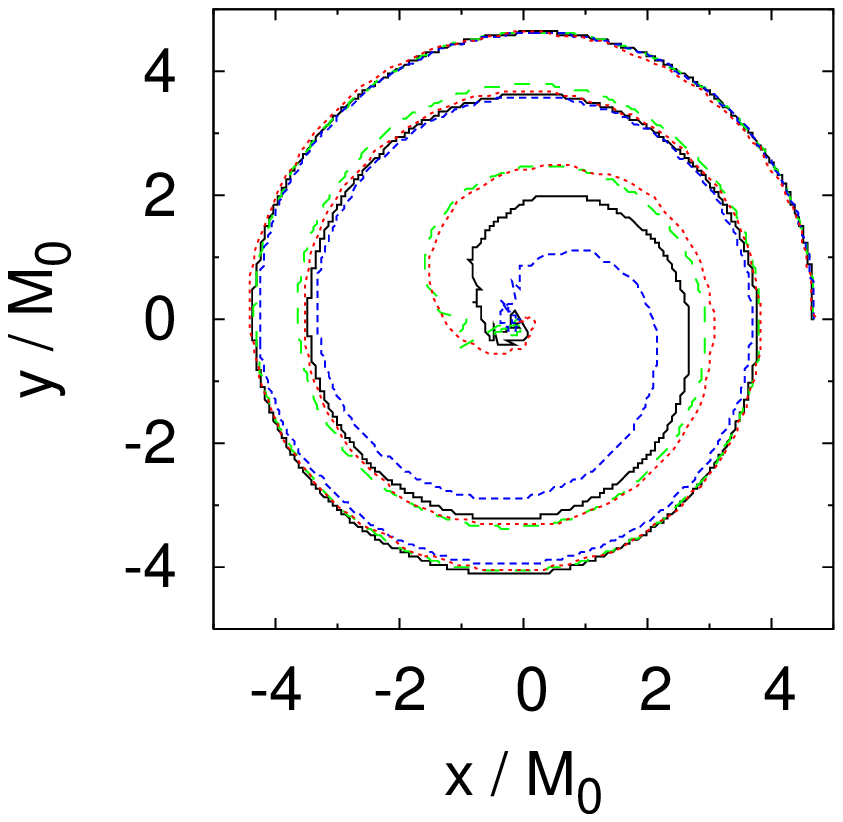} \\
\epsfxsize=3.05in
\leavevmode
(c)~~~~~~~~~\epsffile{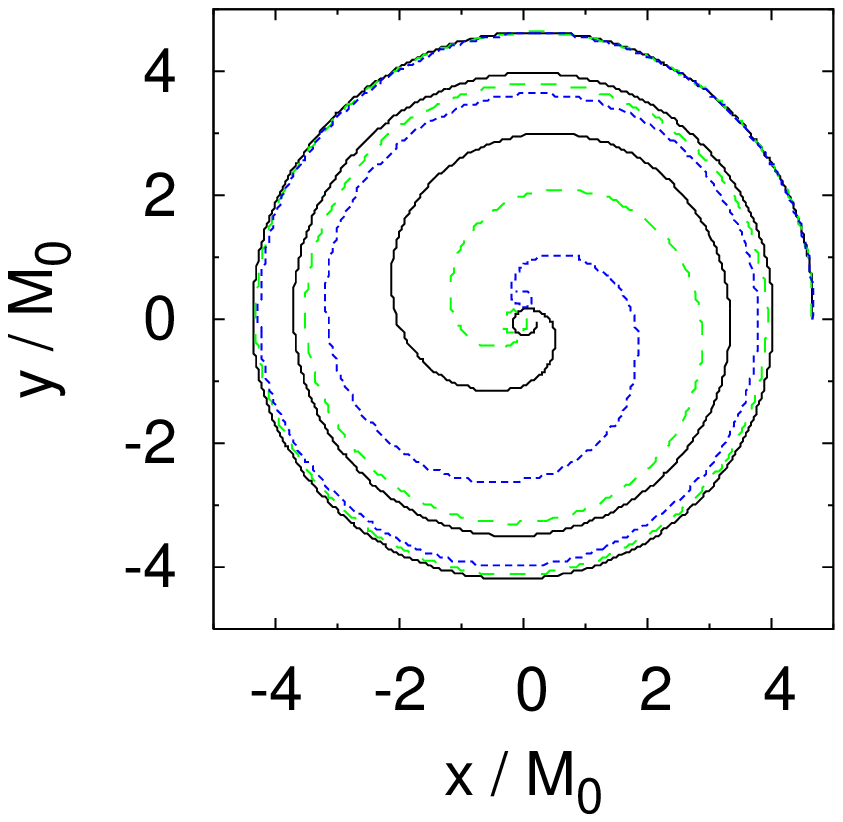} 
\vspace{-2mm}
\caption{(a) Orbital trajectories of an NS to formation of apparent
horizon for runs NS1616s (thick solid curve), NS1616a (solid curve),
NS1616b (long-dashed curve), and NS1616c (dashed curve). (b) The same
as (a) but for runs NS1616d (solid curve), NS1616b (long-dashed
curve), NS1616c (dashed curve), and for run by Shibata's code (dotted
curve).  (c) The same as (a) but for runs NS1416a (solid curve),
NS1416b (long-dashed curve), and NS1416c (dashed curve).  In this
case, the trajectories for more massive NS are plotted.
\label{FIGNS1}}
\end{figure}

For validating our new hydrodynamic code with the AMR algorithm, we first
performed simulations for NS-NS binaries.


\subsubsection{Initial condition}

Following our previous works \cite{NSNS1,NSNS2}, we adopt NS-NS
binaries of the irrotational velocity field in quasiequilibrium
circular orbits as initial conditions.  The quasiequilibrium state is
computed in the so-called conformally flat formalism for the Einstein
equations \cite{WM}. The irrotational velocity field is assumed
because it is considered to be a good approximation of the velocity
field for coalescing binary NSs in nature \cite{KBC}. We employ the
numerical solutions computed by Taniguchi and Gourgoulhon, which are
involved in the LORENE library \cite{GGTMB,TG,LORENE}.  Specifically,
we pick up two models computed in the polytropic EOS with $\Gamma=2$
\cite{TG}; one is an equal-mass binary for which compactness of each
NS is 0.16 and coordinate separation between two centers of mass is 45
km in the LORENE unit.  The other is an unequal-mass binary for which
compactness are 0.14 and 0.16, and coordinate separation between two
centers of mass is 45 km in the LORENE unit. We simulate this to
demonstrate that our code can follow unequal-mass binaries as well as
equal-mass ones. We list several key parameters for these models in
units of $c=G=\kappa=1$ in Table \ref{TNSNS}.

\subsubsection{Setting}

Simulations were performed for a variety of grid structures and grid
resolutions (see Table \ref{NSNSGRID}). For model NS1616, we also
performed a simulation using Shibata's code in which a nonuniform
uni-grid is adopted. This code is the same as that presented in
Ref.~\cite{BHNS1,BHNS2}; the Einstein evolution equations are solved in the
$F_i$-BSSN formalism with a fourth-order finite differencing in space
and the hydrodynamic equations are solved in the same scheme as SACRA. 
The third-order Runge-Kutta scheme is employed for evolution forward in
time.

Grid resolutions and grid sizes in the simulations with SACRA are
listed in Table \ref{NSNSGRID}. For all the cases, the NSs are covered
by the finest and second-finest levels (central region of each NS is
covered by the finest level and the region near the surface is covered
by the second-finest level).  For models NS1616a--NS1616c, the
simulations were performed both in the $F_i$-BSSN and $\tilde
\Gamma^i$-BSSN formalisms, whereas the simulations for models NS1616s,
NS1616d, NS1616e, and NS1416a--c were done in the $\tilde
\Gamma^i$-BSSN formalism. The best-resolved runs for models NS1616 and
NS1416 are NS1616s and NS1416a, respectively.

For the initial values of $\alpha$ and $\beta^i$, we employ those of
the quasiequilibrium solutions.  Even when such initial condition is
adopted in the $\Gamma$-freezing gauge condition, the orbital
eccentricity appears to be not as large as that in the BH-BH-binary
case. The value of $\eta_s$ in the $\Gamma$-freezing gauge is set to
be $\approx 1.7/M_0$ irrespective models. ($\eta_s=0.5$ in units of
$c=G=\kappa=1$.)

\begin{table*}[t]
\caption{Numerical results for simulations of NS-NS binaries.  We list
the time at which apparent horizon is first formed ($T_{\rm AH}$),
approximate final value of the irreducible mass of the apparent 
horizon ($M_{\rm irr}$), ratio of the polar circumferential length to the equatorial
one for the apparent horizon ($C_p/C_e$), BH mass estimated from the
equatorial circumferential length ($C_e/4\pi$), BH spin parameter
estimated from $C_p/C_e$ ($a$), energy and angular momentum carried
away by gravitational waves ($\Delta E$ and $\Delta J$), and rest mass
and angular momentum of disk formed around the BH ($M_{\rm disk}$ and
$J_{\rm disk}$).  The state of the disk is determined when we stopped
the simulation (cf. Fig. \ref{FIGNS3}). For model NS1616, the mass and
angular momentum of disk are determined only for run NS1616a.
\label{NSNSRES}}
\begin{tabular}{cccccccccc} \hline
~Run~ & ~$T_{\rm AH}/M_0$~ & ~$M_{\rm irr}/M_0$~ & ~$C_p/C_e$~
& ~$C_e/(4\pi M_0)$~ & ~$a$~ & ~$\Delta E/M_0$~
& ~$\Delta J/J_0$~ & ~$M_{\rm disk}/M_0$~ & ~$J_{\rm disk}/J_0$~\\ \hline
NS1616s  & 478 & 0.86--0.87  & 0.81--0.83 & 0.993 
& 0.83--0.86 & 0.7\% &12\% & --- & --- \\ \hline
NS1616a  & 454 & 0.85--0.87  & 0.79--0.83 & 0.995
& 0.83--0.89 &0.7\% & 11\% & $\sim 0.01\%$ & $\sim 0.03\%$ \\ \hline
NS1616aF & 448 & 0.85--0.87  & 0.79--0.83 & 0.995
& 0.83--0.89 & 0.7\% & 12\% & --- & --- \\ \hline
NS1616b & 410 & ~0.84--0.88~  & ~0.78--0.84~ & ~$0.997 \pm 0.001$~
& ~0.81--0.91~    & 0.7\% & 11\% & --- & --- \\ \hline
NS1616bF& 399 & 0.85--0.88  & 0.78--0.84 & $0.995 \pm 0.001$
&0.81--0.91    & 0.7\% & 11\% & --- & --- \\ \hline
NS1616c & 357 & 0.84--0.88  & 0.78--0.84 & $0.997 \pm 0.002$
&0.81--0.91    & 0.6\% & 10\% & --- & --- \\ \hline
NS1616cF& 349 & 0.84--0.89  & 0.78--0.84 & $0.996 \pm 0.001$
&0.81--0.91    & 0.6\% & 10\% & --- & --- \\ \hline
NS1616d & 386 & 0.84--0.89  & 0.78--0.84 & $0.995 \pm 0.001$
&0.81--0.91    & 0.7\% & 11\% & --- & ---\\ \hline
NS1616e & 325 & 0.84--0.89  & 0.78--0.84 & $0.995 \pm 0.003$
&0.81--0.91    & 0.7\% & 8\%  & --- & ---\\ \hline
Shibata & 423 & 0.874 & 0.83 & $0.989 \pm 0.001$ 
& 0.83 & 0.8\%  & 12\% & --- & --- \\ \hline \hline
NS1416a  & 469 & 0.83--0.85  & 0.80--0.85 & 0.971  & 0.80--0.88
& 0.5\% & 9\% & 2.4\% & 6.0\% \\ \hline
NS1416b  & 419 & 0.82--0.85  & 0.78--0.85 & $0.976 \pm 0.001$ & 0.80--0.90
& 0.4\%  & 8\% & 2.3\% & 5.9\% \\ \hline
NS1416c  & 364 & 0.80--0.85  & 0.76--0.86 & $0.982 \pm 0.002$ & 0.78--0.93
& 0.4\%  & 7\% & 2.0\% & 5.1\% \\ \hline
\end{tabular}
\end{table*}

\subsubsection{Evolution of NSs and the final outcome}

Figure \ref{FIGNS1}(a) plots orbital trajectories for one of two NSs
for runs NS1616s, a, b, and c. Here, the trajectories of the NSs are
determined by tracking location of the maximum value of $\rho_*$. Note
that for these runs, the locations of outer and refinement boundaries
are the same, although the grid resolution is different. This figure
shows that the binary experiences $\approx 9/4$, 10/4, 11/4, and $\alt
3$ orbits before the onset of merger for runs NS1616c, b, a, and s,
respectively. For finer grid resolutions, the number of orbits is
systematically larger, because numerical dissipation of angular
momentum and energy is smaller.  Figure \ref{FIGNS1}(a) indicates that
convergence of numerical results for $h_{L-1} \rightarrow 0$ appears
to be not very fast, and for $h_{L-1} \rightarrow 0$, the number of
orbits would be larger than 3 (see Fig. \ref{FIGNS5} and related
discussion below).

Figure \ref{FIGNS1}(b) compares orbital trajectories for runs NS1616b--d, 
and run by Shibata's code.  For these runs, the
finest grid spacing is $\Delta x/M_{\rm NS}=0.101$, 0.122, 0.135, and
0.147, respectively.  Although the grid resolution of the finest level
for NS1616c is better than that for NS1616d (and also for run by
Shibata's code), the merger time for NS1616c is shortest among four
runs, and hence, the numerical dissipation is most serious in this
run.  (The merger time, $T_{\rm AH}$, is defined in the same manner as
that in the BH-BH binary case.) The reason for this is that for
NS1616d, most part of the NS is covered by the finest level (for run
by Shibata's code, the whole region of the NS is covered by the finest
grid), whereas for NS1616c, a relatively wide region of the NS is
covered by the second-finest level: Dissipative effects of the
second-finest level is much larger than that of the finest one, and
hence, they spuriously enhance the decrease rate of orbital
separation. This suggests that it is desirable to cover the whole
region of the NSs by the finest level. However, to do this with a
sufficient grid resolution, it is necessary to take a large number of
grid points in the finest level. This is not desirable from the
viewpoint of computational cost. We tried to perform simulation using
several grid structures and found that an optimistic choice is that
the finest level approximately covers about two third of the NSs, 
from the viewpoints of grid-resolution and computational-cost. 
The grid structure for runs NS1616a--c and NS1416a--c is selected due
to this reason. 

Figure \ref{FIGNS1}(c) plots orbital trajectories for more massive NS 
for runs NS1416a--c. This figure is similar to Fig. \ref{FIGNS1}(a), 
and indicates that slight mass difference does not change qualitative 
properties for the orbital trajectories and convergence. As in the 
case of model NS1616, the merger time depends strongly on the grid 
resolution and convergence is not achieved even with the best-resolved 
run. 

\begin{figure*}[t]
\vspace{-6mm}
\caption{Snapshots of density contour curves and density contrasts
from merger phase to formation of a BH for run NS1416a. The contour
curves are plotted for $\rho w=10^{-i}$ where $i=2, 3, \cdots, 6$ (the
outermost short-dashed and dashed curves always denote $\rho w
=10^{-6}$ and $10^{-5}$).  In the first panel, the NS located for $y <
0$ is more massive one. The filled circle near the origin in the last
panel shows the region inside the apparent horizon.
\label{FIGNS2}}
\end{figure*}

For models NS1616 and NS1416, a BH is soon formed after the onset of
merger. This is reasonable because the total rest mass of these
systems is more than 1.6 times as large as the maximum rest mass of
nonrotating NSs ($\approx 0.180$) for the given EOS. We note that this
result agrees well with our previous result obtained in the simulation
with the $\Gamma$-law EOS and $\Gamma=2$ \cite{NSNS2} (see also a
recent work by the Illinois group which confirms our result 
\cite{NSNS9}). 

\begin{figure}[thb]
\epsfxsize=3.2in
\leavevmode
\epsffile{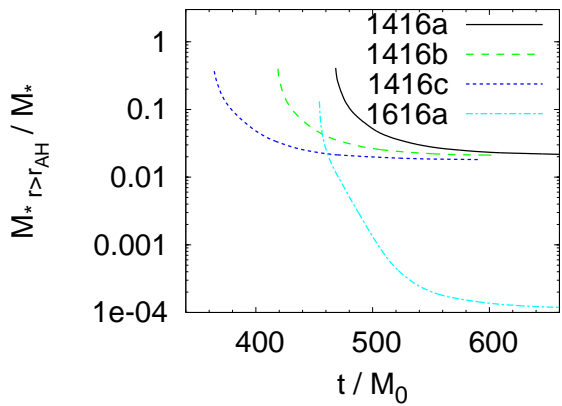}
\vspace{-2mm}
\caption{Evolution of rest mass of material located outside 
apparent horizon for NS1416a--c, and NS1616a. 
\label{FIGNS3}}
\end{figure}

The present code can follow the evolution of the formed BH for a long time
stably. We find that $\approx 99.99\%$ of the total rest mass is
swallowed by the BH for model NS1616 (cf. Fig. \ref{FIGNS3}). This is
due to the facts that (i) specific angular momentum for most of the
material at the onset of merger is not large enough to escape from
capturing by the BH and (ii) there is no mechanism for transporting
angular momentum outward in the merger of equal-mass binaries. This
result agrees again with our previous result \cite{NSNS2}, and also,
with a recent result by Illinois group \cite{NSNS9}.

By contrast, a disk is formed for model NS1416. The rest mass for run
NS1416a is $\sim 2\%$ of the total rest mass when we stopped the
simulation (see Figs. \ref{FIGNS2} and \ref{FIGNS3} and Table
\ref{NSNSRES}). The disk formation results primarily from the mass
difference of two NSs: Just before the merger, the smaller-mass NS is
tidally disrupted by the larger-mass companion (see
Fig. \ref{FIGNS2}). Because of asymmetry in the mass distribution,
angular momentum is subsequently transported and the tidally disrupted
material can spread outward.  Because the specific angular momentum of
such material is $\approx 2.5J_0/M_0$ and larger than that at the
innermost stable circular orbit around the formed rotating BH, a
compact disk is formed (see the last panel of Fig. \ref{FIGNS2}). The
maximum density of the disk is $\approx 10^{-4}$ in the present units
which is $\sim 1/1000$ of the maximum density of the NSs before merger
(i.e., $\sim 10^{12}~{\rm g/cm^3}$ in the cgs units if we assume that
the maximum density of the NS is $10^{15}~{\rm g/cm^3}$). Figure
\ref{FIGNS2} shows that the material of the disk is located in a small
region whose coordinate radius is $\sim 3$--$6M_0$. This is a
result of small averaged specific angular momentum of the disk, $2.5J_0/M_0
\approx 2.4M_{\rm BHf}$ where $M_{\rm BHf}$ is the mass of the BH
finally formed which is $\approx 0.97M_0$ (see below and Table
\ref{NSNSRES}). Such compact disk can be formed due to the fact that
the formed BH is rapidly rotating with the spin parameter $a \agt
0.8$; the specific angular momentum for the innermost stable
circular orbit around a Kerr BH is $j_{\rm ISCO} \approx 2.38M_{\rm
BHf}$ for $a=0.8$. (Note that $j_{\rm ISCO}/M_{\rm BHf} \approx 3.46$,
2.59, and 2.10 for $a=0$, 0.7, and 0.9, respectively.)

Figure \ref{FIGNS3} plots time evolution of the total rest mass of
material located outside apparent horizon. As mentioned above, it
settles down to $\sim 2\%$ of the total rest mass at the end of the
simulations for runs NS1416a--c, indicating that a disk of substantial
mass is formed around the formed BH. The disk mass gradually decreases
with time even at the end of the simulations, but the decrease time
scale is much longer than the orbital period of the material. For a
hypothetical value of $M_*=3M_{\odot}$, the disk mass is about $0.06
M_{\odot}$.  By contrast, for run NS1616a, it decreases to $\sim
10^{-4}M_*$ implying that disk of substantial mass is not formed. It
is worth noting that the disk mass for model NS1416 is much larger
than that for the case that stiff realistic EOSs are used for modeling
NSs \cite{NSNS2}, for the same value of mass ratio: For the stiff EOSs
used in the previous works \cite{NSNS2}, the disk mass around BH is
less than $0.01M_{\odot}$ for mass ratio of $\agt 0.9$ \cite{NSNS2}.
The possible reason for this difference is that the smaller-mass NS
for the present EOS are less compact than those in the stiff EOSs, and
hence, its outer part can spread outward more extensively. Because the
EOS adopted here is not very realistic (in the realistic EOSs the
radius depends on the mass in a much weaker manner), one should not
consider at face value that a massive disk would be formed after the
merger of unequal-mass NS-NS binaries.

\begin{figure}[thb]
\epsfxsize=3.1in
\leavevmode
\epsffile{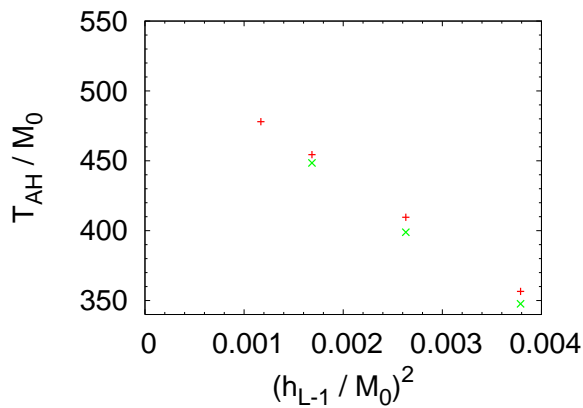}
\vspace{-2mm}
\caption{$T_{\rm AH}$ as a function of $h_{L-1}^2$ for
runs NS1616a--c and s (plus) and NS1616aF--cF (cross). 
\label{FIGNS4}}
\end{figure}

\begin{figure}[thb]
\epsfxsize=3.2in
\leavevmode
\epsffile{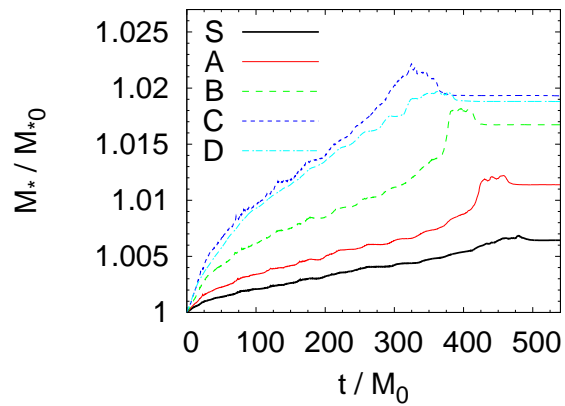}
\vspace{-2mm}
\caption{Change in total rest mass due to numerical error with time
for runs NS1616s (thick solid curve), NS1616a (solid curve), NS1616b
(long-dashed curve), NS1616c (dashed curve), and NS1616d
(dotted-dashed curve).
\label{FIGNS5}}
\end{figure}

In Table \ref{NSNSRES}, we summarize quantities extracted from the
apparent horizon of the formed BH. We note that the area and the ratio
of circumferential lengths decrease by $\sim 5\%$ for the time duration
of $150M_0$ after formation of the apparent horizon, and thus, the
irreducible mass and spin of the BH are not determined within $\sim
5\%$ error in contrast to the case for BH-BH and BH-NS (see the next
subsection) binaries even for the best-resolved run.  Currently, the
reason is not clear (a possible reason is that the spin parameter is
too large and hence radius of the apparent horizon is too small for
the chosen grid resolution to resolve the BH accurately).
Nevertheless, the final state of the BH is determined with an accuracy
which is acceptable for quantitative discussion: For model NS1616, the
final mass of the formed BH is $\approx 0.99 M_0$ and the spin is
$\sim 0.81$--0.84 irrespective of the grid resolution, grid structure,
and chosen formalism. These results are also in good agreement with
those in the simulation by Shibata's code.  For model NS1416, the
final mass of the BH is evaluated to be $\approx 0.97 M_0$ and the spin is
$\sim 0.8$--0.85. The BH mass is smaller than that for model NS1616.
We understand this fact as follows: A disk is formed in this case and
a part of mass and angular momentum are distributed to it.

\begin{figure*}[thb]
\epsfxsize=3.2in
\leavevmode
(a)\epsffile{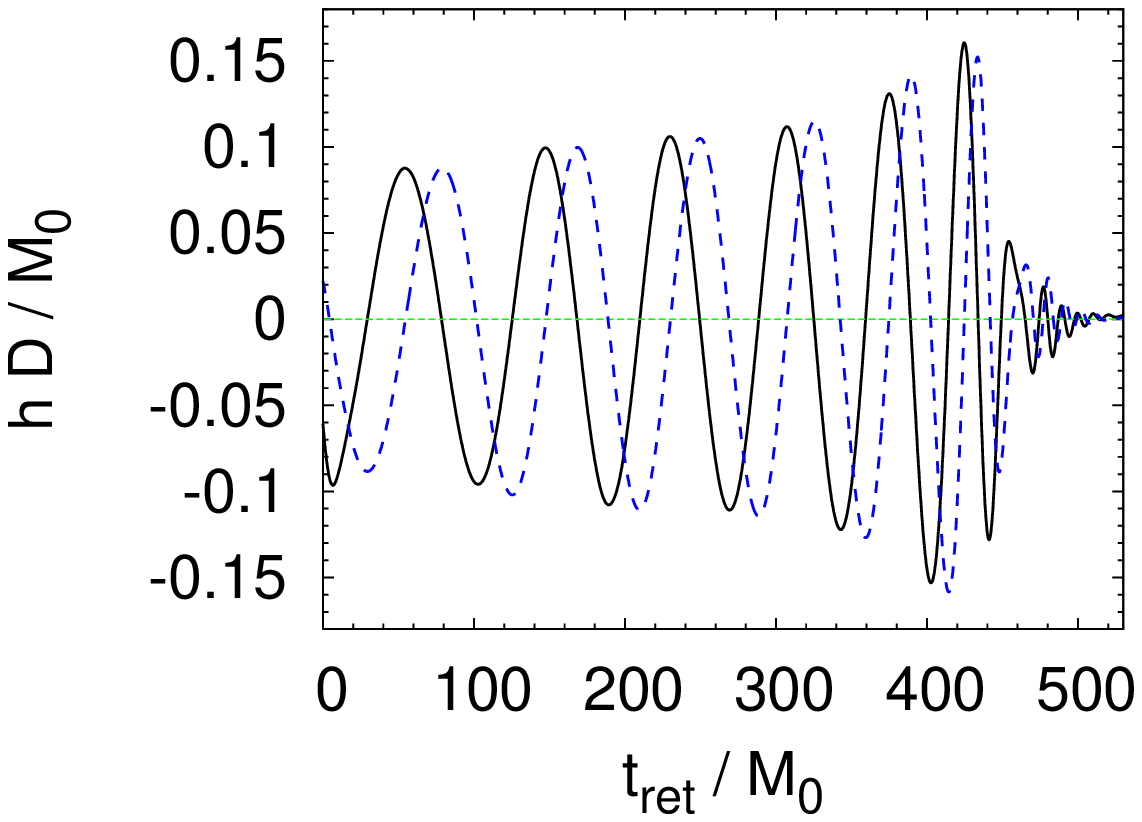}
\epsfxsize=3.2in
\leavevmode
~~~(b)\epsffile{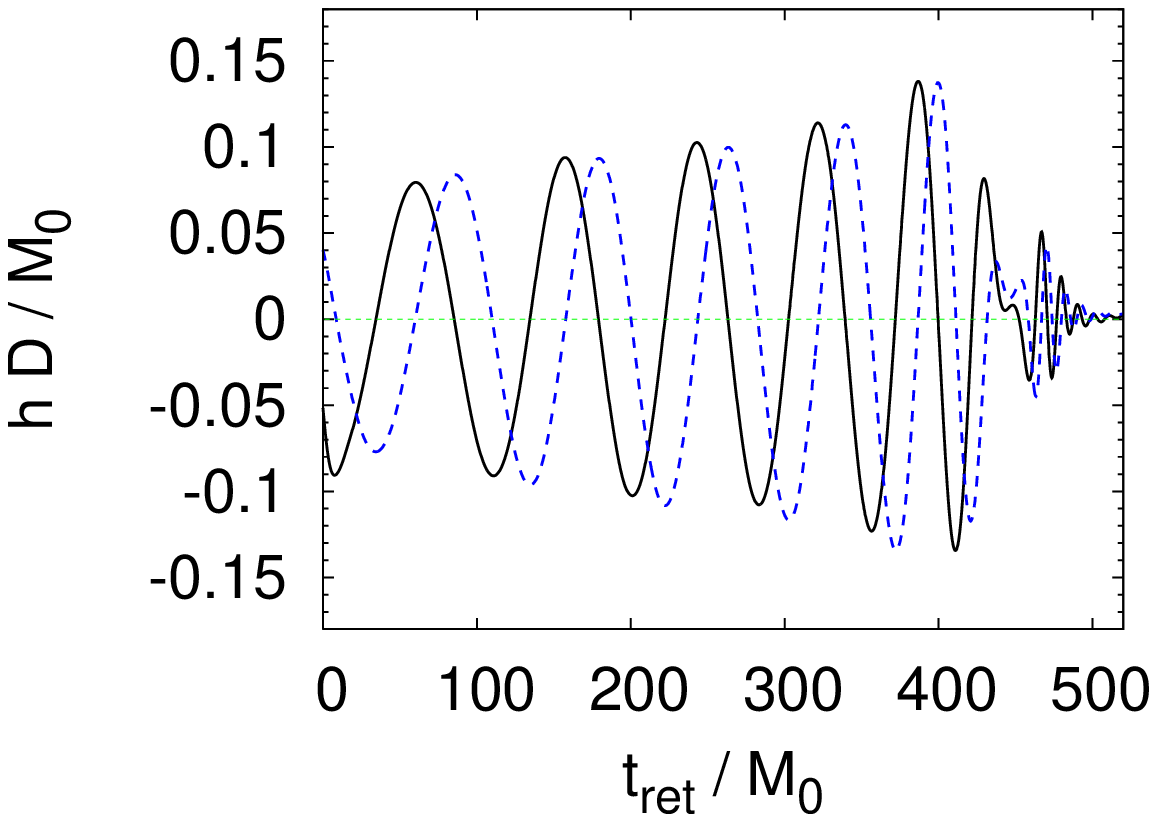}\\
\epsfxsize=3.2in
\leavevmode
(c)\epsffile{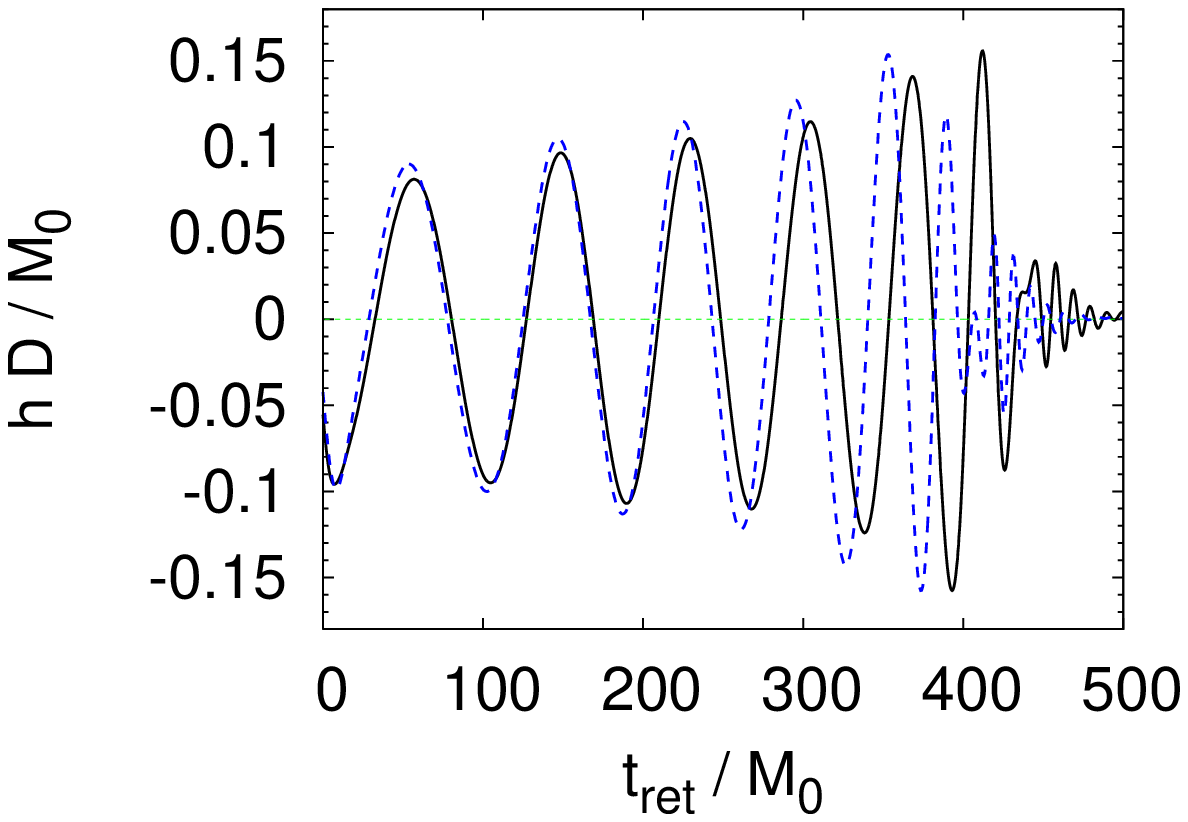}
\epsfxsize=3.2in
\leavevmode
~~~(d)\epsffile{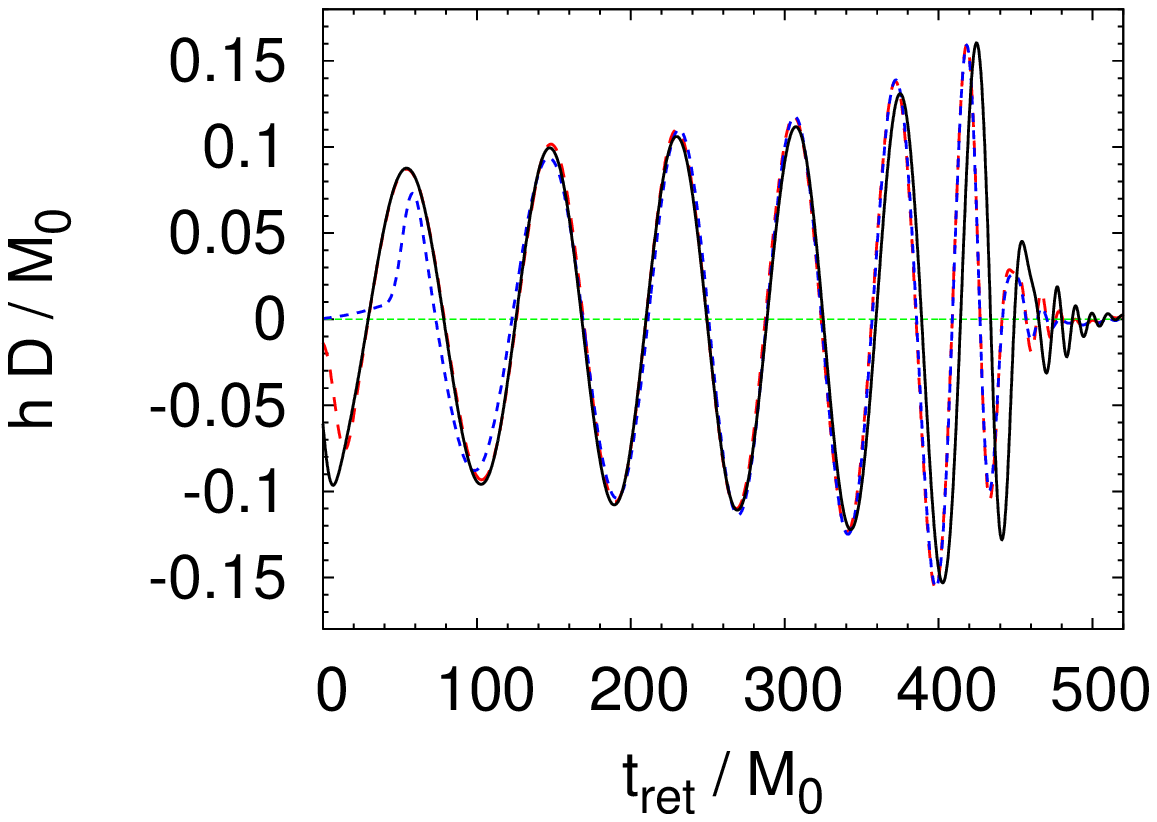}\\
\vspace{-2mm}
\caption{(a) Gravitational waveforms for run NS1616s. $l=m=2$ mode is
plotted. The solid and dashed curves denote the plus and cross modes,
respectively.  (b) The same as (a) but for run NS1416a. (c) $h_+$ for
run NS1616a (solid curve) and run by Shibata's uni-grid code (dashed
curve).  (d) $h_+$ for runs NS1616s, a, and b (solid, long-dashed, and dashed
curves). For runs NS1616a and NS1616b, the gravitational waveforms
are plotted as functions of $t_{\rm ret}+10M_0$ and $t_{\rm
ret}+54M_0$, respectively. In addition, for NS1616a and b,
$h_+ \cos(0.2\pi)-h_{\times}\sin(0.2\pi)$ and
$-[h_+ \cos(0.2\pi)-h_{\times}\sin(0.2\pi)]$ are plotted 
to align the phase.
\label{FIGNS6}}
\end{figure*}

For BHs formed after the merger of equal-mass BH-BH binaries, the
final spin parameter is $\approx 0.7$. For the best-resolved run, the
spin parameter is $0.85 \pm 0.02$ for model NS1616 and $0.83\pm 0.03$
for model NS1416. Thus, the spin parameter of the BHs formed after the
merger of equal-mass NS-NS binary is by $\sim 0.1$--0.15 larger. The
primary reason for this difference is that the angular momentum
carried away by gravitational waves in the merger of NS-NS binaries is
much smaller than that in the merger of BH-BH binaries: For BH-BH
binaries, $\sim 30\%$ of the initial angular momentum is dissipated by
gravitational radiation, whereas for NS-NS binaries, it is $\sim
10\%$.  This difference comes primarily from difference in amplitude
of gravitational waves emitted in the final merger phase.  BH-BH
binaries can take much closer orbital separations than NS-NS binaries
can because BHs are more compact than NSs.  Thus, gravitational waves
of a higher amplitude are emitted at the final inspiral orbit in the
former cases.  In addition, ringdown gravitational waves associated
with quasinormal-mode oscillation of fundamental $l=m=2$ mode 
is excited more significantly in the merger phase for BH-BH
binaries. Indeed, the amplitude is as high as that emitted at the last
inspiral orbit (cf. Fig. \ref{FIG4}). By contrast, in the case of
NS-NS binaries, it is not excited as significantly as in the case of
BH-BH binaries (cf. Fig. \ref{FIGNS6}), because of smaller degree of
nonaxisymmetric deformation of the spacetime curvature at the merger.
In another paper \cite{NSNS8}, we performed simulations for NS-NS
binaries using a realistic stiff EOS, which is highly different from
the $\Gamma$-law EOS with $\Gamma=2$, and found that the final spin
parameter is $\approx 0.8$ for the BH-formation case. Thus, the value
of $\sim 0.8$ for the spin parameter seems to be a universal outcome
for the BHs formed after the merger of NS-NS binaries.

\subsubsection{Conservation of energy and angular momentum}

Validity of the results about mass and spin of BHs finally formed
is checked by examining whether the following conservation relations hold: 
\beqn
&&M_{\rm BHf} + M_{\rm disk} + \Delta E = M_0, \label{conE}\\
&&M_{\rm BHf}^2a + J_{\rm disk} + \Delta J = J_0. \label{conJ}  
\eeqn
Here, $M_{\rm disk}$ and $J_{\rm disk}$ are the rest mass and the
angular momentum of disk, respectively. As in Refs.~\cite{BHNS1,BHNS2},
$J_{\rm disk}$ is calculated approximately by
\beqn
J_{\rm disk} \equiv \int_{r> r_{\rm AH}}
\rho \alpha h u^t u_{\varphi}\sqrt{\gamma} d^3x,
\eeqn
where the $\varphi$-coordinate is defined for an origin determined
from the maximum of $\rho_*$ which is approximately equal to
the center of the BH. For model NS1616 for which disk mass is
$\sim 10^{-4}M_*$ and negligible, the energy conservation holds within
0.2--0.3\% error and the angular momentum one holds with 2--3\%
error. For model NS1416, errors in the energy and angular momentum
conservations are $\approx 0.3\%$ and $\sim 5\%$, respectively.
Here, the error is defined, respectively, by
\beqn
&&1-(M_{\rm BHf} + M_{\rm disk} + \Delta E) /M_0, \\
&&1-(M_{\rm BHf}^2 a + J_{\rm disk} + \Delta J)/J_0. 
\eeqn
The magnitude of the error is approximately the same as that for BH-BH
binaries. For both models, the primary error source in the angular
momentum conservation comes from the fact that $C_p/C_e$ is not
determined in a good accuracy. 

\subsubsection{Merger time}

The merger time, defined as the time at formation of apparent horizon
($T_{\rm AH}$), systematically increases with improving grid
resolution.  Figure \ref{FIGNS4} plots $T_{\rm AH}$ as a function of
$h_{L-1}^2$ for runs NS1616s, NS1616a--c, and NS1616aF--cF. This
figure shows a systematic behavior for convergence of $T_{\rm AH}$
irrespective of the chosen formalism and spatial gauge. For models
NS1416a--c, the similar relation holds, and hence, we do not present
the figure. It is worth noting that runs with the $\tilde
\Gamma^i$-BSSN and $F_i$-BSSN formalisms give approximately the same
values of $T_{\rm AH}$. This indicates that in the absence of BHs,
difference in the spatial gauge does not affect the orbital evolution
of compact stars significantly. Another point to be noted is that the
convergence is relatively slow, although the order of convergence
appears to be second order. Extrapolating the results to the limit
$h_{L-1} \rightarrow 0$ under the assumption of the second-order
convergence, a realistic time of $T_{\rm AH}$ is determined to be
$\approx 530M_0$ for sequences of both formalisms. Thus, even for the
best-resolved run NS1616s, the value of $T_{\rm AH}$ is underestimated
by $\approx 50M_0$ (by $\sim 10\%$ of $T_{\rm AH}$), which is
approximately a half orbital period for an innermost stable circular
orbit with $M_0\Omega \sim 0.06$ \cite{USE,TG}. This indicates that
for obtaining an orbital evolution and gravitational waveforms with a
small phase error (say within $10M_0$ error), the grid resolution
should be by a factor of $\sim 2$ finer than that in the best-resolved
run in the current code, or we should employ a hydrodynamic scheme in
which numerical dissipation is not as large as that in the present
code (but see discussion related to gravitational waves described
below). 

The Taylor T4 formalism predicts the approximate merger time as
$650M_0$, which is obtained by integrating Eq. (\ref{T4}) from the
orbit of $\Omega=\Omega_0$ to the orbit of $M_0\Omega=0.1$ at which
the merger should already proceed. This value is much larger than the
extrapolated numerical result for the merger time.  We explain this
discrepancy as follows: In the Taylor T4 formalism, effects due to
tidal deformation of the NSs are not included. The tidal-deformation
effect increases attraction force between two NSs, and as a result,
the inspiral phase is significantly shortened \cite{LRS}, in
particular for orbits with $M_0 \Omega \agt 0.04$. Indeed, numerical
study for quasiequilibrium NS-NS binaries indicates that tidal effect
plays an important role for $M_0 \Omega \agt 0.04$ (e.g.,
\cite{USE,TG}).  Thus, it is natural that the Taylor T4 formalism
significantly overestimates the merger time.

\begin{figure*}[thb]
\epsfxsize=3.2in
\leavevmode
(a)\epsffile{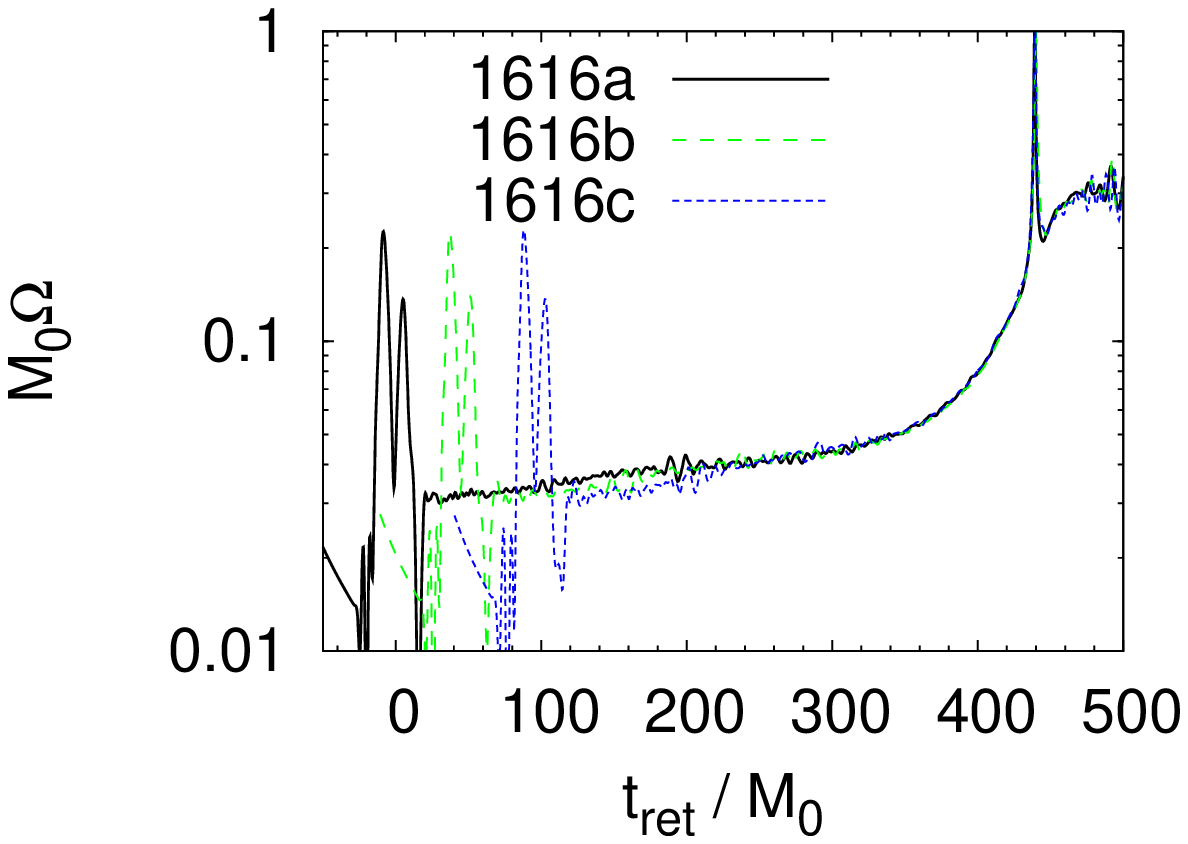}
\epsfxsize=3.2in
\leavevmode
~~~(b)\epsffile{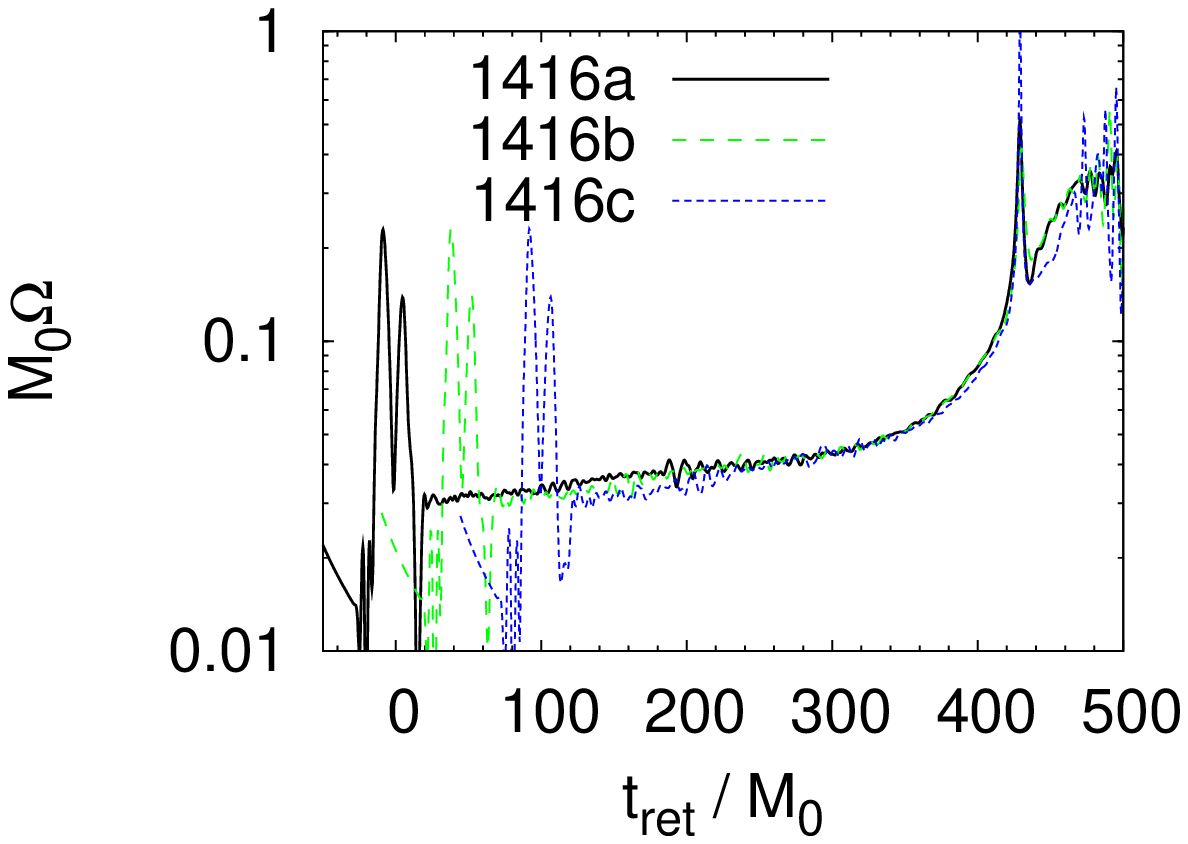}
\vspace{-2mm}
\caption{(a) Angular velocity of gravitational waves as a function of
retarded time for runs NS1616a--c. For runs NS1616b and NS1616c, the
curves are plotted as functions of $t_{\rm ret}+46M_0$ and $t_{\rm
ret}+97M_0$, respectively.  (b) The same as (a) but for runs
NS1416a--c.  For runs NS1416b and NS1416c, the curves are plotted as 
functions of $t_{\rm ret}+47M_0$ and $t_{\rm ret}+101M_0$, 
respectively.
\label{FIGNS7}}
\end{figure*}

\subsubsection{Rest-mass conservation}

Figure \ref{FIGNS5} plots change in total rest mass with time for runs
NS1616a--d and NS1616s. (Similar relations also hold for runs
NS1416a--c; e.g. the maximum violation of the rest-mass conservation
is $\approx 1.5\%$ for run NS1416a). Although the total rest mass
should be conserved, this is not guaranteed in our AMR code (note that
in Shibata's code in which uni-grid is employed, the rest mass is
conserved in a much better accuracy).  The reasons for this are as
follows: (i) Numerical flux determined at refinement boundaries of a
child level does not exactly agree with that determined for the
corresponding parent level. This mismatch of the flux generates slight
violation of the rest-mass conservation. (ii) At the moment of the
regridding, values for a part of the child level are given by
interpolating the values of its corresponding parent level. This process
does not guarantee the rest-mass conservation.  However,
Fig. \ref{FIGNS5} shows that the magnitude of violation is small.  For
the best-resolved runs NS1616s, the violation is at most 0.7\%, and
furthermore, the magnitude of the violation systematically converges with
improving the grid resolution.  (The value of $|M_*/M_{*0}-1|$
converges approximately at second order.)  Therefore, we conclude that
in the well-resolved simulations, the violation of the rest-mass
conservation only gives a minor effect for the numerical results.

\subsubsection{Gravitational waves}

Figure \ref{FIGNS6}(a) and (b) plot gravitational waveforms for runs
NS1616s and NS1416a, respectively.  In the early phase with $t_{\rm
ret} \alt 400 M_0$, the waveforms are characterized by the inspiral
waveforms, and in the final phase, ringdown gravitational waveforms
associated with a quasi-normal mode of the formed BH are seen. In these
simulations, the BH is not immediately formed at the onset of merger
because thermal energy generated by shock heating and/or centrifugal
force due to large angular momentum halt the collapse of the
merged object to a BH for a short time scale. The transient object emits
quasiperiodic gravitational waveforms just before gravitational waves
associated with a quasinormal mode are emitted. Amplitude of
gravitational waves associated with the quasiperiodic oscillation and
the quasinormal mode is by about one order of magnitude smaller than
that emitted in the final inspiral phase. This feature is different
from that in the merger of BH-BH binaries. Due to this small
amplitude, total energy and angular momentum carried away by
gravitational waves are much smaller than those in the merger of BH-BH
binaries (see Table \ref{NSNSRES}).  Due to the relatively small 
emitted angular momentum, the spin parameter of the BH finally formed 
is by a factor of 0.1--0.15 larger than that in the merger of BH-BH
binaries, as already mentioned.

Figure \ref{FIGNS6} (c) compares gravitational waveforms (plus mode)
for run NS1616a and run by Shibata's code. It is seen that two results
agree qualitatively well besides a phase error caused by the
difference in the grid resolution. This shows that the results by
SACRA and Shibata's code agree in a reasonable manner. Figure
\ref{FIGNS6} (d) compares gravitational waveforms (plus mode) for runs
NS1616s, a, and b. For comparing gravitational waveforms for the last
inspiral and merger phases, the data for NS1616a and NS1616b are
plotted as functions of $t_{\rm ret}+10M_0$ and $t_{\rm ret}+54M_0$
for $h_+ \cos(0.2\pi)-h_{\times} \sin (0.2\pi)$ and $-[h_+
\cos(0.2\pi)-h_{\times} \sin (0.2\pi)]$, respectively.  It is found
that the waveforms in the late $\sim 2$ inspiral orbits (about for
3--4 wavelengths) agree well among three models. From the orbit just
before the merger, the difference in the wave phases becomes
outstanding.  This is because the evolution in such phase depends
sensitively on the degree of tidal deformation which is sensitive to
the grid resolution. The agreement of the waveforms in the
intermediate phase also indicates that the strong dependence of
$T_{\rm AH}$ on the grid resolution is primarily due to the fact that
the inspiral orbit at a large orbital separation depends on the grid
resolution. Thus, to follow at least only the late $\sim 2$ orbits,
the grid resolution used in the present work is acceptable.

Figure \ref{FIGNS7} plots angular velocity of gravitational waves as a
function of time for runs NS1616a--c. The angular velocity (and
frequency) of gravitational waves gradually increases in the inspiral
phase. Then, at the onset of merger, it forms a spiky peak. This
appears simply due to the fact that the amplitude of gravitational
waves remains approximately a constant for a moment soon after merger
sets in, and the denominator of Eq. (\ref{gwangv}) approaches to zero. 
In such moment, collapse of the merged object to a BH is halted for a
short time scale and a very compact object of a relatively small
nonsphericity is temporally formed. However, this phase is short and
the compact object soon collapses to a BH. Then, gravitational waves
associated with a quasinormal mode are emitted, and therefore, the
angular velocity eventually reaches to $M_0 \Omega \approx 0.3$. This
value agrees approximately with the angular velocity of the
fundamental $l=m=2$ quasinormal mode of a BH with spin parameter $\sim
0.8$ and the final ADM mass $\sim M_0$.

Figure \ref{FIGNS7} also indicates slow convergence of the merger time
with improving the grid resolution; for improving the grid
resolutions, the inspiral time increases by a large factor. This makes
us reconfirm that for an accurate longterm simulation of inspiraling
NS-NS binaries, a high grid resolution is required. However, this
figure also shows that for $t_{\rm ret} \agt 200M_0$, the curves for
runs NS1616a and NS1616b and for runs NS1416a and NS1416b
approximately agree with each other. This reconfirms that for computing
gravitational waveforms for the late $\sim 2$ orbits, the grid
resolution for NS1616a and NS1416a is acceptable. 

Gravitational waveforms for model NS1416 are very similar to those for
model NS1616. One difference worth noting is that energy and angular
momentum carried away by gravitational waves for model NS1416 are
smaller than those for model NS1616. The reason for these small values
is that tidal disruption occurs at a relatively large orbital
separation.  For the case of NS-NS binaries, gravitational waves are
emitted most effectively at the final inspiral phase just before the
merger.  Thus, absence of such phase due to the tidal disruption
significantly decreases the total amount of gravitational wave
emission.

\subsection{BH-NS Binaries} \label{sec:res-bhns}

\begin{table*}[t]
\caption{List of several quantities for quasicircular states of BH-NS
binaries. We show the mass parameter of puncture ($M_{\rm p}$),
irreducible mass of the BH ($M_{\rm BH}$), rest mass of the NS
($M_{*}$), mass ($M_{\rm NS}$) and compactness defined by ratio of
$M_{\rm NS}$ to circumferential radius ($R_{\rm NS}$) of the NS in
isolation, mass ratio ($q=M_{\rm NS}/M_{\rm BH}$), ADM mass at $t=0$
($M_0$), total angular momentum at $t=0$ in units of $M_0$
($J_0/M_0^2$), and $M_0\Omega_0~(m_0\Omega_0)$ where $\Omega_0$ is
orbital angular velocity at $t=0$ and $m_0=M_{\rm BH}+M_{\rm NS}$.
The BH irreducible mass is computed from area of apparent horizon $A$
as $(A/16\pi)^{1/2}$. All these quantities are normalized by $\kappa$
appropriately to be dimensionless. \label{TBHNS}}
\begin{tabular}{cccccccccc} \hline
~~~Model~~~ & ~~$M_{\rm p}$~~ & ~~$M_{\rm BH}$~~ & ~~$M_{*}$~~ 
& ~~$M_{\rm NS}$~~ & ~$M_{\rm NS}/R_{\rm NS}$~ 
& ~~~~$q$~~~~ & ~~~~$M_{0}$~~~~ & ~~$J_0/M_0^2$~~ &~~$M_0\Omega_0~(m_0\Omega_0)$ ~~
\\ \hline
BHNS-A & ~~0.4185~~ & ~~0.4260~~ & ~~0.1500~~
& ~0.1395~ & 0.145 & 0.327 & 0.5604 & 0.662 & 0.0403~(0.0408)  \\ \hline
BHNS-B & ~~0.4185~~ & ~~0.4250~~ & ~~0.1500~~
& ~0.1395~ & 0.145 & 0.328 & 0.5598 & 0.687 & 0.0337~(0.0340) \\ \hline
\end{tabular}
\end{table*}

\begin{figure}[thb]
\epsfxsize=2.8in
\leavevmode
\epsffile{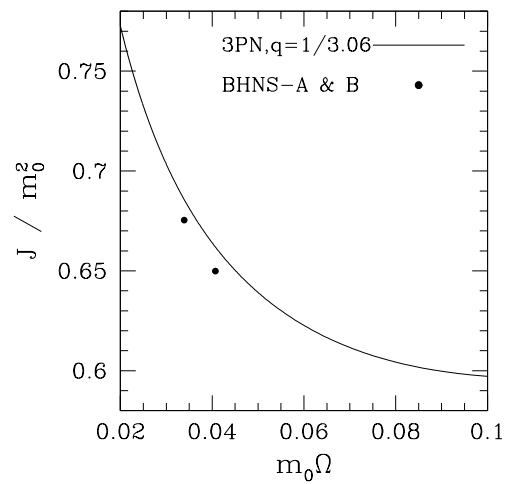}
\vspace{-5mm}
\caption{Angular momentum ($J/m_0^2$) vs angular velocity
($m_0\Omega$) for models BHNS-A and B. For comparison, the same relation
for a binary of $q=1/3.06$ in the third post-Newtonian theory (the
solid curve) is shown together. 
\label{FIG0}}
\end{figure}


\begin{figure}[thb]
\epsfxsize=3.05in
\leavevmode
(a)~~~~~~~~~\epsffile{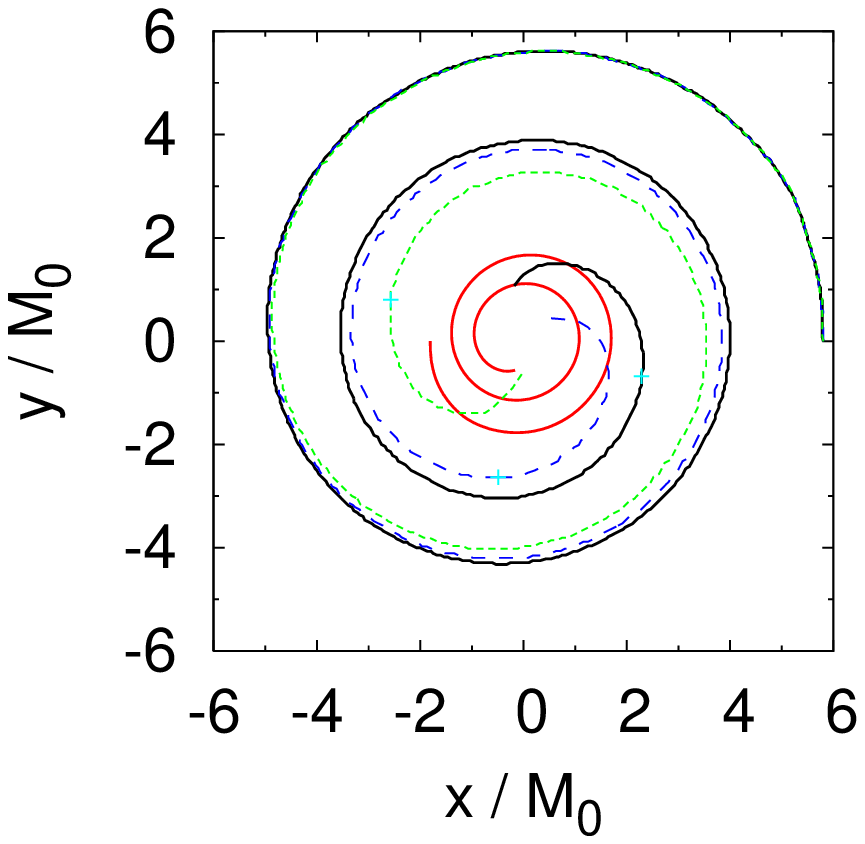} \\
\epsfxsize=3.05in
\leavevmode
(b)~~~~~~~~~\epsffile{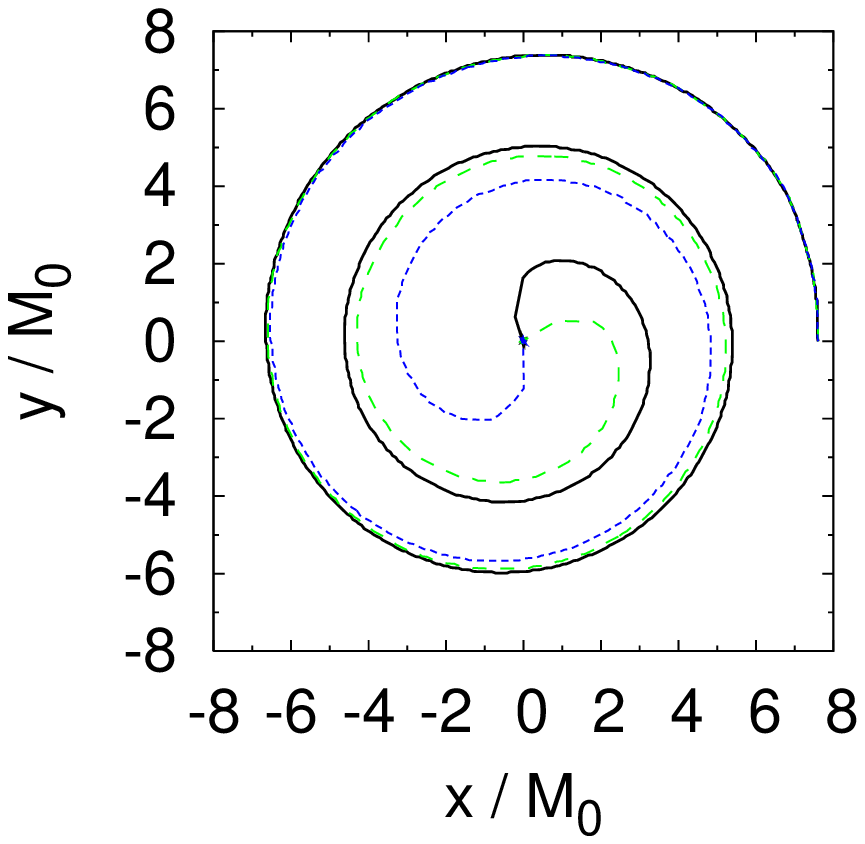} \\
\epsfxsize=3.05in
\leavevmode
(c)~~~~~~~~~\epsffile{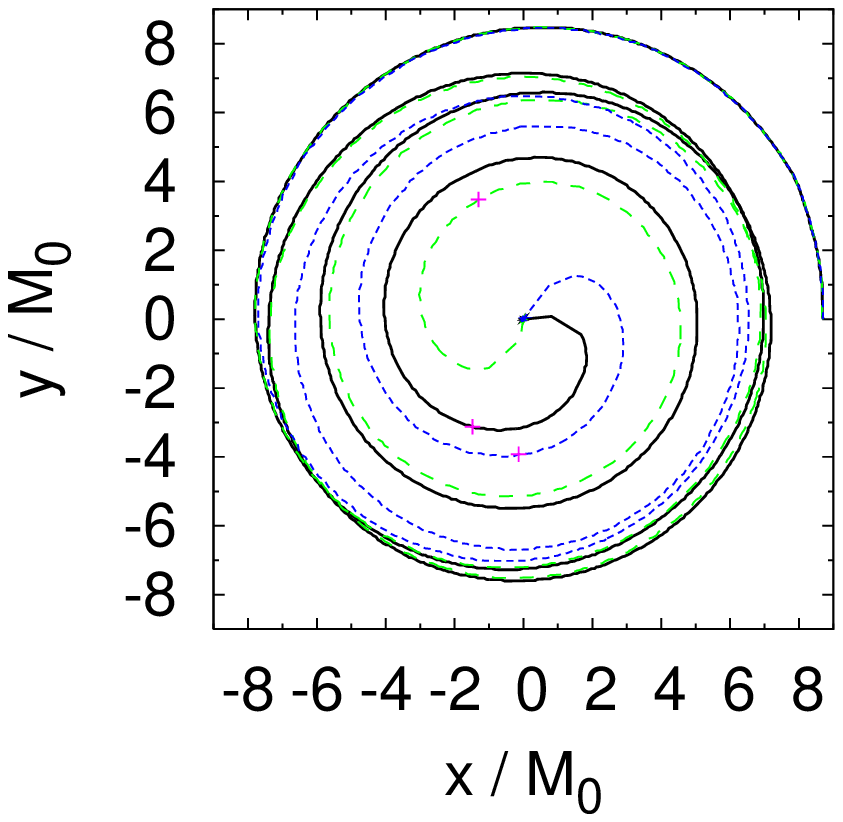} 
\vspace{-2mm}
\caption{(a) Orbital trajectories of NSs for runs BHNS-A1 (solid curve),
BHNS-A2 (long-dashed curve), and BHNS-A3 (dashed curve).  Trajectory
of the BH for run BHNS-A1 (inner solid curve) is also plotted. (b) The same
as (a) but for $x^i_{\rm NS}-x^i_{\rm BH}$. (c) The same as (b)
but for runs BHNS-B1 (solid curve), BHNS-B2 (long-dashed curve), and
BHNS-B3 (dashed curve).  For (a) and (c), the plus mark denotes 
location of the NSs at the onset of tidal disruption (at $t=T_{\rm disr}$). 
\label{FIGBHNS1}} 
\end{figure}

As the last test, we performed simulations for BH-NS binaries.


\subsubsection{Initial condition} \label{sec:bhns-ini}

We adopt BH-NS binaries in quasiequilibrium circular orbits computed
in the moving puncture framework as initial conditions, following our
previous works \cite{BHNS1,BHNS2} (see these references for basic
equations and methods for solving them). We pick up two models in this
work. In both models, the BH is nonspinning and the NS has the
irrotational velocity field with its compactness $\approx
0.145$. Ratio of irreducible mass of the BH to gravitational mass of
the NS in isolation is $\approx 3.05$--3.06. Several key quantities
are listed in Table \ref{TBHNS}. Model BHNS-A is the same as model A
of Ref.~\cite{BHNS2} in which the initial value of the angular
velocity satisfies $M_0 \Omega_0 \approx 0.040$. In the previous paper
\cite{BHNS1}, we find that for the best-resolved run, the binary
orbits for about 1.7 times before the onset of tidal disruption for
this model. We compare numerical results obtained by SACRA with those in the
previous simulation. The other model, referred to as BHNS-B, has a
smaller initial orbital angular velocity as $M_0 \Omega_0 \approx
0.034$. Third post-Newtonian equations of motion for two point masses
predict that the binary orbits $\sim 4$ times before the onset of
merger. We will show that our code can stably and accurately follow
such a longterm orbit.

As we indicated in the previous paper \cite{BHNS2}, the
quasiequilibria used in this paper have a nonzero eccentricity (see
Ref.~\cite{BIW} for the related topic). The primary reason seems to be
a slight deficit of angular momentum in the quasiequilibria. Figure
\ref{FIG0} plots angular momentum ($J/m_0^2$) as a function of angular
velocity ($m_0\Omega$) for models BHNS-A and B as well as that
calculated for two point particles in the third Post-Newtonian theory
\cite{Blanchet}.  This shows that the angular momenta for models
BHNS-A and B are by $\sim 1\%$ smaller than those in the third
post-Newtonian results for a given value of $m_0 \Omega$. Because of
this deficit, the orbital separation quickly decreases soon after the
simulations are started.  More detailed analysis of the
quasiequilibria in the moving puncture framework as well as comparison
of the results with those in the excision framework
\cite{GRAN,TBFS,TBFS2} will be presented in a separate paper
\cite{KST}.

\subsubsection{Setting}

Following previous papers \cite{BHNS1,BHNS2}, the initial condition
for $\alpha$ is modified from the solution of the quasiequilibrium
state so as to satisfy the condition of $\alpha >0$ everywhere. For
the shift, we adopt the quasiequilibrium solution with no change.

Simulations were performed for three grid resolutions (see Table
\ref{BHNSGRID}). For all the cases, a numerical domain is composed of
eight refinement levels (four finer and coarser levels) and locations
of outer and refinement boundaries are chosen to be the same. The NSs
are covered by the finest and second-finest levels. The finest grid
resolution for run BHNS-A2 is approximately the same as that for run
A0 in Ref.~\cite{BHNS2}. The $\tilde \Gamma^i$-BSSN formalism is used
for all the runs performed in the AMR code, and the $F_i$-BSSN
formalism is used for the simulations of model BHNS-A1.  In the
following, results with the $\tilde \Gamma^i$-BSSN formalism are
basically presented, because they depend very weakly on the chosen
formalism, as in the case of NS-NS binaries. 

\begin{table*}[t]
\caption{The same as Table \ref{NSNSGRID} but for simulations of
models BHNS-A and BHNS-B. $\Delta x$ is the minimum grid spacing,
$R_{\rm diam}$ the coordinate length of semi-major diameter of the NS,
$L$ the location of outer boundaries along each axis, $\lambda_0$ the
gravitational wavelength at $t=0$, and $\Delta x_{\rm gw}$ the grid
spacing at which gravitational waves are extracted. $M_{\rm p}$
denotes the mass parameter of puncture BH \cite{BHNS2}.
\label{BHNSGRID}}
\begin{tabular}{ccccccc} \hline
Run & ~~Levels~~ & ~~$N$~~ & $\Delta x/M_0~(\Delta x/M_{\rm p})$ &
~$R_{\rm diam}/\Delta x$~
& ~$L/M_0~(L/\lambda_0)$~ & ~$\Delta x_{\rm gw}/M_0$~ \\ \hline
BHNS-A1, A1F & 8~(4+4) & 30 & 0.036~(0.048) & 76 & 138~(1.8) &
0.58--2.32 \\ \hline
BHNS-A2, A2F & 8~(4+4) & 24 & 0.045~(0.060) & 62 & 138~(1.8) &
0.72--2.88 \\ \hline
BHNS-A3, a3F & 8~(4+4) & 20 & 0.054~(0.072) & 51 & 138~(1.8) &
0.86--3.46 \\ \hline
Ref.~\cite{BHNS2} & --- & --- & 0.047~(0.063) & 59 & 66.7 (0.86)  & 1.05 
\\ \hline
BHNS-B1 & 8~(4+4) & 30 & 0.036~(0.048) & 76 & 138~(1.5)  & 0.58--2.32 \\ \hline
BHNS-B2 & 8~(4+4) & 24 & 0.045~(0.060) & 61 & 138~(1.5)  & 0.72--2.88 \\ \hline
BHNS-B3 & 8~(4+4) & 20 & 0.054~(0.072) & 51 & 138~(1.5)  & 0.86--3.46 \\ \hline
\end{tabular}
\end{table*}

\begin{table*}[t]
\caption{Numerical results for simulations of BH-NS binaries.  We list
the approximate time at the onset of tidal disruption ($T_{\rm
disr}$), irreducible mass of apparent horizon formed after merger
($M_{\rm irr}$), ratio of polar circumferential length to the
equatorial one for the apparent horizon formed after merger
($C_p/C_e$), final BH mass estimated from the equatorial
circumferential length ($C_e/4\pi$), final BH mass estimated from
$M_{\rm irr}$ and $C_p/C_e$ ($M_{\rm BHf}$), final spin parameter of
the BH estimated from $C_p/C_e$, and energy ($\Delta E$) and angular
momentum ($\Delta J$) carried away by gravitational waves.
\label{BHNSRES}}
\begin{tabular}{ccccccccc} \hline
~Run~ & ~$T_{\rm disr}/M_0$~ & ~$M_{\rm irr}/M_0$~ & ~$C_p/C_e$~
& ~$C_e/(4\pi M_0)$~& ~$M_{\rm BHf}/M_0$~ & ~$a$~ & ~$\Delta E/M_0$~
& ~$\Delta J/J_0$ \\ \hline
BHNS-A1 &206 &0.942 &0.938 &0.982 &0.983 & 0.55 & 0.9\% & 14\%\\ \hline
BHNS-A1F&202 &0.942 &0.938 &0.983 &0.983 & 0.55 & 0.9\% & 14\%\\ \hline
BHNS-A2 &186 &0.942 &0.938 &0.983 &0.983 & 0.55 & 0.8\% & 12\% \\ \hline
BHNS-A2F&182 &0.943 &0.939 &0.983 &0.984 & 0.55 & 0.8\% & 13\% \\ \hline
BHNS-A3 &158 &0.943 &0.935 &0.985 &0.986 & 0.56 & 0.7\% & 11\% \\ \hline
BHNS-A3F&156 &0.945 &0.937 &0.986 &0.987 & 0.55 & 0.7\% & 11\% \\ \hline
Ref.~\cite{BHNS2} & 179 
& 0.935 & 0.939 & 0.975 & 0.976 & 0.55 & 0.7\%  & 11\% \\ \hline \hline
BHNS-B1 &472 &0.940 & 0.937 & 0.982 & 0.982 & 0.56 &0.9\% & 17\% \\ \hline
BHNS-B2 &433 &0.940 & 0.936 & 0.983 & 0.983 & 0.56 &0.8\% & 15\%\\ \hline
BHNS-B3 &353 &0.941 & 0.933 & 0.985 & 0.985 & 0.57 &0.8\% & 15\%\\ \hline
\end{tabular}
\end{table*}

\subsubsection{Evolution of the BH and the NS, and final outcome}

Figure \ref{FIGBHNS1}(a) plots orbital trajectories of the NS for
model BHNS-A. Figures \ref{FIGBHNS1}(b) and (c) plot $x^i_{\rm
NS}-x^i_{\rm BH}$ for models BHNS-A and BHNS-B, respectively.  For the
best-resolved run, orbital trajectory of the BH is also plotted in
Fig. \ref{FIGBHNS1}(a). Here, the trajectories of the NSs are
determined from the location of the maximum value of $\rho_*$, and
that of the BHs is from the location of the moving puncture. For both
models BHNS-A and BHNS-B, the NS is tidally disrupted by the companion
BH before it is swallowed by the BH. Before the onset of tidal
disruption, models BHNS-A and BHNS-B spend about 2 and about $3+3/4$
orbits, respectively, for the best-resolved runs. Here, the
approximate time for the onset of tidal disruption (referred to as
$T_{\rm disr}$) is determined from the time at which 1\% of the total
rest mass is swallowed into apparent horizon of the BH.

For runs BHNS-A1, A2, and A3, the tidal disruption starts at $\sim
1.95$, 1.75, and 1.45 orbits, respectively (see also Table
\ref{BHNSRES} for the time in units of $M_0$). In the run A0 of
Ref.~\cite{BHNS2}, the tidal disruption starts approximately at the
same time as that for run BHNS-A2. This is quite reasonable because
the grid resolution around the BH and the NS for run BHNS-A2 agrees
approximately with that of run A0 of Ref.~\cite{BHNS2}. The values of
$T_{\rm disr}$ depend weakly on the chosen formalism for a given grid
resolution. This illustrates that the numerical results depend weakly
on the formalism and gauge.

For runs BHNS-B1, B2, and B3, the tidal disruption starts at $\sim
3.7$, 3.3, and 2.75 orbits, respectively.  Because the time spent in
the inspiral phase for model BHNS-B is longer than that for model
BHNS-A, numerical error is accumulated more, resulting in a larger
dispersion in $T_{\rm disr}$. A characteristic feature for model
BHNS-B is that its orbital eccentricity is initially very large: Soon
after the simulation is started, the orbital separation decreases by a
large factor, and then, it significantly increases.  For the first two
orbits, the orbit is obviously different from circular orbits. Gravitational
waveforms shown later also illustrate that the orbit is eccentric. As
mentioned in Sec.~\ref{sec:bhns-ini}, the primary reason for the
presence of the eccentricity is that the angular momentum of the 
quasiequilibrium initially given is likely to be by $\sim 1\%$ smaller
than that for the true quasiequilibrium. However, in the last $\sim 2$
orbits, the orbital separation gradually and monotonically decreases
to merger (see the trajectory for run BHNS-B1), suggesting that the
eccentricity is reduced by emission of gravitational waves.

\begin{figure}[thb]
\epsfxsize=3.2in
\leavevmode
(a)\epsffile{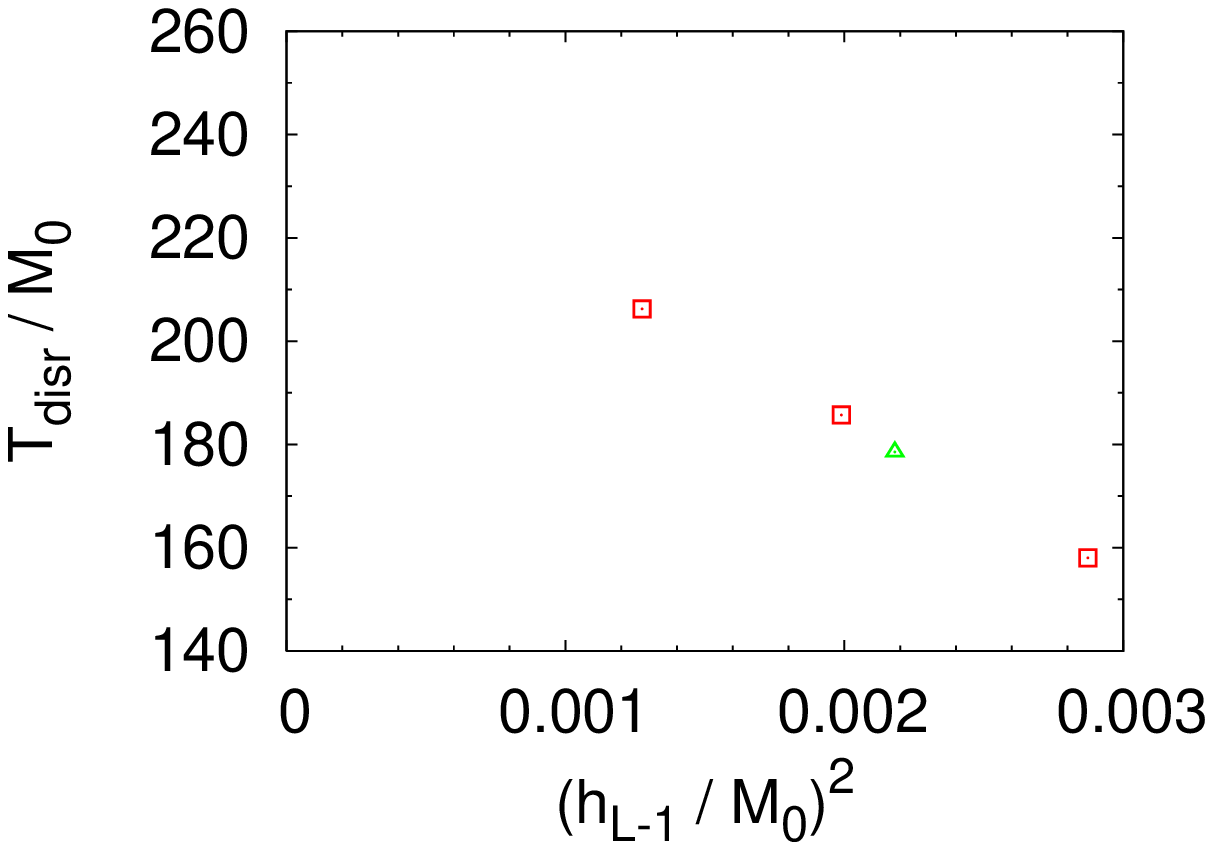} \\
\epsfxsize=3.2in
\leavevmode
(b)\epsffile{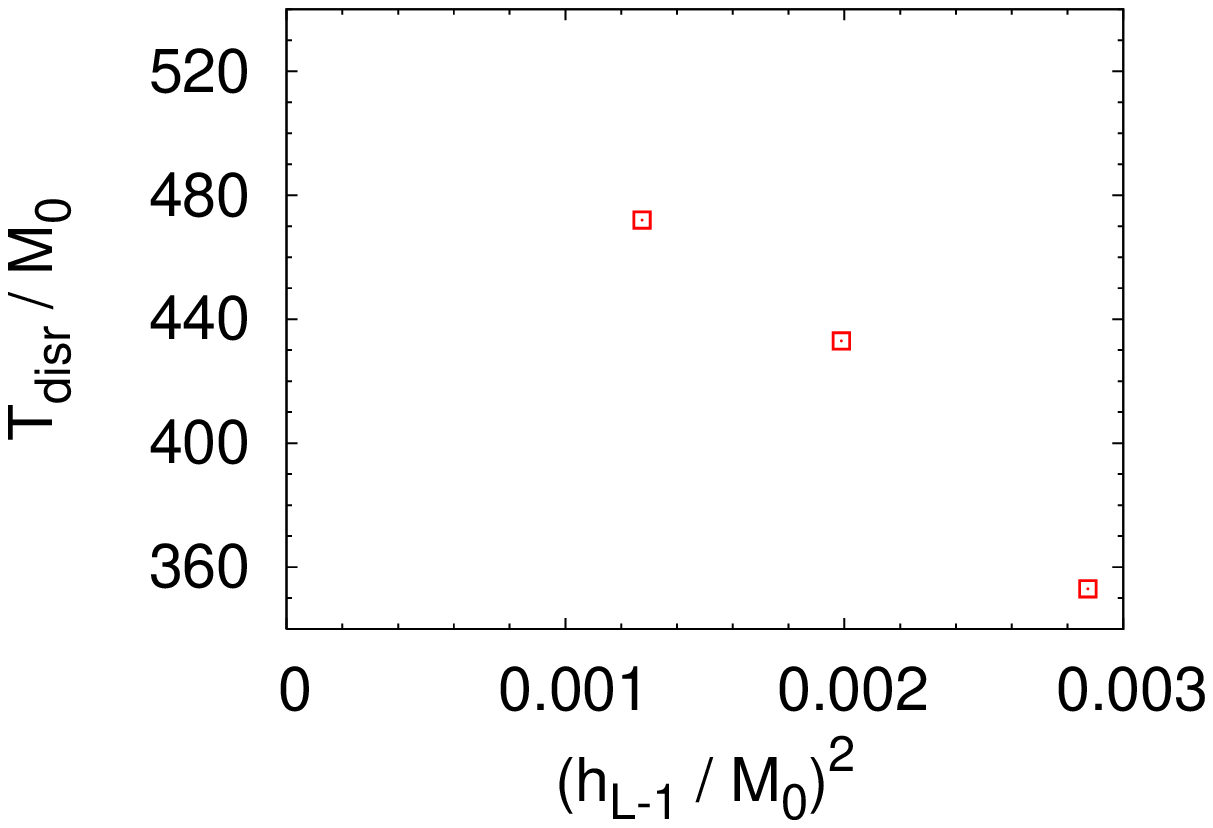}
\vspace{-1mm}
\caption{$T_{\rm disr}$ as a function of $h_{L-1}^2$.  (a) The open
squares denote the data by runs BHNS-A1--A3, and the triangle denotes
the result obtained in Ref.~\cite{BHNS2}. (b) The same as
(a) but for the data by runs BHNS-B1--B3. 
\label{FIGBHNS2}}
\end{figure}

As reported above, the time at the onset of tidal disruption, $T_{\rm
disr}$, systematically increases with improving the grid resolution.
Figure \ref{FIGBHNS2}(a) and (b) plot $T_{\rm disr}$ as a function of
$h_{L-1}^2$ for models BHNS-A and BHNS-B, respectively.  Figure
\ref{FIGBHNS2}(a) shows that $T_{\rm disr}$ is approximately
proportional to $h_{L-1}^2$.  Extrapolating this relation to $h_{L-1}
\rightarrow 0$, it is found that the converged value for $T_{\rm
disp}$ is $\approx 240M_0$. This suggests that by the onset of tidal
disruption, the binary would orbit for $\sim 9/4$ times.  Thus, the
results in run BHNS-A1 are near the convergent ones, although the
phase error of $\approx 30M_0$ (about a quarter orbit) would be still
present. The value determined by the extrapolation is much smaller
than the value predicted by the Taylor T4 formalism which gives
$\approx 315M_0$. This is reasonable because the Taylor T4 formalism
neglects effects associated with tidal deformation of the NS and BH,
which accelerates the inward motion and shortens $T_{\rm disr}$ (e.g.,
see Ref.~\cite{LRS}).

Figure \ref{FIGBHNS2}(b) shows that the results of $T_{\rm disr}$
converge with improvement of the grid resolution at an order better
than the second order. The results for three grid resolutions appear
to be approximately fourth-order convergent, but such a high order is
unlikely for the chosen scheme for hydrodynamics. This may be due to a
too small value of $T_{\rm disr}$ for run BHNS-B3 in which convergence
may not be achieved.  Thus, we assume the second-order convergence, as
inferred from the result for model BHNS-A, and use the values for runs
BHNS-B1 and B2 for extrapolation. Then, the extrapolation gives
$T_{\rm disr} \approx 540M_0$. Thus, in run BHNS-B1, the value of
$T_{\rm disr}$ is underestimated by $\approx 70M_0$. (Note that if we
assume the fourth-order convergence, the predicted value of $T_{\rm
disr}$ is $\approx 500M_0$.) The primary reason for this
underestimation is that numerical dissipation spuriously shortens the
inspiral time.  The extrapolated value of $T_{\rm disr} \approx
540M_0$ is again smaller than the value predicted by the Taylor T4
formalism which gives $\approx 585M_0$.  As mentioned above, this
disagreement is reasonable because tidal effects are neglected in the
Taylor T4 formalism. 

\begin{figure*}[thb]
\epsfxsize=3.2in
\leavevmode
(a)\epsffile{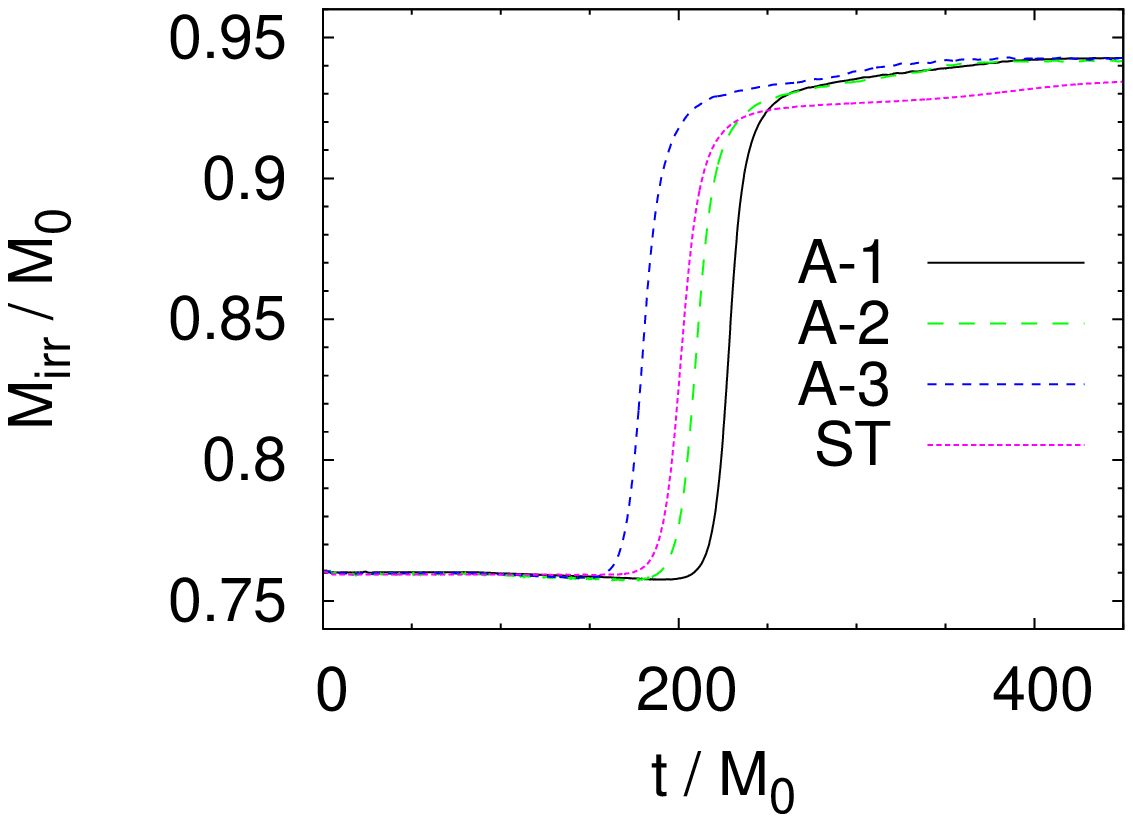} 
\epsfxsize=3.2in
\leavevmode
(d)\epsffile{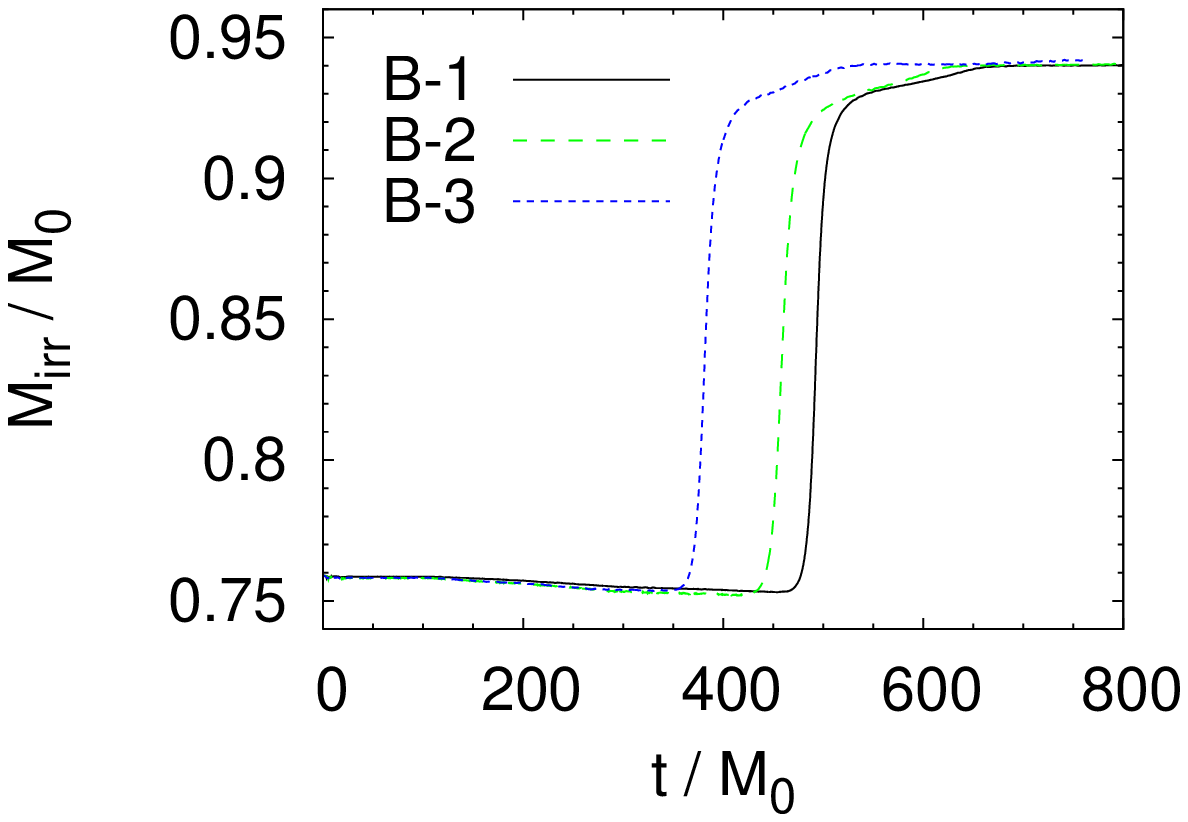} \\
\epsfxsize=3.2in
\leavevmode
(b)\epsffile{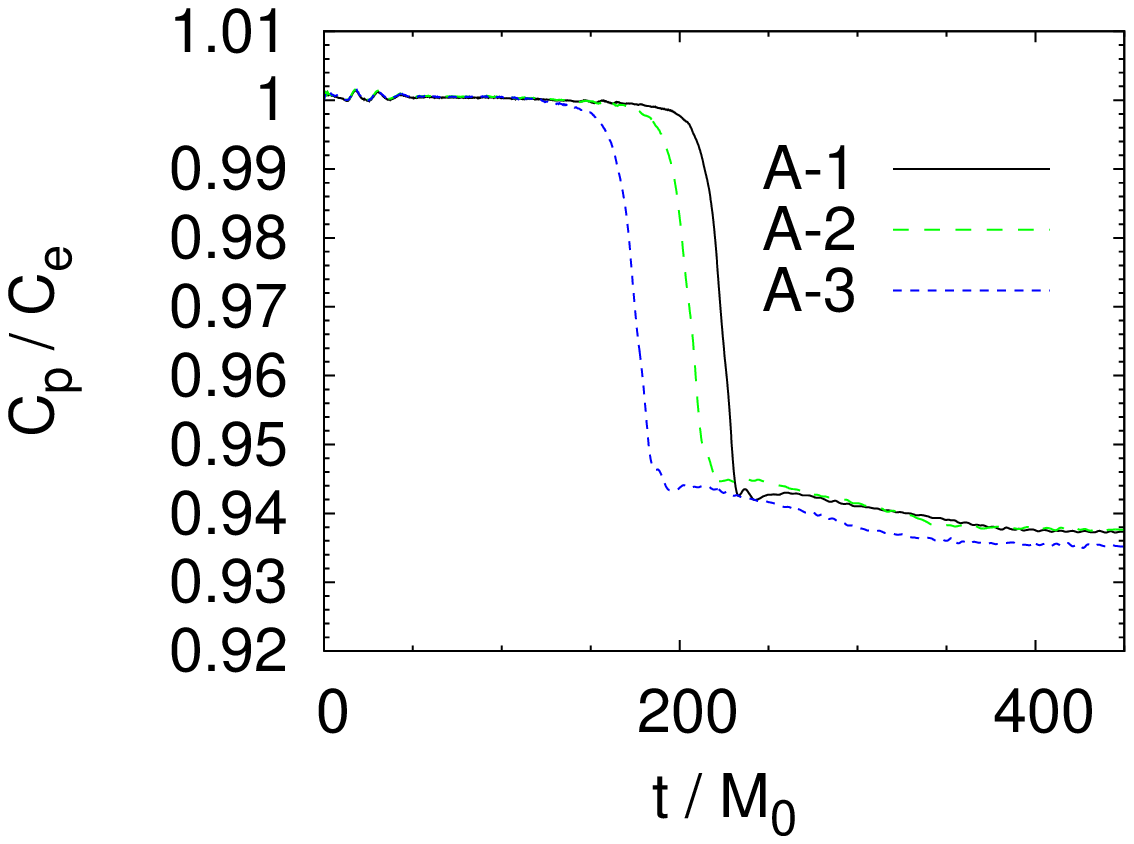} 
\epsfxsize=3.2in
\leavevmode
(e)\epsffile{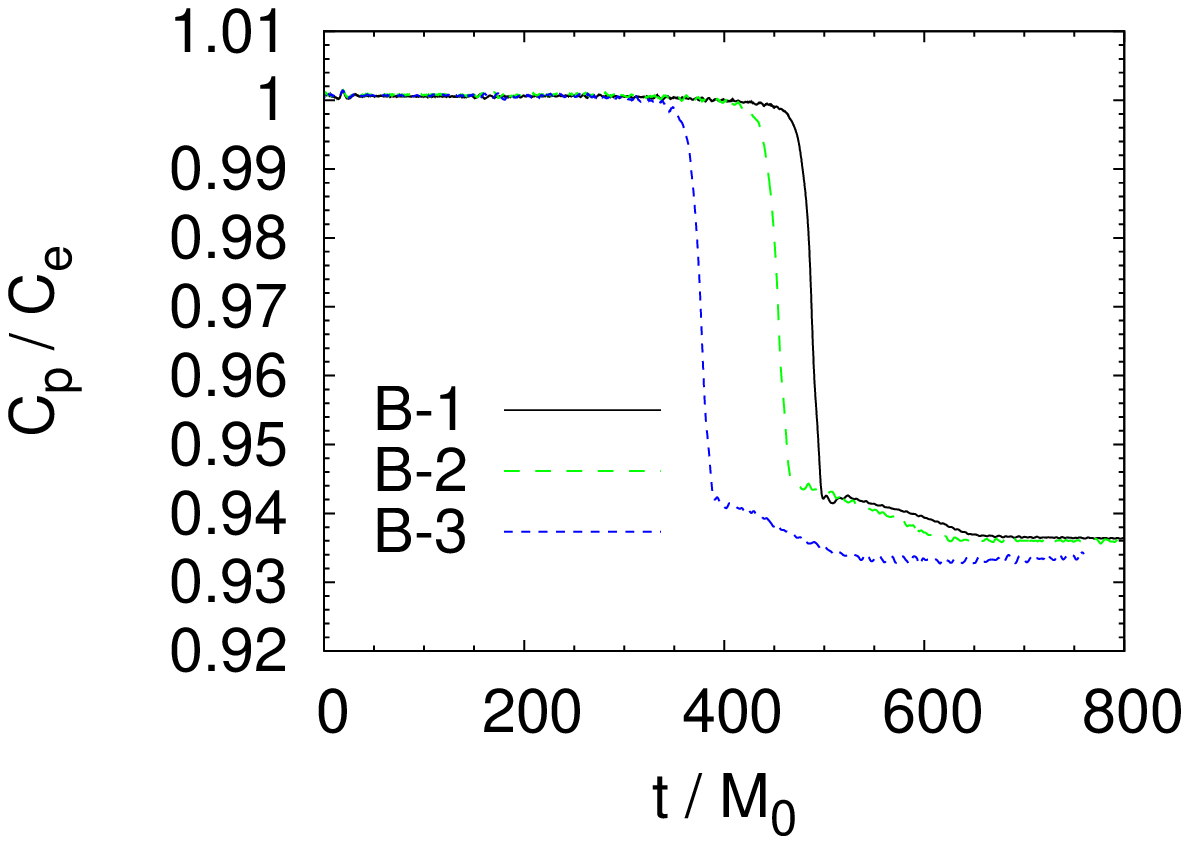} \\
\epsfxsize=3.2in
\leavevmode
(c)\epsffile{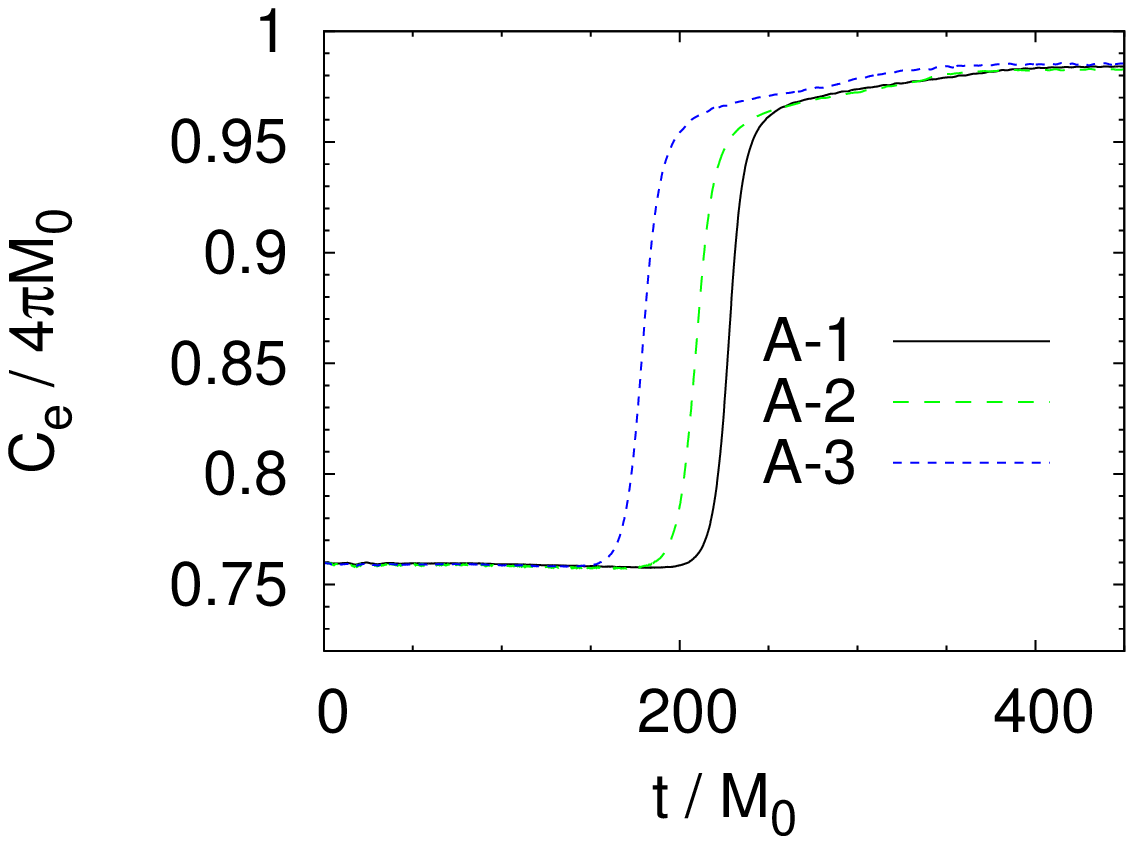}
\epsfxsize=3.2in
\leavevmode
(f)\epsffile{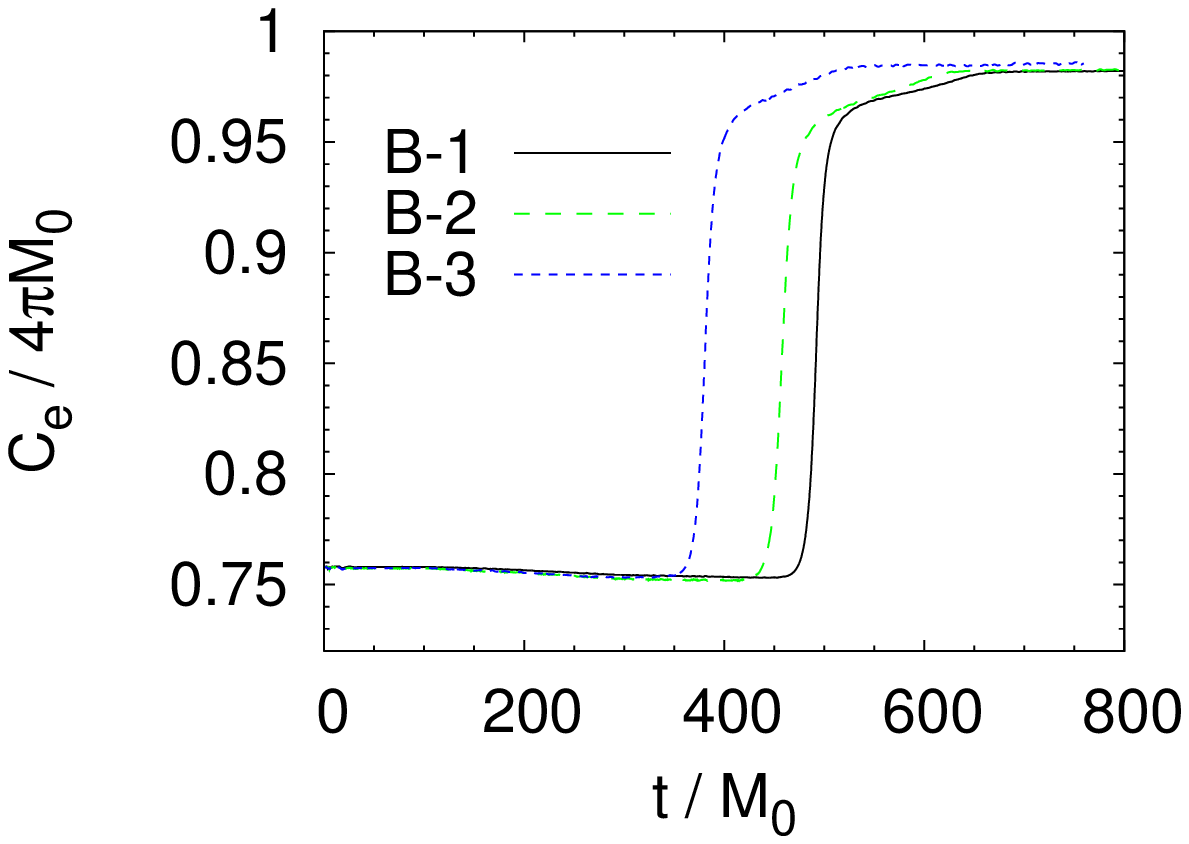}
\vspace{-4mm}
\caption{(a) $M_{\rm irr}$ of the BH as a function of time
for runs BHNS-A1--A3. ``ST'' denotes the results of run A0
in Ref.~\cite{BHNS2}. 
(b) The same as (a) but for $C_p/C_e$. 
(c) The same as (a) but for $C_e/(4\pi M_0)$.
(d) The same as (a) but for BHNS-B1--B3. 
(e) The same as (d) but for $C_p/C_e$. 
(f) The same as (d) but for $C_e/(4\pi M_0)$. 
\label{FIGBHNS3}}
\end{figure*}

Figure \ref{FIGBHNS3} plots evolution of $M_{\rm irr}/M_0$, $C_p/C_e$,
and $C_e/(4\pi M_0)$ as functions of time.  This shows that BHs are
approximately in stationary states before and after tidal disruption
of the companion NS. By contrast, they quickly evolve during tidal
disruption and subsequent accretion process, irrespective of the
initial condition.  The figure for $C_p/C_e$ shows that the BH is
approximately nonrotating before the onset of tidal disruption because
it is approximately unity.  However, as the mass accretion proceeds,
its value decreases, reflecting the fact that the BH spins up by
getting angular momentum from the infalling material. The mass
accretion is also reflected in the figures of $M_{\rm irr}/M_0$ and
$C_e/4\pi M_0$ because they increase after the onset of tidal
disruption. The values of these quantities are approximately the same
before the onset of tidal disruption, reflecting that the spin of the
BH is approximately zero. After the onset of tidal disruption, these
are different, because the final state is a spinning BH for which
$M_{\rm irr} \not= C_e/4\pi$. 

The final values of $M_{\rm irr}$, $C_p/C_e$, and $C_e/(4\pi M_0)$ for
both models BHNS-A and BHNS-B depend only weakly on the grid
resolution. In particular, the results for models BHNS-A1 and BHNS-A2
and for BHNS-B1 and BHNS-B2 show approximate convergence. This
indicates that with the present numerical simulation, the final state
of the BH is determined with a good accuracy, although the merger time
depends strongly on the grid resolution.

The final value of the BH spin for model BHNS-A agrees approximately
with the result in Ref.~\cite{BHNS2}. As reported in
Ref.~\cite{BHNS2}, the final value of the BH spin is smaller than the
initial spin of the system. The reason is that gravitational waves
carry away a substantial fraction of angular momentum. (For the
previous result in Ref.~\cite{BHNS2}, a part of the angular momentum is
distributed to disk, and this is also a part of the reason.)  The
final value of the BH mass for model BHNS-A slightly disagrees with the
previous result \cite{BHNS2}. The reason for this difference is that a
disk of $\sim 0.017M_0$ is formed around the BH in the previous result
(see discussion in Sec.~\ref{sec5c4}).

\begin{figure*}[t]
\vspace{-6mm}
\caption{Snapshots of density contour curves and density contrasts as
well as location of the BH, from the onset of tidal disruption to the
semi-final state of the BH for run BHNS-B1. The contour curves are plotted for
$\rho w=10^{-i}$ where $i=2, 3, \cdots, 6$ (the outermost curve always
denotes $\rho w =10^{-6}$). The first panel denotes the state at about
3.2 orbits. The filled circles show the region inside apparent horizon.
\label{FIGBHNS4}}
\end{figure*}

\subsubsection{Conservation of energy and angular momentum}

The numerical results for the final outcome are checked by examining
whether or not the conservation relations, Eqs. (\ref{conE}) and
(\ref{conJ}), hold.  As shown below, the contribution of disk formed
around the BH is negligible in this case. From Table \ref{BHNSRES}, we
find that errors in the conservation for the best-resolved runs are
$\approx 1\%$ for the energy and $\approx 5\%$ for the angular
momentum; $1-\Delta E/M_0 \approx 0.01$ and $1-\Delta J/J_0 \approx
5\%$. Namely, the final values of mass and angular momentum of the BH
are smaller than those expected from the conservation relation. The
possible reasons are either (i) the energy and angular momentum
carried away by gravitational radiation might be underestimated or
(ii) during the evolution, the energy and angular momentum might be
dissipated by some spurious numerical effects. We note that the
conservation relations hold in a much better manner in the previous
result \cite{BHNS2} (in this case, contribution of disk plays an
important role). Thus, the reason for the violation of conservation
may be associated with interpolation and extrapolation performed in
the AMR algorithm, which are absent in the previous simulation
\cite{BHNS2}. Currently, the source for the error is not
specified. Improving the accuracy for the conservation is an issue for
the future work.

\subsubsection{Disk mass}\label{sec5c4}

\begin{figure*}[thb]
\epsfxsize=3.2in
\leavevmode
(a) \epsffile{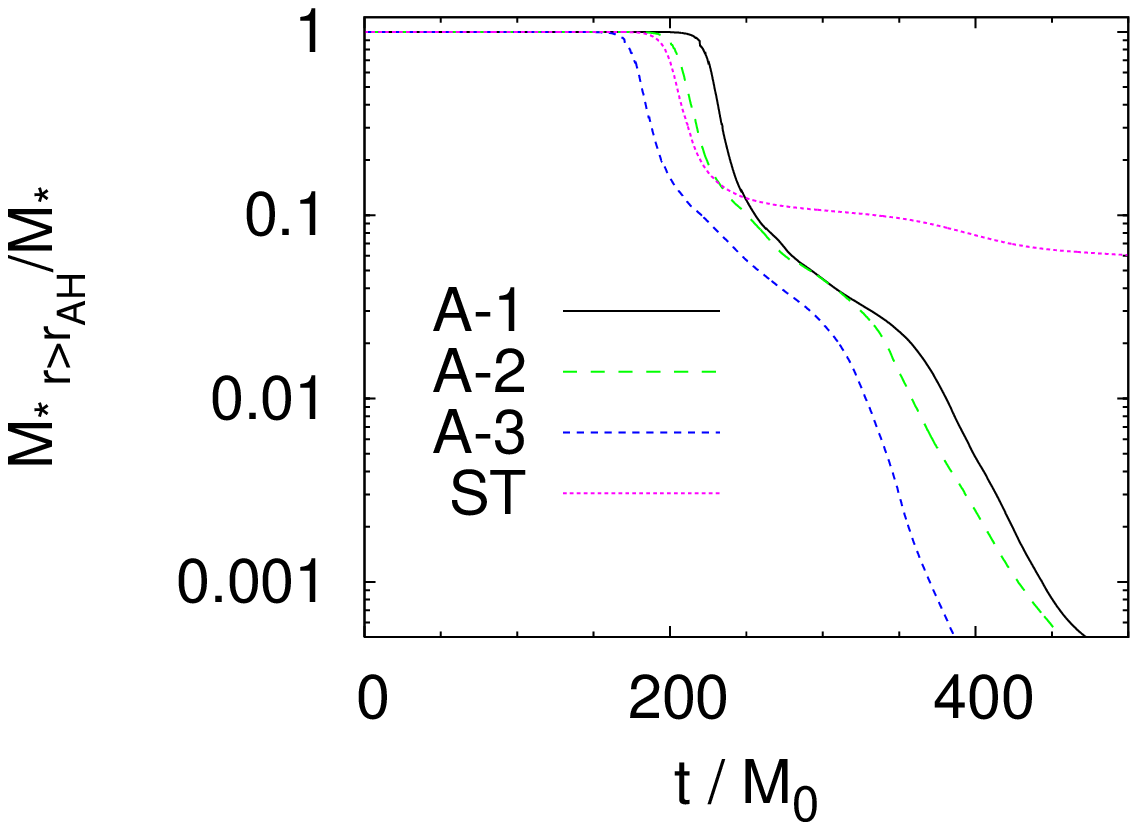}
\epsfxsize=3.2in
\leavevmode
(b) \epsffile{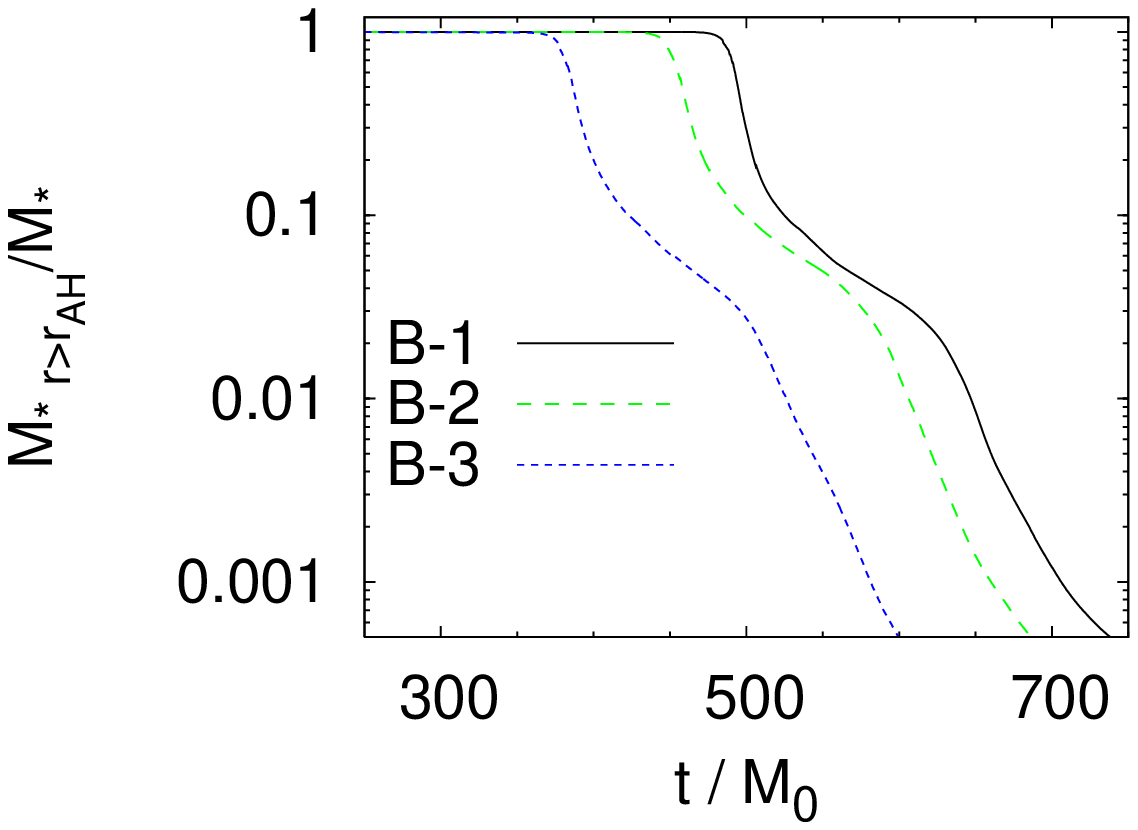}
\vspace{-4mm}
\caption{Total rest mass of material located outside apparent horizon as a
function of time (a) for model BHNS-A and (b) for model BHNS-B. ``ST''
in panel (a) denotes the result of Ref.~\cite{BHNS2}.
\label{FIGBHNS5}}
\end{figure*}

Figure \ref{FIGBHNS4} displays snapshots of density contour curves and
density contrasts as well as location of the BH, from tidal-disruption
phase to the semi-final state of the BH for run BHNS-B1.  The tidal
disruption sets in when the coordinate separation between BH and NS
centers becomes $\sim 5M_0$ (see the first panel of
Fig. \ref{FIGBHNS4}). Because the separation is small and the orbit is
close to the innermost stable circular orbit, the radial approaching
velocity induced by gravitational radiation reaction is not small at
the onset of tidal disruption. This implies that the NS is disrupted
while it is approaching to the BH with a high speed which is a
substantial fraction of the orbital velocity. Due to this large
approaching velocity, most of the NS material is swallowed by the BH
soon after the onset of tidal disruption. However, the material in the
outer part of the NS still spreads outward and subsequently forms a
spiral arm around the BH (see the second and third panels). Mass of
the spiral arm is $\sim 0.1M_*$ initially and the spiral arm spreads
to a large radius with $r \agt 5M_0$ (see the third panel). These
properties are qualitatively the same as those found in the previous
paper \cite{BHNS2}.  However, most of the material in the spiral arm
subsequently falls toward the BH and only a tiny fraction of the
material can escape from the BH (see the fourth panels of
Fig. \ref{FIGBHNS4}). This result disagrees with the previous one for
a given NS radius and mass ratio \cite{BHNS2,footBHNS}.

The present result indicates that the material in the spiral arm does
not obtain specific angular momentum large enough for forming a disk
around the BH.  Although the material in the outer part receives
angular momentum from the material in the inner part during tidal
disruption and subsequent spiral-arm formation via an angular-momentum
transport process, this effect may not play a significant
role. Alternatively, some mechanisms for dissipation and/or anti
transportation of the angular momentum may work during the evolution
of the spiral arm, which might not be accurately computed in the
previous work \cite{BHNS2} due to some computational problems.  For
example, (i) the grid structure might not be appropriate for
accurately following the angular momentum transport and (ii) in the
previous simulation \cite{BHNS2}, we evolved $\phi$ (instead of $W$)
which has large magnitude and gradient near the moving puncture, and
hence, trajectory of the BH which sensitively depends on $\phi$ might
not be accurately computed to follow the BH orbit after the tidal
disruption sets in (e.g., see Fig.~4 of Ref. \cite{pedro}).  For
example, if the BH spuriously moves away from the spiral arm, disk
formation would be spuriously enhanced. Completely alternative
possibility is that the grid resolution far from the BH may not be
high enough in the present grid structure to follow the evolution of
the spiral arm accurately (see discussion below).

For more specific discussion about the fate of material after the
tidal disruption, we generate Figs. \ref{FIGBHNS5}(a) and (b), which
plot the total rest mass of material located outside apparent horizon
as a function of time for runs BHNS-A1--A3 and BHNS-B1--B3,
respectively. Irrespective of models and grid resolutions, this
decreases monotonically after the tidal disruption sets in.  However,
there are two phases after the tidal disruption. For the first
100--$150M_0$, the infall rate of the material into the BH is
relatively low.  In such phase, a part of the tidally disrupted
material spreads outward and subsequently a spiral arm is formed
around the BH (cf. the third panel of Fig. \ref{FIGBHNS4}).  The
presence of this phase agrees qualitatively with our previous result
\cite{BHNS2}.  In the second phase, the infall rate increases and the
fraction of the rest mass around the BH decreases quickly to be much
smaller than 1\%, implying that a disk or torus with substantial mass
is not formed.  The presence of this later phase disagrees with our
previous result \cite{BHNS2,footBHNS}. 

A possible reason for the small disk mass is that in the present
simulation, the grid resolution for following the formation of disk or
torus around the BH might not be sufficient: During tidal disruption,
the NS is elongated and then a fraction of material escapes from the
finest-refinement domain.  The motion of such material around the BH
might not be accurately computed in relatively coarser levels. As a
consequence, spurious dissipation or transportation of the angular
momentum by numerical viscosity would happen, and the material might
subsequently fall into the BH spuriously.  To improve this situation,
it is necessary to prepare a fine grid which covers a larger region
around the BH (say within a radius of $\sim 10M_0$).  To perform such
simulation, modification of the present AMR scheme may be necessary,
e.g. to increase the grid number $N$ for the finer refinement levels
while fixing it for the coarser levels. Such improvement is an issue
for the future.

Assume that the present result for no disk formation is correct. Then, 
the typical life time of the accreting material with mass larger than
$10^{-2}M_* \sim 0.01M_{\odot}$ is $\sim 100M_0 \approx 2.5
(M_0/5M_{\odot})$ ms. Here, we assume that a hypothetical mass of the
NS is $\sim 1.4M_{\odot}$ and as a result $M_0 \sim 5M_{\odot}$. Such
short life time is not appropriate for explaining generation of
gamma-ray bursts from the accretion disk, for which the duration is
longer than at least 10 ms. 

\subsubsection{Gravitational waves}

\begin{figure*}[thb]
\epsfxsize=3.2in
\leavevmode
(a)\epsffile{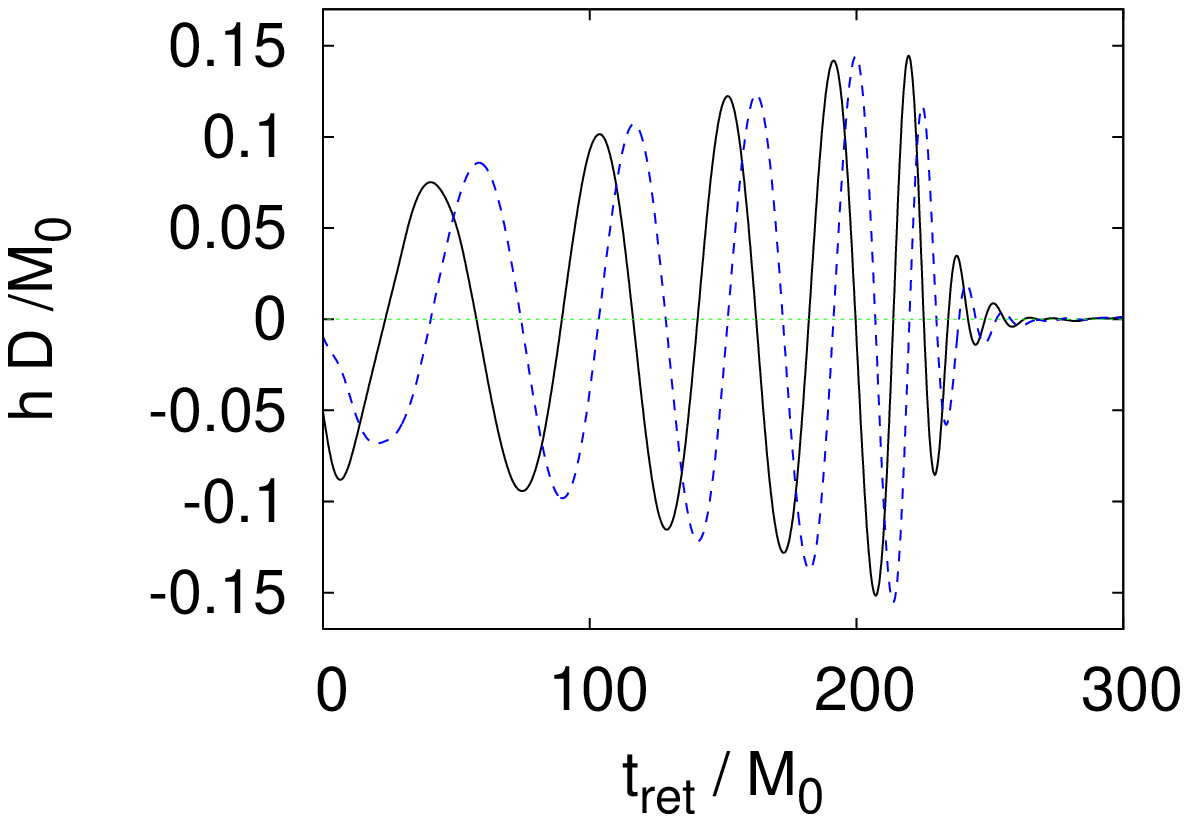}
\epsfxsize=3.2in
\leavevmode
(b)\epsffile{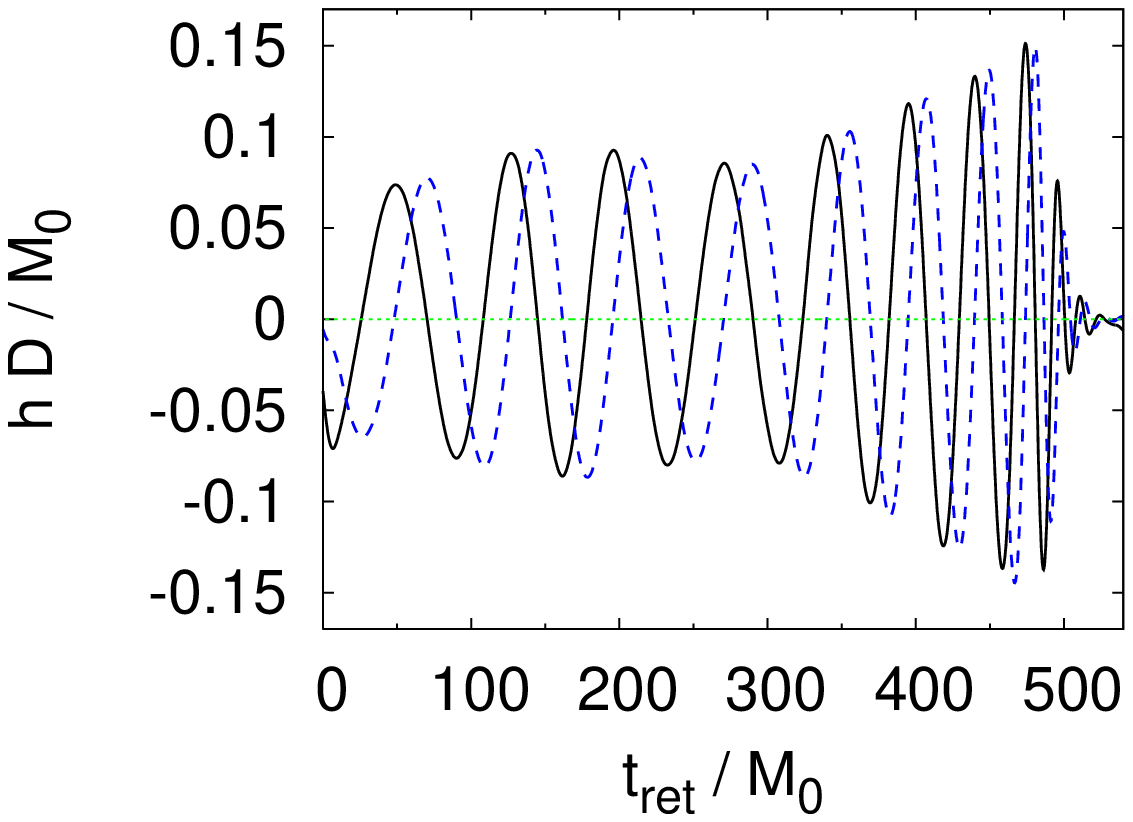} \\
\epsfxsize=3.2in
\leavevmode
(c)\epsffile{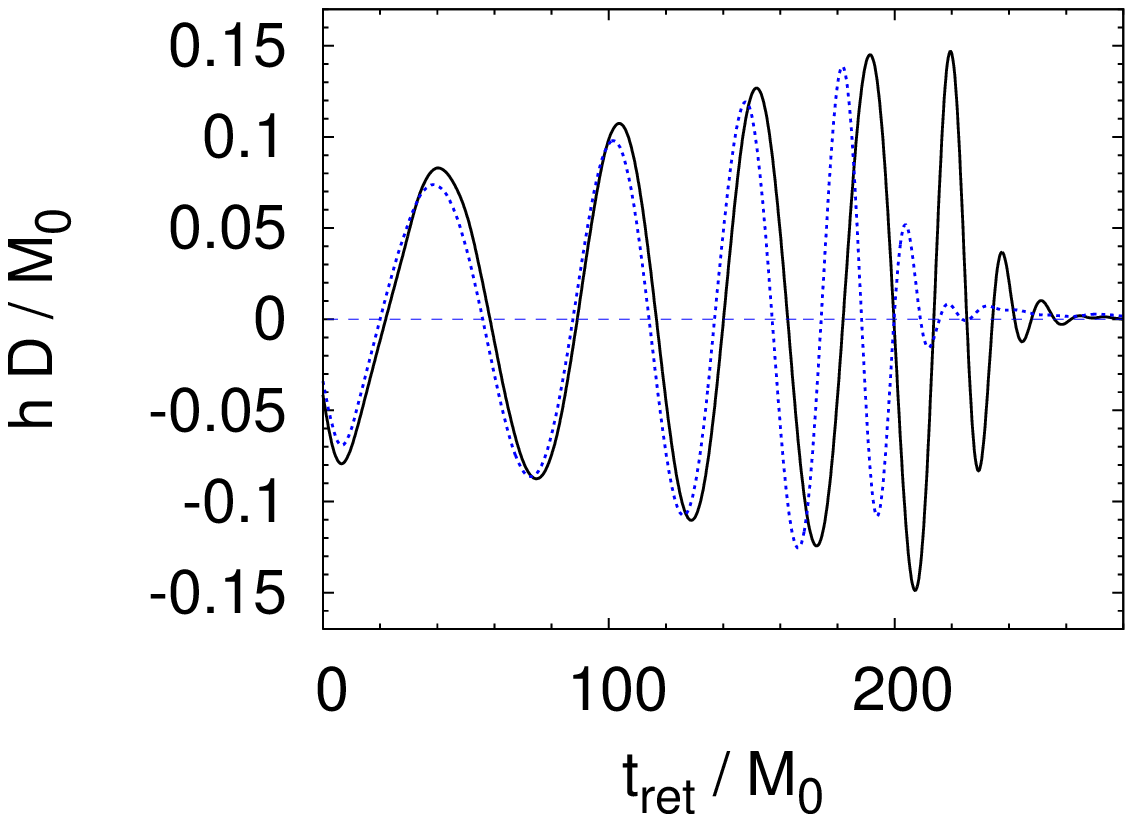}
\epsfxsize=3.2in
\leavevmode
(d)\epsffile{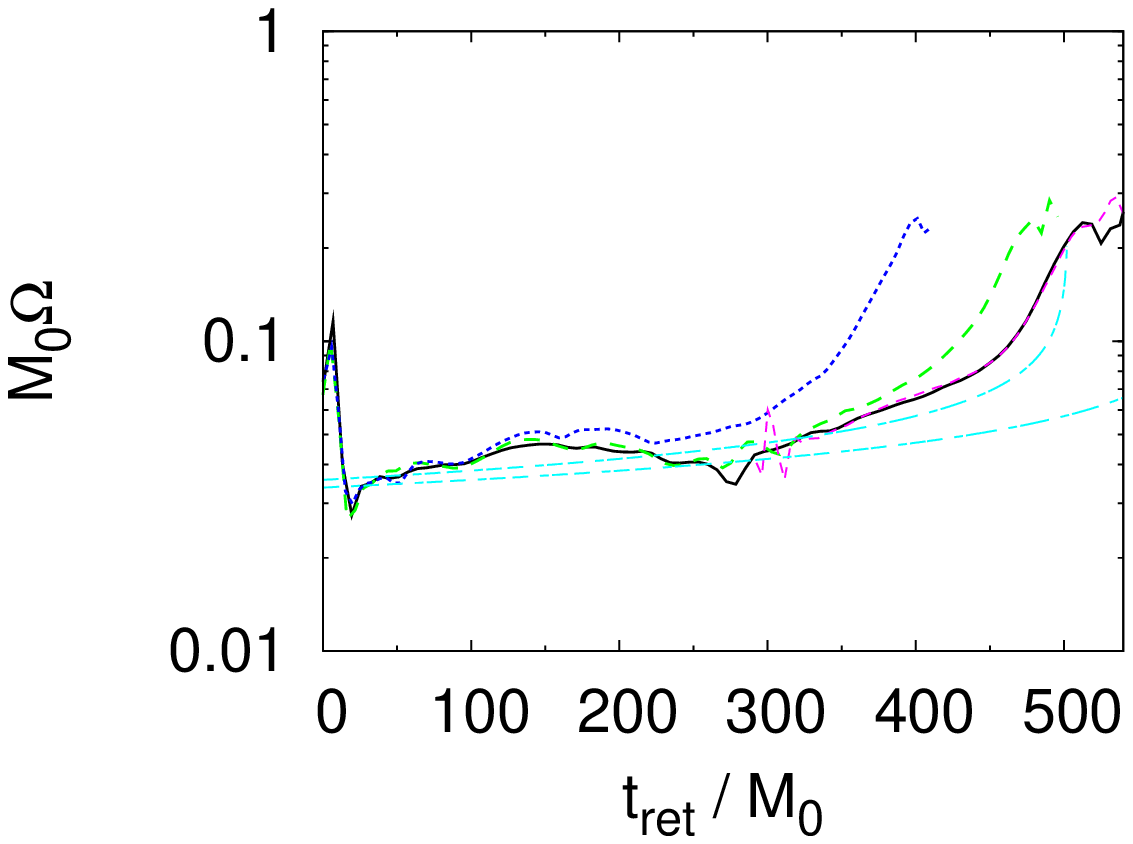}
\vspace{-2mm}
\caption{Gravitational waveforms (a) for run BHNS-A1 and (b) for run
BHNS-B1. The solid and dashed curves denote the plus and cross modes,
respectively. (c) Plus mode of gravitational waveforms for run BHNS-A1
(solid curve) and for run A0 in Ref.~\cite{BHNS2} (dashed
curve). (d) Angular velocity of gravitational waves for BHNS-B1--B3
(solid, long-dashed, dotted curves) and BHNS-A1 (dashed curve). For
BHNS-A1, the curve is plotted as a function of $t_{\rm
ret}+266M_0$. The dotted-dashed curves denote the results predicted
by the Taylor T4 formalism for $m_0 \Omega_0=0.0340$ (lower curve)
and 0.0360 (upper curve).
\label{FIGBHNS6}}
\end{figure*}

Figure \ref{FIGBHNS6} plots gravitational waveforms for runs BHNS-A1
and BHNS-B1.  As shown in Ref.~\cite{BHNS2}, the waveforms are
composed of two components.  One is the inspiral waveform, and the
other is the merger waveform. The amplitude quickly decreases after
the onset of tidal disruption. The reason for this behavior is
explained as follows: At the tidal disruption, material of the NS
spreads, and then, the matter density as well as degree of nonaxial
symmetry quickly decrease. Hence, the amplitude of gravitational
waves, which depends strongly on the compactness and degree of
nonaxial symmetry, damps. In the final phase, the ringdown
gravitational waveform associated with quasinormal mode oscillation is
seen. As pointed out in Ref.~\cite{BHNS2}, the amplitude is not as
large as that in the merger of BH-BH binaries; the amplitude is $\sim
10\%$ of that at the last inspiral orbit.  We explain the reason as
follows: The material does not coherently fall into the BH because of
tidal disruption, and the resulting phase cancellation suppresses
coherent excitation of the quasinormal mode oscillation.

Because of a large eccentricity of the binary orbit for model BHNS-B,
the gravitational waveforms are modulated in the early inspiral phase.
To derive more realistic waveforms emitted during quasicircular
orbits, it is necessary to prepare better initial condition. This
issue is left for the future work. However, in the last $\sim 2$
inspiral orbits, the BH and the NS for run BHNS-B1 appear to be
approximately in a circular orbit.  Thus, even if a simulation is
started from an eccentric orbit, the orbit is eventually circularized.
To pay attention only to gravitational waves emitted from the final
inspiral to the merger phases, the present initial condition may be
acceptable. 


Figure \ref{FIGBHNS6}(c) compares the plus mode of gravitational
waveform for run BHNS-A1 with that of run A0 in
Ref.~\cite{BHNS2}. Because the tidal disruption time ($T_{\rm disr}$)
is different between two simulations, the phase does not
agree. However, the qualitative feature is very similar; in the early
phase, gravitational waves associated with inspiral motion are
seen. After the onset of tidal disruption, the amplitude quickly
damps, and eventually, waveforms are characterized by a ringdown
oscillation of small amplitude associated with a quasinormal mode of
the BH. This qualitative agreement confirms the conclusion about
gravitational waveforms in the previous paper \cite{BHNS2}.

Figure \ref{FIGBHNS6}(d) plots angular velocity of gravitational
waves for runs BHNS-B1--B3 and BHNS-A1.  For BHNS-A1, the curve is
plotted as a function of $t_{\rm ret}+266M_0$. The dotted curve
denotes the result predicted by the Taylor T4 formalism for
$m_0\Omega_0=0.034$ and 0.036. This also shows that the binary for
model BHNS-B is in an eccentric orbit, because $\Omega$ considerably
modulates with time.  The plots for runs BHNS-B1--B3 clarify again
that the value of $T_{\rm disr}$ depends strongly on the grid
resolution. The results for runs BHNS-A1 and BHNS-B1 agree
approximately for the late inspiral phase. This is expected because
the waveforms for both models are similar as shown in
Figs. \ref{FIGBHNS6}(a) and (b).  The curve for run BHNS-B1 does not
agree with that derived by the Taylor T4 formalism for
$m_0\Omega_0=0.034$, but agree relatively with that for
$m_0\Omega_0=0.036$. A part of the reason is that the inspiral time is
spuriously shortened by numerical dissipation.  Another possible
reason is that the binary has a large eccentricity initially, which
enhances gravitational wave emission and shortens the inspiral time;
namely, an averaged orbital velocity of the initial condition does not
satisfy $m_0\Omega_0=0.034$ but may be close to $m_0\Omega_0=0.036$.

\section{Summary} \label{sec:summary}

We have reported our new numerical relativity code, named SACRA,
in which an AMR algorithm is implemented. In this code, the Einstein
evolution equations are solved in the BSSN formalisms with a
fourth-order spatial finite-differencing scheme, the hydrodynamic
equations are solved by a third-order high-resolution central scheme,
and the time integration is done in the fourth-order Runge-Kutta
scheme. Both $F_i$-type and $\tilde \Gamma^i$-type BSSN formalisms are
implemented. In both cases, $W =e^{-2\phi}$ is evolved instead of
evolving $\phi$. This enables us to adopt grid-center-grid coordinates. 

\subsection{Technical Points and Issues for the Future}

To check feasibility of SACRA, we performed simulations for
coalescence of BH-BH, NS-NS, and BH-NS binaries. All the simulations
were performed on personal computers using at most 5 GBytes
memory. The required CPU time is at most 1 month even for the
best-resolved runs. For simulating BH-BH binaries, we employed the
same initial conditions as those adopted by Buonanno et
al. \cite{BHBH12}. Our results agree with theirs in a reasonable
manner except for a slight disagreement possibly associated with the
difference in choices of gauge conditions and numerical scheme. We
also show that our code can follow inspiraling BH-BH binaries at least
for about 4.5 orbits even in the absence of dissipation term such as
Kreiss-Oliger-type dissipation term. This implies that even in the AMR
code, the dissipation term is not always necessary, if appropriate
schemes for interpolation and extrapolation are employed for the
procedures at refinement boundaries.

Our numerical results for BH-BH binaries indicate that for
accurately evolving final 2.5 orbits before the merger, relatively
small number of grid points is sufficient. The orbit of the BH is
computed accurately and, as a result, gravitational waveforms are
computed with a small phase error. In the present simulations, the
used memory is at most 3 GBytes, and a personal computer of 4 GBytes
memory is sufficient for accurate evolution of the final phase of
BH-BH binaries.

By contrast, numerical results, in particular the merger time, depend
strongly on the grid resolution, grid structure, and gauge condition
for evolving $\approx 4.5$ inspiral orbits. The estimated phase error
in gravitational waveforms for such cases is about $40m_0$ even in the
finest-resolution simulation in this paper. For obtaining convergent
results within the phase error of, say, $10m_0$ for the whole
evolution, the grid resolution has to be finer by a factor of $\sim
2$. However, we note the following: Numerical results for the final
state of the BH formed after merger do not depend on the grid
resolution as strongly as the merger time and gravitational wave
phase.  It should be noted that for determining the final state of the
BH within 1\% error, it is not necessary to take a high-grid
resolution. We find that the present choice is appropriate.


We find that the merger time and gravitational wave phase could
depend on spatial gauge conditions.  The reason for this is explained
as follows: The physical grid spacing and grid structure depend on the
spatial gauge condition, in particular, around BHs. Thus, the
magnitude of numerical dissipation also depends on the spatial gauge
and may be reduced for a simulation performed with an appropriate
choice for the spatial gauge, even if the same grid structure is
employed. Therefore, an appropriate choice of the spatial gauge
condition may reduce computational costs, and a careful choice is
required.

We also show that our code can evolve NS-NS and BH-NS binaries.
Numerical results obtained by SACRA agree with those in the previous
simulations, if we resolve the NSs and BHs by approximately the same
accuracy. However, the computational cost is at most 5\% of the
previous uni-grid simulations and robustness of the AMR scheme is
confirmed. Simulations with much better accuracy than those in the
previous simulations can be performed by less computational
costs. Because we performed the simulations for a wide range of grid
resolutions, we can also estimate the magnitude of the phase error of
gravitational waveforms in the present and previous numerical results
\cite{BHNS2} in an inexpensive computational cost.

We followed inspiral phase of BH-NS binaries for a long time ($\sim 4$
orbits) for the first time. In the best-resolved simulation, the
inspiral orbit up to the onset of tidal disruption is followed for
about 3.7 orbits.  Subsequent merger and ringdown phases are also
computed well for producing gravitational waveforms. However, we find
that the prepared quasicircular initial condition has a large
eccentricity, and the inspiral orbit is highly eccentric for the first
$\sim 2$ orbits, although the eccentricity for a few orbits just before
the merger is reduced by the emission of gravitational waves. To
perform a realistic simulation for the inspiral phase with small
eccentricity, it is necessary to improve the initial condition (see,
e.g., Ref.~\cite{FKPT} for a method). This is an issue for the future. 

We compare the duration spent in the inspiral phase obtained by
numerical simulations with that predicted by the Taylor T4 formalism
for BH-BH and BH-NS binaries.  For longterm runs with the merger time
$\agt 500M_0$, the merger time determined by extrapolation of the
numerical results agree with the prediction by the Taylor T4 formalism
within an error of $\sim 10\%$. This makes us reconfirm that the
Taylor T4 formalism provides a good semi-analytical estimate for the
time spent in the inspiral phase.  We also find that the Taylor T4
formalism always provides an overestimated value of the merger time
for NS-NS and BH-NS binaries. The reason for this overestimation is
that in this formalism, tidal effects of NSs, which accelerate the
infalling process to merger, are not included. Nevertheless, the error
is not extremely large because tidal effects play a crucial role only
for close orbits. Therefore, for validating a numerical result, it is
useful to compare the merger time with the result derived by the
Taylor T4 formalism.

We find that the convergence of the merger time for NS-NS binaries is
relatively slow. For this case, the evolution of NSs in the late
inspiral phase depends on the effects of tidal deformation of each NS,
which in general shortens the merger time.  Thus, to accurately
determine the orbital evolution, the tidal deformation of each NS has
to be followed accurately in hydrodynamics. The degree of tidal
deformation is in general larger near the surface of the NS because
the tidal force is approximately proportional to the distance from the
center of each NS. In our AMR scheme, the grid resolution around the
surface region is not as high as that in the central region.
Consequently, the tidal deformation is not followed as accurately as
that in the central region. A simple way to overcome this problem is
to resolve the surface region as accurately as the central region,
i.e., to cover each NS in the finest level. However, doing this in our
present scheme is computationally expensive because we have to choose
a large value of $N$ for the finest level. There may be a better grid
structure to overcome this problem, e.g. to change the cube size in
each refinement level. Improving our AMR scheme is an issue in the
next step. A completely alternative possibility is to employ a
different hydrodynamic scheme which is less dissipative. Improving
this scheme is also an issue in the future.

We note that the final state of the BH and surrounding disk after
merger of NS-NS binaries do not depend on the grid resolution as
strongly as the merger time.  This property is the same as that in the
case of BH-BH binaries. Thus, for studying the final state,
the present choice of the grid resolution is acceptable. 


We check whether or not the conservation relations of energy and
angular momentum denoted by Eqs. (\ref{cons}) and (\ref{consJ}) or by
Eqs. (\ref{conE}) and (\ref{conJ}) hold. The energy conservation holds
within $\sim 1\%$ error irrespective of the binary components for the
best-resolved run. The error of angular momentum conservation is
larger: The error is $\sim 3\%$--5\%. The resulting total energy and
angular momentum of BHs are always smaller than the values predicted
by the conservation relations, and hence, numerical dissipation is the
most likely source of the error. The error size for the angular
momentum conservation may not be negligible, in particular, for
studying disk formation around the BH formed after merger. In our
results, the disk mass is likely to be underestimated. Indeed, the
result for the disk mass in this paper does not agree with the
previous result of a BH-NS binary \cite{BHNS2}.  In the previous
result, the angular momentum conservation holds in a much better
manner. Thus, the small disk mass in the present results might be
partly due to the spurious loss of angular momentum \cite{footend}. 


Another possible drawback in our present AMR scheme is that we might
not be able to accurately follow material that spreads around the BH
after tidal disruption of the NS.  The reason is that a large fraction
of material escapes from the finest level soon after the onset of
tidal disruption.  The motion of such material orbiting the BH is
located at relatively coarser levels and hence it may not be followed
accurately. The material, which forms a spiral arm around the BH,
subsequently falls into the BH in a short time scale in the present
result. This may be in part due to the fact that its angular momentum
is spuriously dissipated. In the present simulations, we found that
the resulting mass of accretion disk is much smaller than $10^{-3}M_*$
for $q \approx 0.33$ and $M_{\rm NS}/R_{\rm NS}=0.145$. This result
totally disagrees with our previous results \cite{BHNS2,footBHNS} as mentioned
above. Note that the evolution of binaries up to tidal disruption
agrees well indicating that the grid structure is appropriate at least
up to the onset of tidal disruption.  This suggests that the grid
structure in our AMR code might not be well-suited only for following
the material orbiting the BH of a distant orbital separation. To
improve this situation, it may be necessary to prepare a fine grid
which covers a larger region around the BH.  To perform such a
simulation, it will be necessary to change the grid structure, e.g. to
increase the grid number for the finer levels while fixing that for
the coarser levels. Such improvement of our current AMR scheme is an
issue in the next step.

\subsection{Comparison of Numerical Results for Three Types of Binaries}

We performed simulations for three types of binaries. Because of the
presence of strong equivalence principle, the orbital evolution and
gravitational waveforms in the inspiral phase with a large orbital
separation depend very weakly on the components of the binaries.  By
contrast, the final outcome and gravitational waveforms in the merger
phase depend strongly on the components. As already found in the
previous studies (e.g., \cite{BHBH12}), we found that after merger of
slowly spinning two equal-mass BHs, a rotating BH with spin $\approx
0.7$ is formed.  However, the magnitude of the spin parameter is much
higher for a BH formed after merger of NS-NS binaries: The present
results show that the spin is $\sim 0.8$--0.85. This disagreement
comes primarily from the difference in amplitude of gravitational
waves emitted in the final merger phase. In the case of BH-BH
binaries, the BHs can have a closer orbit than the NSs because the BHs
are more compact. As a result, gravitational waves are significantly
emitted in the final inspiral orbit.  In addition, the quasinormal
mode oscillation of fundamental $l=m=2$ mode is excited significantly
in the merger phase. Indeed, the gravitational wave amplitude is as
high as that emitted at the last inspiral orbit (cf. Fig. \ref{FIG4}). 
By these gravitational wave emissions, the angular momentum is
significantly dissipated in the final phase. By contrast, in the case
of NS-NS binaries, the merger sets in at a relatively distant orbit
because NSs are not as compact as BHs, and moreover, the quasinormal
mode is not excited as significantly as in the case of BH-BH binaries
because of smaller degree of nonaxisymmetric deformation of the
spacetime curvature at the merger.

Because of the difference in amplitude of ringdown gravitational
waveforms, the property of gravitational waveforms in the final merger
phase depends strongly on the binary components. As mentioned above,
the amplitude of ringdown gravitational waves is as high as that in
the last inspiral phase for the merger of BH-BH binaries.  By
contrast, the amplitude is $\sim 10\%$ as high as that in the last
inspiral phase for the merger of NS-NS binaries. Thus, the wave
amplitude quickly decreases in this case.

We also study the merger of BH-NS binaries. In the present paper, we
focus on the case that the NS is tidally disrupted before it is
swallowed by the companion BH.  In this case, the quasinormal mode is not
significantly excited as in the case of NS-NS binaries, and hence, the
amplitude of ringdown gravitational waves is also much smaller than
that in the last inspiral orbit. However, this may not be always the
case. If the mass ratio, $q(=M_{\rm NS}/M_{\rm BH})$, is small enough,
the NS will not be tidally disrupted before swallowing by the BH. In
such case, a quasinormal mode may be excited significantly at a moment
that the NS falls into the BH. This topic should be investigated in
the future work. 

As summarized in this section, gravitational waveforms at merger phase
depend strongly on the binary components. This makes us reconfirm that
gravitational waves at merger phase will carry information about the
properties of binary components. As reviewed in Sec. I, a number of
simulations have been performed in the past decade. However, there are
a huge parameter space for which numerical study has not been done
yet, in particular for NS-NS and BH-NS binaries. Obviously, further
study is required. Our new code SACRA will be able to make a 
contribution to this purpose. 

\begin{acknowledgments}
We thank L. Baiotti and Y. Sekiguchi for helpful discussion, and
K. Kyutoku and H. Okawa for their help. We also would like to thank H. 
Pfeiffer and his collaborators for providing initial data of BH-BH
binaries for open use (see
http://www.black-holes.org/researchers3.html/).  Numerical
computations were in part performed on the NEC-SX8 at Yukawa Institute
of Theoretical Physics of Kyoto University. This work was supported by
a Monbukagakusho Grant (No. 19540263).
\end{acknowledgments}

\appendix

\section{Apparent horizon finer} \label{app:ahfinder}

Apparent horizon is defined as a marginally outermost
trapped two-surface on which the following equation is
satisfied:
\beqn
K - K_{ij} s^i s^j - D_i s^i =0. \label{A1}
\eeqn
Here, $s^i$ denotes a unit normal vector, orthogonal to
the two-surface of the apparent horizon. Denoting the
location of the apparent horizon as $r=h(\theta, \varphi)$
for an appropriately chosen coordinate center, $s_i$ is written as
\beqn
s_i= C W^{-1}(1, -h_{,\theta}, -h_{,\varphi}),\label{A2}
\eeqn
where $C=(\tilde \gamma^{ij}s_i s_j)^{-1/2}$. Substituting Eq. (\ref{A2})
into Eq. (\ref{A1}), the equation for $h(\theta,\varphi)$ is derived, 
and its schematic form is
\beqn
h_{,\theta\theta}+\cot\theta h_{,\theta}+{h_{,\varphi\varphi} \over
\sin^2\theta}-2 h =S, \label{A3}
\eeqn
where $S$ denotes the source term composed of $\gamma_{ij}$, $K_{ij}$,
$h$, and its derivatives. In the method of Ref.~\cite{AH}, we write the
source term in a straightforward manner and solve the 2D elliptic-type
equation (\ref{A3}) iteratively.

In SACRA, first of all, we slightly change the form of the basic equation,
simply rewriting Eq. (\ref{A1}) as 
\beqn
&&h_{,\theta\theta}+\cot\theta h_{,\theta}+{h_{,\varphi\varphi} \over
\sin^2\theta}-2h  \nonumber \\
&&=h_{,\theta\theta}+\cot\theta h_{,\theta}+{h_{,\varphi\varphi} \over
\sin^2\theta}-2h \nonumber \\
&&~~~~~~~~~~~~-(K - K_{ij} s^i s^j - D_i s^i)/(C W),
\label{A4}
\eeqn
where on the right-hand side, we input trial values for $h$ in 
each iteration step. In our previous method, $D_i s^i$ is
calculated to be a complicated function of
$h$, $h_{,\theta}$, $h_{,\theta\theta}$,
$h_{,\varphi}$, $h_{,\varphi\varphi}$, $h_{,\theta\varphi}$,
and $\gamma^{ij}$. In the present method, 
we simply use a finite-differencing for evaluating $D_i s^i$.
Namely, we write it as
\beqn
&&W^3 \biggl[ {1 \over r^2} \pa_r (r^2 W^{-3} s^r)
+{1 \over \sin\theta} \pa_{\theta} (\sin\theta W^{-3} s^{\theta})
\nonumber \\
&&~~~~~~~~~~~~~~~~~+\pa_{\varphi} (W^{-3} s^{\varphi}) \biggr],
\eeqn
and evaluate each term by the second-order finite-differencing.  Here,
$\pa_{\theta}$ and $\pa_{\varphi}$ are evaluated on the apparent
horizon and hence the evaluation is straightforward, whereas $\pa_r$
cannot be evaluated on the apparent horizon. To compute it, we prepare
two dummy points for each point of $(\theta, \varphi)$ which are
located at slightly outside and inside of the apparent horizon along
an orthogonal direction with respect to the two-surface. The
coordinate distance from those dummy points to the apparent horizon is
chosen to be $\approx h/20$. By this method, the source term of
Eq. (\ref{A4}) is significantly simplified.

With this setting, the solution of $h$ is obtained by solving 2D
elliptic-type equation (\ref{A4}). The method for solving this
equation is the same as that described in detail in Ref.~\cite{AH}. We
compare the results of the apparent horizon mass obtained in the
present and previous methods and find that both results agree well.

\section{Numerical results for BH-BH binary with different initial condition
of gauge variables} \label{app:res}

\begin{table*}[th]
\caption{The same as Table \ref{BHBHRES}, but for runs 19a' and 19aF'.  
\label{BHBHRES2}}
\begin{tabular}{cccccccccc} \hline
~Run~ & ~$T_{\rm AH}/m_0$~ & ~$M_{\rm irr}/m_0$~ & ~$C_p/C_e$~
& ~$C_e/(4\pi m_0)$~& ~$M_{\rm BHf}/m_0$~ & ~$a$~ & ~$\Delta E/m_0$~
& ~$\Delta J/J_0$~ & ~Level~ \\ \hline
19a' & 518.5&0.879 & 0.891 & 0.949 & 0.950 &0.702 &0.035&0.28& $L$-1
\\ \hline
19aF'& 507.3&0.879 & 0.891 & 0.949 & 0.950 &0.702 &0.035&0.28& $L$-1
\\ \hline
\end{tabular}
\end{table*}

\begin{figure}[thb]
\epsfxsize=3.1in
\leavevmode
~~~~~~~~~~~~~\epsffile{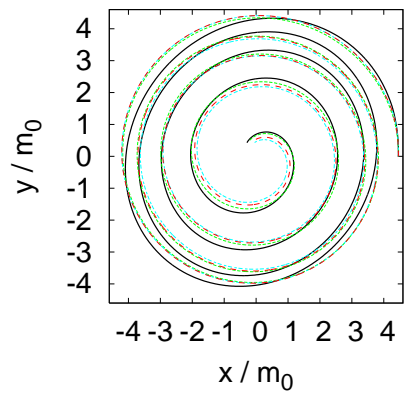} 
\vspace{-2mm}
\caption{The same as Fig. \ref{FIG1}, but for runs 19a'(solid curve)
and 19aF' (dashed curve).  The trajectories for runs 19a and 19aF are
shown together by the dotted and dotted-dashed curves. 
\label{FIGAPP1}}
\end{figure}

\begin{figure}[thb]
\epsfxsize=3.2in
\leavevmode
\epsffile{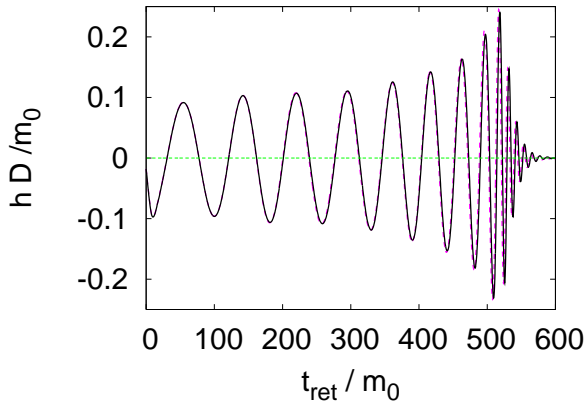} 
\vspace{-2mm}
\caption{Gravitational waveforms (plus mode) for runs 19a (dashed
curve) and 19a'(solid curve), respectively.
\label{FIGAPP2}}
\end{figure}

As we mentioned in Sec.~\ref{sec:res-bhbh}, coordinate trajectory
of BHs in inspiraling BH-BH binaries depends strongly on the initial
condition for $\beta^k$, although gravitational waveforms depend only
very weakly on it.  This appendix is denoted to a summary of the
results. Specifically, we performed two simulations for $d=19$ using
the solution of the quasiequilibrium condition for the gauge variables
as the initial condition. We prepared the same grid structure as that
of runs 19a and 19aF. The simulations were performed both in the
$\tilde \Gamma^i$- and $F_i$-BSSN formalisms. Hereafter these
simulations are referred to as runs 19a' and 19aF', respectively.

Figure \ref{FIGAPP1} plots the coordinate trajectories for runs 19a'
and 19aF' together with those for runs 19a and 19aF. We find that for
run 19a', the trajectory is highly elliptical and different from that
of run 19a.  By contrast, the trajectory of run 19aF' is approximately
the same as that of run 19aF. For runs 19a' and 19aF', the merger time
is slightly longer than that for runs 19a and 19aF. However, the
parameters of BHs finally formed depend very weakly on the initial
condition for the gauge variables (see Table \ref{BHBHRES2}).

Although the coordinate trajectory depends on the initial condition
for gauge variables, this is purely a gauge effect. One evidence is
found from the fact that the merger time for runs 19a and 19a' is
approximately the same (see Table \ref{BHBHRES2}). To further show 
the evidence for this, we generate Fig. \ref{FIGAPP2} which compares
gravitational waveforms of runs 19a and 19a'. This figure shows that
two waveforms agree approximately with each other besides a small
phase error in the final phase. This indicates that the waveform for
run 19a' is not as elliptic as the trajectory suggests, and hence, the
orbit is physically circular.


\begin{thebibliography}{99}
\bibitem{footend}
However, it is also possible that the previous result on the
disk mass might be rather inaccurate and better conservation of
angular momentum might be purely by chance. Also, worth noting is that
viscosity-induced dissipation does not change total angular
momentum. Thus, it is possible that viscosity associated with
numerical error in the previous simulation might be rather large.
\end{thebibliography}
\end{document}